\documentclass[twoside,12pt]{article}
\usepackage[pdftex]{graphicx}
\usepackage{dcolumn}  
\usepackage{bm}       
\usepackage{color}
\usepackage{amsmath,epsfig,cite}
\newcommand{\nonett}[9]
{\setlength{\unitlength}{0.65mm}
\begin{picture}(150.00,90.00)
\put(10.00,45.00){\vector(1,0){70.00}}
\put(45.00,10.00){\vector(0,1){70.00}}
\put(110.00,45.00){\vector(1,0){30.00}}
\put(125.00,30.00){\vector(0,1){30.00}}
\put(85.50,42.50){\makebox(5.00,5.00){$I_3$}}
\put(145.50,42.50){\makebox(5.00,5.00){$I_3$}}
\put(127.50,60.00){\makebox(5.00,5.00){$S$}}
\put(47.50,80.00){\makebox(5.00,5.00){$S$}}
\put(45.00,45.00){\circle*{2.00}}
\put(70.00,45.00){\circle*{2.00}}
\put(20.00,45.00){\circle*{2.00}}
\put(32.50,70.00){\circle*{2.00}}
\put(57.50,70.00){\circle*{2.00}}
\put(32.50,20.00){\circle*{2.00}}
\put(57.50,20.00){\circle*{2.00}}
\put(125.00,45.00){\circle*{2.00}}
\put(45.00,45.00){\circle{0.00}}
\put(45.00,45.00){\circle{5.00}}
\put(32.50,70.00){\line(1,0){25.00}}
\put(57.50,70.00){\line(1,-2){12.50}}
\put(70.00,45.00){\line(-1,-2){12.50}}
\put(57.50,20.00){\line(-1,0){25.00}}
\put(32.50,20.00){\line(-1,2){12.50}}
\put(20.00,45.00){\line(1,2){12.50}}
\put(60.00,70.00){\makebox(15.00,5.00)[l]{#2}}
\put(15.00,70.00){\makebox(15.00,5.00)[r]{#1}}
\put(15.00,15.00){\makebox(15.00,5.00)[r]{#3}}
\put(6.75,37.50){\makebox(13.25,5.00)[r]{#5}}
\put(60.00,15.00){\makebox(15.00,5.00)[l]{#4}}
\put(70.00,37.50){\makebox(13.75,5.00)[l]{#6}}
\put(47.50,37.50){\makebox(12.50,5.00)[l]{#7}}
\put(47.50,47.50){\makebox(12.50,5.00)[l]{#8}}
\put(127.50,47.50){\makebox(12.50,5.00)[l]{#9}}
\end{picture}
}
\begin{document}
%
%
%
%
\title{The Experimental Status of Glueballs}
%
%
%
%
\author{V.\ Crede\,$^{1}$ and C.\ A.\ Meyer\,$^{2}$\\
$^{1}$\,Florida State University, Tallahassee, FL 32306 USA\\
$^{2}$\,Carnegie Mellon University, Pittsburgh, PA 15213 USA} 
\maketitle
\begin{abstract}
Glueballs and other resonances with large gluonic components are predicted as bound
states by Quantum Chromodynamics (QCD). The lightest (scalar) glueball is estimated
to have a mass in the range from 1 to 2 GeV/$c^2$; a pseudoscalar and tensor glueball 
are expected at higher masses. Many different experiments exploiting a large variety 
of production mechanisms have presented results in recent years on light mesons with 
$J^{PC} = 0^{++}$, $0^{-+}$, and $2^{++}$ quantum numbers. This review looks at the 
experimental status of glueballs. Good evidence exists for a scalar glueball which 
is mixed with nearby mesons, but a full understanding is still missing. Evidence for
tensor and pseudoscalar glueballs are weak at best. Theoretical expectations of 
phenomenological models and QCD on the lattice are briefly discussed.
\end{abstract} 
\clearpage
\tableofcontents
\clearpage
\section{Introduction}
While we believe that Quantum Chromodynamics (QCD) is the correct description
of the interactions of quarks and gluons, it is a theory that is very
difficult to solve in the low-energy regime -- that which describes the 
particles of which the universe is made. This is changing with advances
that have been made in Lattice QCD, and the access to ever faster computers.
Within QCD, one of the perplexing issues has been the existence of gluonic 
excitations. In the meson sector, nearly all the observed states can be 
explained as simple $q\bar{q}$ systems, with the naive quark model both 
providing a very good explanation for these particles, as well as providing 
a nice framework in which they can be described. 

However, both phenomenological models and lattice calculations predict that
there should exist additional particles in which the gluons themselves can
contribute to the quantum numbers of the states. These include the pure-glue 
objects known as glueballs as well as $q\bar{q}$ states with explicit glue, 
known as hybrid mesons. Some of these latter states are expected to have 
quantum numbers which are forbidden to $q\bar{q}$ systems -- exotic quantum 
numbers which can provide a unique signature for the existence of such particles. 

Over the last decade, a great deal of new experimental data on mesons has 
been collected. This new information bears directly on both the search for, 
and our current understanding of gluonic excitations of mesons, in particular 
glueballs and hybrids. In this review, we will focus on glueballs, rather than 
gluonic excitations in general. This paper will review the new data, and present 
a sampling of the phenomenological work that has been developed based on this. 

In order to be able to discuss glueballs, it is necessary to understand 
conventional quark-antiquark systems (mesons). Over the last decade, there 
have been several relevant reviews on this subject, all of which touch upon
gluonic excitations at some level. A very nice review by Close~\cite{0034-4885-51-6-002} 
looked at \emph{gluonic hadrons} and summarized much of the criteria used to 
classify glueballs. Much of the new data on mesons came from the Crystal Barrel 
experiment. An excellent review by Amsler~\cite{Amsler:1997up} covers the results 
of that experiment. A later review by Godfrey and Napolitano~\cite{Godfrey:1998pd} 
reviewed the status of meson spectroscopy in light of many of the new results at 
the time. A more recent review by Amsler and T\"ornqvist~\cite{Amsler:2004ps} covers 
the status of the mesons beyond the naive quark model. In the pseudoscalar sector,
Masoni, Cicalo and Usai~\cite{Masoni:2006rz} have written an historically complete
review. Most recently, an encyclopedic review by Klempt and Zaitsev~\cite{Klempt:2007cp} 
covers in great detail both meson spectroscopy and gluonic excitations. There is also a
very recent review by Mathieu and colleagues~\cite{Mathieu:2008me} on the theoretical 
situation of glueballs. Finally, many  useful mini-reviews can be found in the 
\emph{Review of Particle Physics}~\cite{Amsler:2008zz}.

\section{Meson Spectroscopy}
\label{spectroscopy}
Before discussing the expectations for gluonic excitations, we will
briefly discuss the simple quark model picture for mesons. A meson 
consists of a $q\bar{q}$ system, which because it contains both a 
particle and an antiparticle, has intrinsic negative parity, $P=-1$.
The total parity of such a system is given as $P=-(-1)^{L}$, where 
$L$ is the orbital angular momentum in the $q\bar{q}$ system. Because
quarks have spin $\frac{1}{2}$, the total spin of such a system can 
be either $S=0$ or $S=1$, which leads to a total angular momentum
$J=L+S$, where the sum is made according to the rules of addition for
angular momentum. In addition to parity, there is also C-parity, or
charge conjugation, which for a $q\bar{q}$ system is $C=(-1)^{(L+S)}$.

The two lightest quarks also carry an additional quantum number: isospin.
Each has total isospin~$\frac{1}{2}$, with the $u$ quark being the 
$+\frac{1}{2}$ part and the $d$ quark being the $-\frac{1}{2}$ part of 
the doublet. If we form a meson out of only these, we can have $I=0$ or 
$I=1$. If one of the quarks is a strange quark, then $I=\frac{1}{2}$ and 
if both are strange, then $I=0$. For a $q\bar{q}$ system, we can define
an additional conserved quantum number, $G$-parity: $G=(-1)^{L+S+I}$.
Using these relationships to build up possible $J^{PC}$'s for $\bar{q}q$ 
mesons, we find the following quantum numbers are allowed:
\begin{equation} 
0^{-+},~0^{++},~1^{--},~1^{+-},~1^{--},~2^{--},~2^{-+},~2^{++},~3^{--},~3^{+-},~3^{--},~\cdots 
\end{equation}
and looking carefully at these, we find that there is a sequence
of $J^{PC}$'s which are not allowed for a simple $q\bar{q}$ system:
\begin{equation}
0^{--},~0^{+-},~1^{-+},~2^{+-},~3^{-+},~\cdots
\end{equation}
These latter quantum numbers are known as \emph{explicitly exotic} quantum 
numbers and, if observed, would correspond to something beyond the simple 
$q\bar{q}$ states of the quark model.

If we consider only the three lightest quarks, $u$, $d$ and $s$, then we can 
form nine $q\bar{q}$ combinations, all of which can have the same $S$, $L$ and 
$J$. We can represent these in spectroscopic notation, $^{2S+1}L_{J}$, or as states 
of total spin, parity and for the neutral states, charge conjugation: $J^{PC}$. 
Naively, these $q\bar{q}$~combinations would simply be a quark and an antiquark. 
However, those states consisting of the same quark and antiquark ($u\bar{u}$, 
$d\bar{d}$ and $s\bar{s}$) are rotated into three other states based on isospin 
and SU(3)~symmetries. The combinations shown in equation~\ref{eq:mesons1} correspond 
the the non-zero isospin states, while those in equation~\ref{eq:mesons2} correspond 
to a pair of isospin zero states. The latter two states are also mixed by SU(3) to 
yield a singlet ($\mid 1\,\rangle$) and an octet ($\mid 8\,\rangle$) state:
\begin{eqnarray}
& \begin{array}{ccccc} 
        & (d\bar{s}) &            & (u\bar{s}) &         \\
(d\bar{u})  & & \frac{1}{\sqrt{2}}\,(d\bar{d}-u\bar{u}) & & (u\bar{d}) \\
        & (s\bar{d}) &       & (s\bar{u}) & \\ 
\end{array} & \label{eq:mesons1} \\[1ex]
& \begin{array}{cc}
  \mid 8\,\rangle\, = \, \frac{1}{\sqrt{6}}\,(u\bar{u}+d\bar{d}-2s\bar{s}) & 
  \mid 1\,\rangle\, = \,\frac{1}{\sqrt{3}}\,(u\bar{u}+d\bar{d}+s\bar{s})   \\
  \end{array} & 
\label{eq:mesons2}
\end{eqnarray}

\begin{figure}[t!]
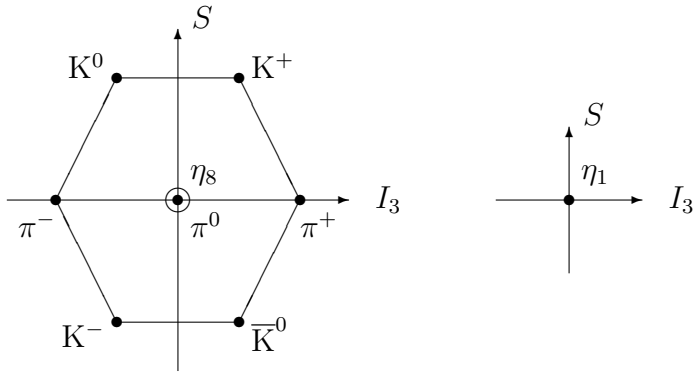

\begin{center}
\nonett{K$^0$}{K$^+$}{K$^-$}{$\overline{\mbox{K}}^0$}{$\pi^-$}
       {$\pi^+$}{$\pi^0$}{$\eta_8$}{$\eta_1$}
\end{center}
\vspace{-1cm}
\caption[]{\label{fig:nonet}SU(3)$_{flavor}$ nonet of the lightest pseudoscalar 
           mesons ($J^{PC}=0^{-+}$). The light~$u$,~$d$~and~$s$~quarks and 
           their corresponding antiquarks~$\bar{u}$,~$\bar{d}$ and $\bar{s}$ 
           form the basis for $9=3\otimes 3$~mesons. These are the illustrated 
           octet (left) and the $\eta_1$~singlet (right).}
\end{figure}

The nominal mapping of these states onto the familiar pseudoscalar mesons is shown 
in Fig.~\ref{fig:nonet}.
However, because SU(3) is broken, the two $I=0$ mesons in a given nonet are usually
admixtures of the singlet (\,$\mid 1\,\rangle=\frac{1}{\sqrt{3}}\,\left(u\bar{u}+d\bar{d}
+s\bar{s}\right)$) and octet (\,$\mid 8\,\rangle=\frac{1}{\sqrt{6}}\,\left(u\bar{u}+d\bar{d}
-2s\bar{s}\right)$) states. In nature, the physical states ($f$ and $f\,^{\prime}$\,) 
are mixtures, where the degree of mixing is given by an angle $\theta$:
\begin{eqnarray}
f  & = & \cos\theta\,\mid 1\,\rangle + \sin\theta\,\mid 8\,\rangle\\[1ex]
f\,^{\prime} & = & \cos\theta\,\mid 8\,\rangle - \sin\theta\,\mid 1\,\rangle\,.
\end{eqnarray}
For the vector mesons, $\omega$ and $\phi$, one state is nearly pure 
light-quark ($n\bar{n}$) and the other is nearly pure~$s\bar{s}$. This is known 
as ideal mixing and occurs when $\tan\theta= 1/\sqrt{2}$ ($\theta=35.3^{\circ}$).
In Table~\ref{tab:pdg2004} is listed our current picture of the ground state 
mesons for several different $L$\,'\,s. The last two columns list the linear 
(equation~\ref{eq:linear}) and quadratic (equation~\ref{eq:quadratic}) calculations 
of the mixing angle for the nonets:
\begin{eqnarray}
\label{eq:linear}
\tan\theta_{l} & = & \frac{4m_{K}-m_{a}-3m_{f\,^{\prime}}}{2\sqrt{2}(m_{a}-m_{K})}
\\[1ex]
\label{eq:quadratic} 
\tan^{2}\theta_{q} & = & \frac{4m_{K}-m_{a}-3m_{f\,^{\prime}}}{-4m_{K}+m_{a}+3m_{f}}\,.
\end{eqnarray}
The mixing angle $\theta$ can also be used to compute relative decay rates to 
final states such as pairs of pseudoscalar mesons, or two-photon widths, for
the $f$ and $f\,^{\prime}$ in a given nonet. Examples of this can be found in
reference~\cite{Amsler:quarkmodel:2006} and references therein. The key feature 
is that for a given nonet, the $f$ and $f\,^{\prime}$ states can be identified 
by looking at the relative decay rates to pairs of particles. 

\begin{figure}[tb!]\centering
\begin{tabular}{cc}
\includegraphics[width=0.47\textwidth]{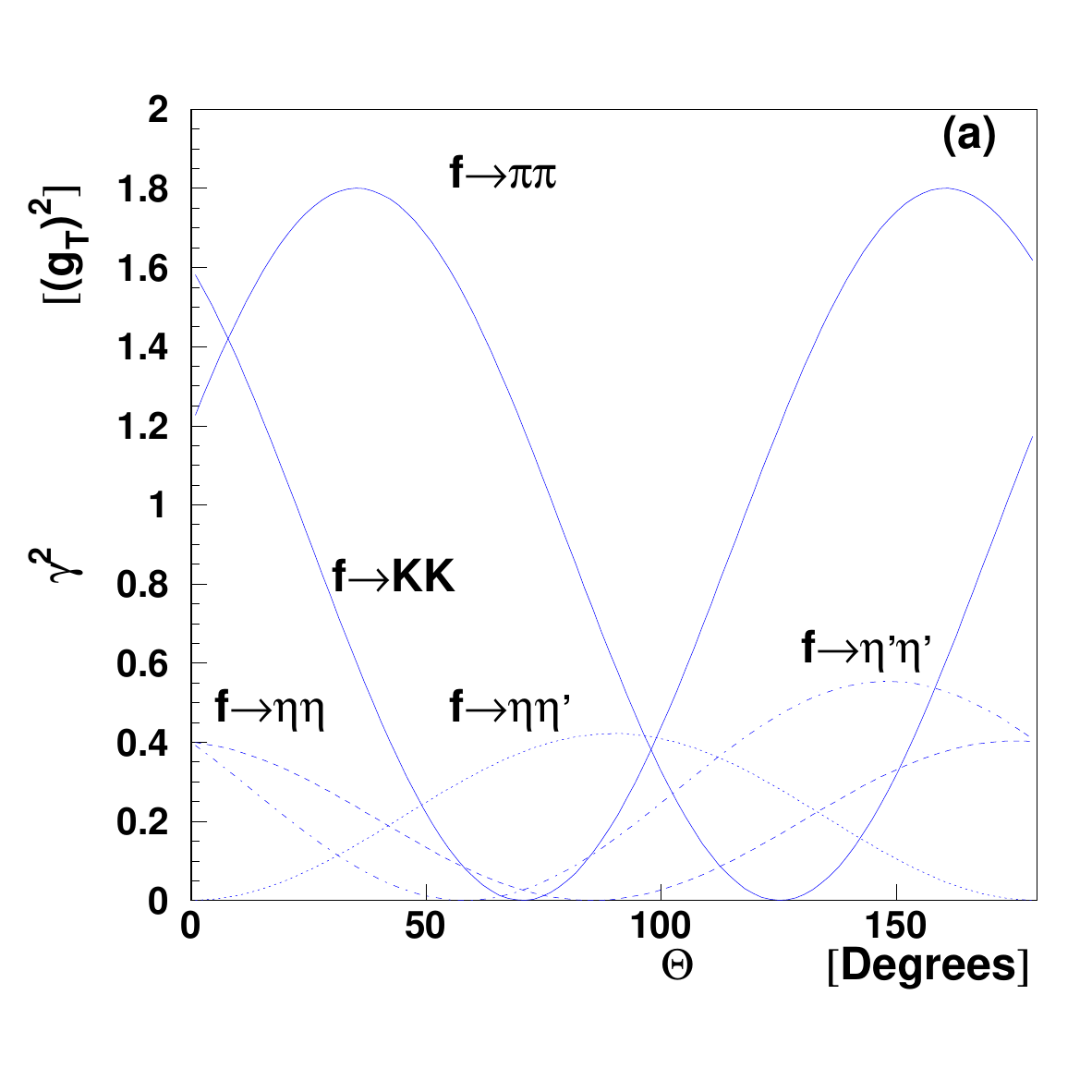} &
\includegraphics[width=0.47\textwidth]{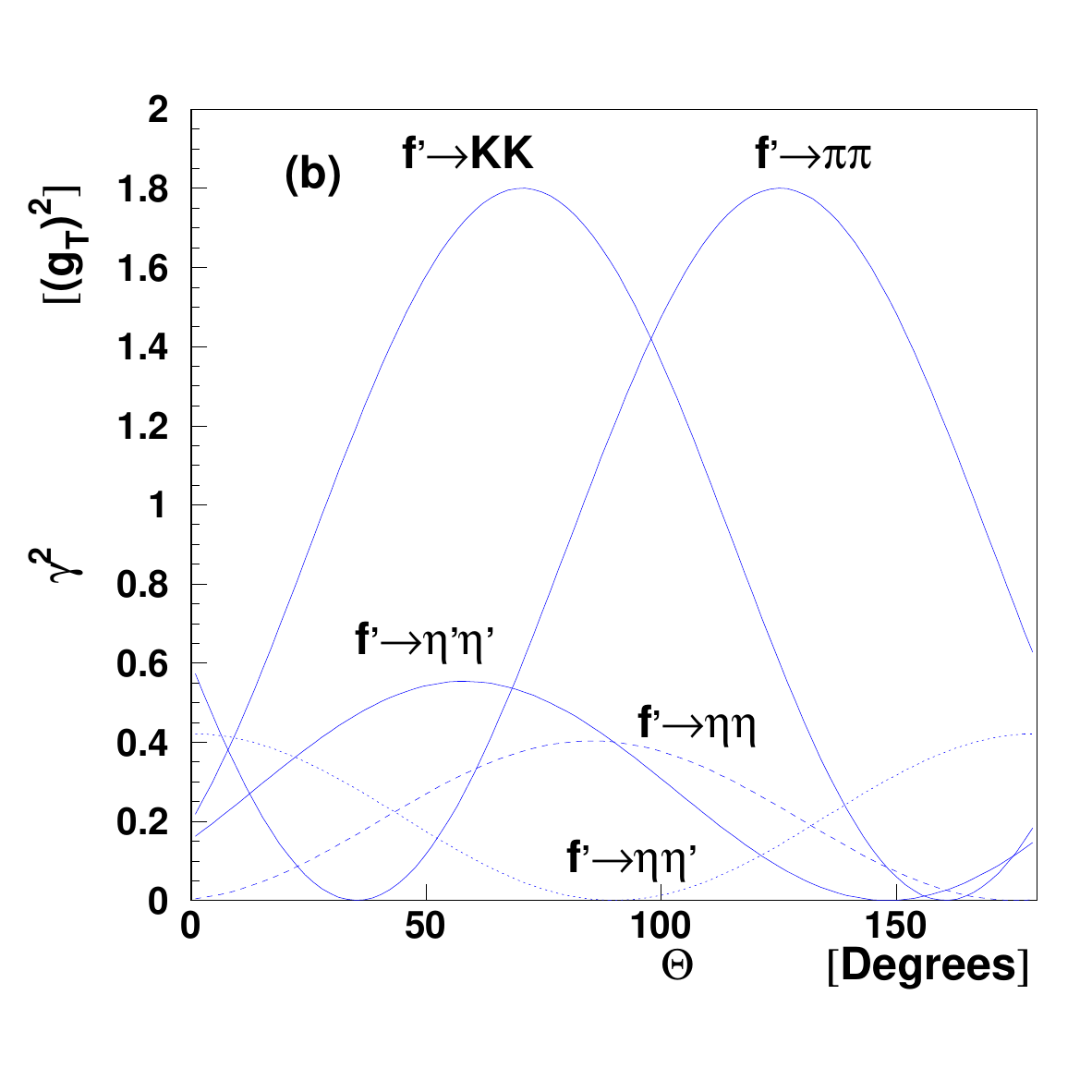}
\end{tabular}
\vspace{-8mm}
\caption[]{\label{fig:decayconstant}The decay amplitude, $\gamma^{2}$, as a 
function of the nonet mixing angle, $\theta$. (a) is for $f$ decays while (b) 
is for $f\,^{\prime}$ decays. In this particular example, the pseudoscalar mixing 
angle is taken as $\theta_{P}=-17^{\circ}$.}
\end{figure}

\begin{table}[h!]\centering
\addtolength{\extrarowheight}{5pt}
\begin{tabular}{|cc|cccc|rr|}\hline
$n^{2s+1}L_{J}$ & $J^{PC}$ & $I=1$ & $I=\frac{1}{2}$ & $I=0$ & $I=0$ 
& $\theta_{q}$ & $\theta_{l}$ \\
  & & $u\bar{d}\cdots$ &  $u\bar{s}\cdots$ &
  $f$ & $f\,^{\prime}$ & & \\ \hline
%
%
$1^{1}S_{0}$ & $0^{-+}$ & $\pi$ & $K$ & $\eta$ & $\eta\,^{\prime}$ & 
$-11.5^{\circ}$ & $-24.6^{\circ}$ \\
$1^{3}S_{1}$ & $1^{--}$ & $\rho$ & $K^{*}$ & $\omega$ & $\phi$ &
$38.7^{\circ}$ & $36.0^{\circ}$ \\
 & & & & & & & \\
$1^{1}P_{1}$ & $1^{+-}$ & $b_{1}(1235)$ & $K_{1B}$ & $h_{1}(1170)$ & $h_{1}(1380)$ & 
 & \\
$1^{3}P_{0}$ & $0^{++}$ & $a_{0}(1450)$ & $K^{*}_{0}(1430)$ & $f_{0}(1370)$ & 
$f_{0}(1710)$ &  & \\
$1^{3}P_{1}$ & $1^{++}$ & $a_{1}(1260)$ & $K_{1A}$          & $f_{1}(1285)$ & 
$f_{1}(1420)$ &  & \\
$1^{3}P_{2}$ & $2^{++}$ & $a_{2}(1320)$ & $K^{*}_{2}(1430)$ & $f_{2}(1270)$ & 
$f_{2}^{\,\prime}(1525)$ & $29.6^{\circ}$ & $28.0^{\circ}$ \\ 
 & & & & & & & \\
$1^{1}D_{2}$ & $2^{-+}$ & $\pi_{2}(1670)$   & $K_{2}(1770)$ & $\eta_{2}(1645)$ 
& $\eta_{2}(1870)$ &  & \\
$1^{3}D_{1}$ & $1^{--}$ & $\rho(1700)$      & $K^{*}(1680)$ & $\omega(1650)$ & 
&  & \\
$1^{3}D_{2}$ & $2^{--}$ &                   & $K_{2}(1820)$ & & &  & \\
$1^{3}D_{3}$ & $3^{--}$ & $\rho_{3}(1690)$ & $K^{*}_{3}(1780)$ & $\omega_{3}(1670)$ & 
$\phi_{3}^{\,\prime}(1850)$ & $32.0^{\circ}$ & $31.0^{\circ}$ \\ 
%
%
\hline
$1^{1}F_{4}$ & $4^{++}$ & $a_{4}(2040)$   & $K^{*}_{4}(2045)$ & $f_{4}(2050)$ 
&  &  & \\
$1^{3}G_{5}$ & $5^{--}$ & $\rho_{5}(2350)$      & & & 
&  & \\
$1^{3}H_{6}$ & $6^{++}$ & $a_{6}(2450)$   & & $f_{6}(2510)$ & &  & \\
%
%
\hline
%
$2^{1}S_{0}$ & $0^{-+}$ & $\pi(1300)$ & $K(1460)$ & $\eta(1295)$ & $\eta(1475)$ & 
$-22.4^{\circ}$ & $-22.6^{\circ}$ \\
$2^{3}S_{1}$ & $1^{--}$ & $\rho(1450)$ & $K^{*}(1410)$ & $\omega(1420)$ & $\phi(1680)$ &
&  \\
%
%
\hline 
\end{tabular}
\vspace{3mm}
\caption[]{\label{tab:pdg2004}A modified reproduction of the table from the 
2006 Particle Data Book~\cite{Amsler:quarkmodel:2006} showing the current 
assignment of known mesons to quark-model states. When sufficient states are 
known, the nonet mixing angle is computed using both the quadratic and linear 
forms.}
\end{table}

\begin{figure}[t!]\centering
\begin{tabular}{c}
\includegraphics[width=0.50\textwidth]{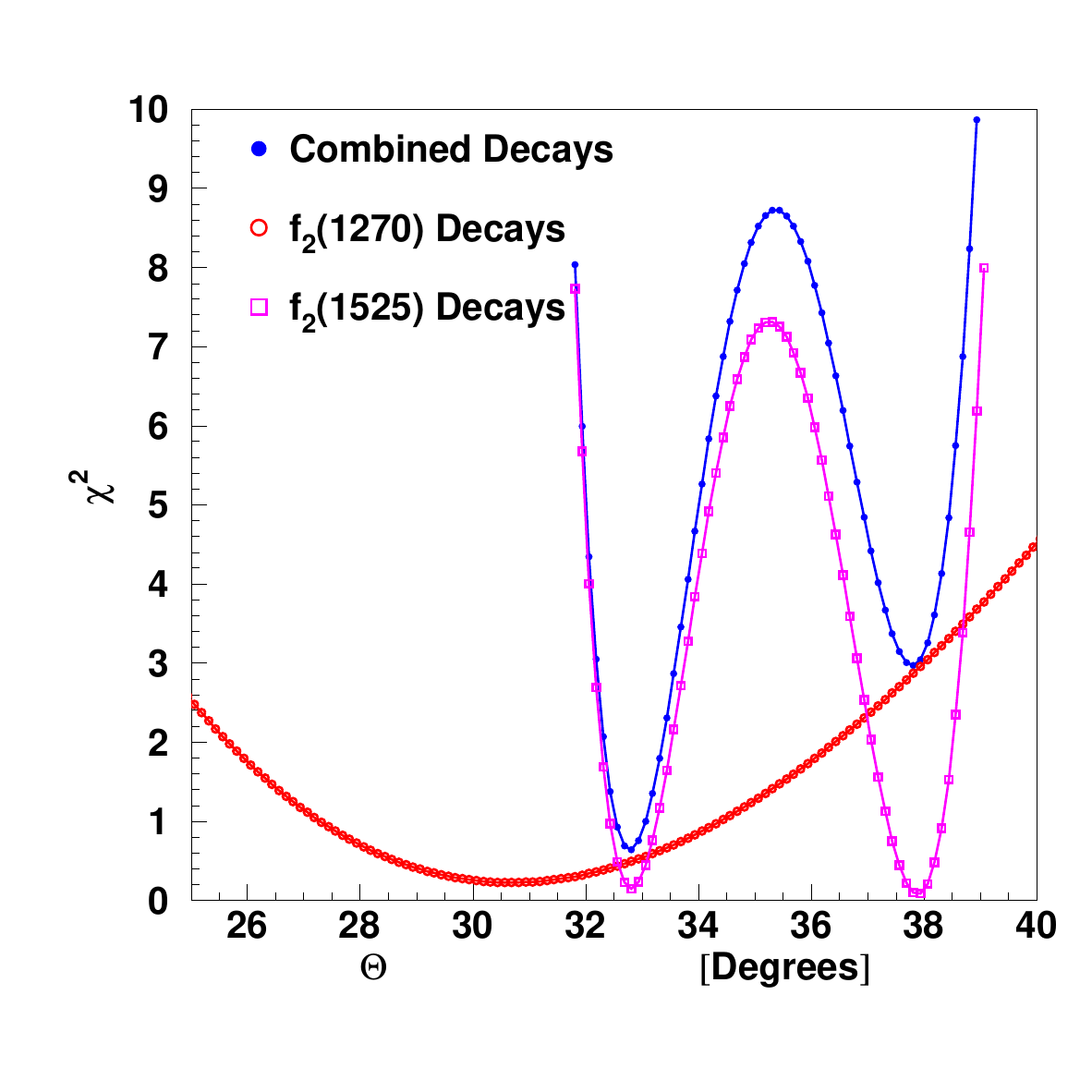}
\end{tabular}
\vspace{-7mm}
\caption[]{\label{fig:decayfits}(color online) The $\chi^{2}$ computed between the 
measured and predicted decay rates for the isoscalar tensor mesons. The open red 
circles are for the $f_{2}(1270)$ alone. The open purple squares are for the 
$f_2^{\,\prime}(1525)$ alone and the solid blue circles are for the combined fit. 
The former two are for one degree of freedom, while the latter is for three.}
\end{figure}

As an example of a decay calculation, we consider the decay of the tensor ($J^{PC}
= 2^{++}$) mesons to pairs of pseudoscalar mesons ($\pi\pi$, $K\bar{K}$ and $\eta\eta$). 
Following the work in references~\cite{Amsler:1995tu,Amsler:1995td,Amsler:quarkmodel:2006} 
and using the decay rates in reference~\cite{Amsler:2008zz}, we can compute a decay 
constant $\gamma$ from the SU(3) algebra corresponding to the decays. This can then 
be turned into a decay rate, $\Gamma$, as in equation~\ref{eq:decay}:
\begin{eqnarray}
\label{eq:decay}
\Gamma & = & \gamma^{2}\cdot f_{L}(q)\cdot q\,,
\end{eqnarray}
where $q$ is the break-up momentum of the meson into the pair of daughter mesons. 
The amplitude $\gamma$ depends on the nonet mixing angle and the pseudoscalar mixing 
angle, $\theta_{P}$. A typical example is given in Figure~\ref{fig:decayconstant} 
which is shown in terms of an arbitrary scale factor. The quantity $f_{L}$ is a form 
factor that depends on the angular momentum, $L$, between the pair of daughter mesons. 
The choice of the form factor is model dependent -- where monopole or dipole forms are 
usually taken. Here, we have taken a very model-dependent form as given in 
equation~\ref{eq:breakup} which is limited in validity to small values of $q$:
\begin{eqnarray}
\label{eq:breakup}
f_{L}(q) & = & q^{2L} e^{-\frac{q^{2}}{8\beta^{2}}}\,.
\end{eqnarray}
In this form, $\beta$ is a constant that is in the range of $0.4$ to $0.5$~GeV/$c$. 
One can fit the ratio of decay rates to pairs of mesons for both the $f_{2}(1270)$ and 
the $f_2^{\,\prime}(1525)$ and fit to the best value of the nonet mixing angle. We can 
compute a $\chi^{2}$ between the measured and predicted decay rates to determine what 
the optimal choice of the mixing angle is. This is shown in Figure~\ref{fig:decayfits},
where the optimal value is at about $32.5^{\circ}$. The location of the optimum does
not depend strongly on either $\theta_{P}$ or $\beta$ and is in good agreement with the 
values from the mass formulas for the tensors in Table~\ref{tab:pdg2004}. In fact, it 
is quite surprising how well this does in describing the data.

Measuring the masses and decay rates of mesons can be used to identify the quark 
content of a particular meson. The lightest glueballs have $J^{PC}$ quantum numbers 
of normal mesons and would appear as an SU(3) singlet state. If they are near a nonet 
of the same $J^{PC}$ quantum numbers, they will appear as an extra $f$-like state. 
While the fact that there is an extra state is suggestive, the decay rates and production 
mechanisms are also needed to unravel the quark content of the observed mesons.

\section{Theoretical Expectations for Glueballs}
\subsection{\emph{Historical}}

One of the earliest models in which glueball masses were computed is the bag
model~\cite{PhysRevD.9.3471}. In these early calculations, boundary conditions 
were placed on gluons confined inside the bag~\cite{Jaffe:1975fd}. The gluon 
fields could be in transverse electric (TE) or transverse magnetic modes (TM).
For a total angular momentum~$J$, the TE modes had parity of $(-1)^{(J+1)}$, 
while the TM modes had parity $(-1)^{J}$. The gluons then had to populate the
bag to yield a color singlet state. This led to predictions for two- and three-gluon
glueballs as given in Table~\ref{tab:bagmass}. 
\begin{table}[b!]\centering
\addtolength{\extrarowheight}{3pt}
\begin{tabular}{|c|rc|} \hline
Gluons     & $J^{PC}$ Quantum Numbers & Mass \\ \hline
(TE)$^{2}$ & $0^{++}$, $2^{++}$                     & $0.96$~GeV/$c^{2}$ \\
(TE)(TM)   & $0^{-+}$, $1^{-+}$, $2^{-+}$           & $1.29$~GeV/$c^{2}$ \\
(TE)$^{3}$ & $0^{+-}$, $1^{++}$, $2^{+-}$, $3^{++}$ & $1.46$~GeV/$c^{2}$ \\
(TM)$^{2}$ & $0^{++}$, $2^{++}$                     & $1.59$~GeV/$c^{2}$ \\
\hline
\end{tabular}
\caption[]{\label{tab:bagmass}Masses of glueballs in the bag model~\cite{Jaffe:1975fd}.}
\end{table}

Because glueballs contain no quarks, the expectation is that they would couple
to all flavors of quarks equally. In a simple SU(3) calculation, the $\gamma^{2}$ 
values from equation~\ref{eq:decay} are given in Table~\ref{tab:glueballdecay}. The 
most significant feature for pure glueball decays to a pair of pseudoscalar mesons 
is the vanishing branching ratio of the process $G\to\eta\eta\,^\prime$. However, 
a missing $\eta\eta\,^\prime$ mode can also be due to interference effects between 
different components in a $q\bar{q}$ 
state~\cite{Amsler:1995tu,Amsler:1995td,Close:2000yk,Close:2005vf}.

In addition to the decay predictions, there are reactions which are expected to 
be \emph{glue rich}.
\begin{table}[t!]\centering
\addtolength{\extrarowheight}{3pt}
\begin{tabular}{|c|ccccc|}\hline
Decay        & $\pi\pi$ & $K\bar{K}$ & $\eta\eta$ & $\eta\eta\,^{\prime}$ & 
               $\eta\,^{\prime}\eta\,^{\prime}$ \\
\hline
$\gamma^{2}$ & $3$ & $4$ & $1$ & $0$ & $1$ \\ \hline
\end{tabular}
\caption[]{\label{tab:glueballdecay}The expected decay amplitudes for a glueball 
normalized to the rate to $\eta\eta$.}
\end{table}
The first of these is radiative decays of the $J/\psi$. Because the $c\bar{c}$ quarks
have to decay via annihilation, the intermediate state must have gluons in it. This 
same argument can be applied to other $q\bar{q}$ reactions such as proton-antiproton 
annihilation and $\Upsilon$ decays. In $J/\psi$ decays, Chanowitz~\cite{Chanowitz:1984cb} 
proposed a variable known as \emph{stickiness} which is the relative rate of production 
of some hadron $h$ in radiative $J/\psi$ decays to its two-photon width (photons only 
couple to electric charge, hence to quarks). The stickiness, $S$, can be defined as in 
equation~\ref{eq:sticky} for a hadron $h$ of mass $M(h)$. The parameter $l$ is the lowest 
orbital angular momentum needed to couple to two vector particles and the photon energy 
$k_{\gamma}$ is that from the radiative decay in the $\psi$ rest frame. The overall 
constant $C$ is chosen such that $S\left[f_{2}(1270)\right]=1$:
\begin{eqnarray}
\label{eq:sticky}
S & = & C \left( \frac{M(h)}{k_{\gamma}}\right )^{2l+1} \frac{\Gamma(\psi\rightarrow\gamma h)}
{\Gamma(h\rightarrow\gamma\gamma)}\,.
\end{eqnarray}
The idea of stickiness is to measure the color-to-electric-charge ratio with phase 
space factored out, and to maximize the effects of the glue dominance inside a glueball. 
Thus, a glueball should have large stickiness in contrast to a conventional $q\bar{q}$ 
state. While this quantity has been computed for many of the glueball candidates, it 
appears to be limited by our detailed knowledge of the two-photon width of the 
states~\cite{Boglione:1997aw}. A recent analysis by Pennington~\cite{Pennington:2008xd}
has looked closely at the world data for this and still finds sizeable uncertainties
in these widths. 

Related to stickiness, Farrar~\cite{Cakir:1994jf} proposed a method of extracting
decay rate of hadrons to gluons based on the radiative decay rate of vector quarkonium
to the state. This was later applied to several mesons by Close~\cite{Close:1996yc}
to try and distinguish glueball candidates. 

There have also been questions raised about the validity of the flavor-blind decay
assumption of glueballs. Lee and Weingarten~\cite{Lee:1999kv} looked at decays of
glueballs and proposed that the decay rates should scale with the mass of the mesons,
thus favoring the heaviest possible meson pairs. Moreover, the idea of a chiral 
suppression mechanism for $J=0$ glueballs has first been pointed out by 
Carlson~\cite{Carlson:1980kh} and has recently been developed by 
Chanowitz~\cite{Chanowitz:2005du}. Glueballs with $J=0$ should have larger couplings 
to $K\bar{K}$ than to $\pi\pi$, for instance. This is due to the fact that in pQCD 
the amplitude is proportional to the current quark mass in the final state. For 
$J\neq 0$, the decay amplitude is flavor symmetric.

\subsection{\emph{Model Calculations}}
The first glueball mass calculation within the flux-tube model was carried out by 
Isgur and Paton~\cite{Isgur:1983wj,Isgur:1984bm}. In this model, the glueball is 
treated as a closed flux tube. Isgur and Paton found that the lightest glueball has 
$J^{PC}=0^{++}$ and a mass of $1.52$~GeV/$c^{2}$. In later flux-tube 
calculations~\cite{PhysRevD.68.074007}, predictions were made for the lightest three 
glueball masses. These were found to be consistent with the lattice calculations in 
reference~\cite{Morningstar:1997ff}. Table~\ref{tab:fluxtubeglueball} summarizes 
several flux-tube calculations of the scalar, tensor and pseudoscalar glueball. 
Finally, a recent article speculates that within the flux-tube model, the scalar and 
pseudoscalar glueball should be degenerate in mass~\cite{Faddeev:2003aw}.

\begin{table}[b!]\centering
\addtolength{\extrarowheight}{3pt}
\begin{tabular}{|lll|} \hline
\multicolumn{1}{|c}{$J^{PC}=0^{++}$} & \multicolumn{1}{c}{$J^{PC}=2^{++}$} & 
\multicolumn{1}{c|}{$J^{PC}=0^{-+}$} \\ \hline\hline  
$1.52$~GeV/$c^{2}$~\cite{Isgur:1983wj,Isgur:1984bm} &
$2.84$~GeV/$c^{2}$~\cite{Isgur:1983wj,Isgur:1984bm} &
$2.79$~GeV/$c^{2}$~\cite{Isgur:1983wj,Isgur:1984bm} \\
$1.68$~GeV/$c^{2}$~\cite{PhysRevD.68.074007} & 
$2.69$~GeV/$c^{2}$~\cite{PhysRevD.68.074007} & 
$2.57$~GeV/$c^{2}$~\cite{PhysRevD.68.074007} \\
$1.60$~GeV/$c^{2}$~\cite{PhysRevD.60.074024} & & \\
\hline\hline
\end{tabular}
\vspace{2mm}
\caption[]{\label{tab:fluxtubeglueball}Flux-tube predictions for masses of the 
lowest-lying glueballs.}
\end{table}

Another tool that has been used to predict glueball masses are QCD sum rules. This
method looks at a two-point correlator of appropriate field operators from QCD
and produces a sum rule by equating a dispersion relation for the correlator to 
an operator product expansion. One of the first such calculations for glueballs
was carried out by Novikov~\cite{Novikov:1979va}, and since then by many other authors.
Kisslinger~\cite{Kisslinger:2001pk} points out that most of these calculations find 
the lightest scalar glueball with a mass in the range of $0.3$ to $0.6$~GeV/$c^{2}$. 
A more recent article by Narison~\cite{Narison:2005wc} looks at all the known scalar 
mesons and finds a scalar meson state at about $1$~GeV/$c^{2}$ and the scalar glueball 
at $1.5\pm0.2$~GeV/$c^{2}$. In fact, the heavier glueball masses appear to come from
the ``two Laplace subtracted sum rule'', while the unsubtracted sum rules give
the lower masses. A discussion of this is found in reference~\cite{Mennessier:2008kk}. 



Swanson and Szczepaniak compute the glueball spectrum using Hamiltonian QCD in the 
Coulomb gauge~\cite{Szczepaniak:2003mr} by constructing a quasiparticle gluon basis. 
They also find results which are in good agreement with lattice calculations. The 
lightest glueball is a scalar with mass of $1.98$~GeV/$c^{2}$, followed by a 
pseudoscalar at $2.22$~GeV/$c^{2}$ and then a tensor at $2.42$~GeV/$c^{2}$. 

\subsection{\emph{Lattice Calculations}}
Lattice QCD discretizes space and time on a four-dimensional Euclidean lattice, 
and uses this to solve QCD numerically. This is done by looking at path integrals 
of the action on the discrete lattice. Quarks and antiquarks live on the discrete 
points of the lattice, while gluons span the links between the points. Depending 
on the problem being solved, and the availability of computing resources, the size 
of the lattice may vary. In addition, there are many different choices for the action, 
each of which has its own advantages and disadvantages. In specifying the calculation, 
the grid and action are specified and a damping factor, $\beta$, is chosen. After the 
calculation has been completed, the physical quantities of interest as well as the 
lattice spacing can be determined. For a given choice of action, a larger $\beta$ 
maps into a smaller lattice spacing. However, the same $\beta$ with different actions 
can lead to quite different lattice spacings. 

\begin{figure}[t!]\centering
\includegraphics[width=0.6\textwidth]{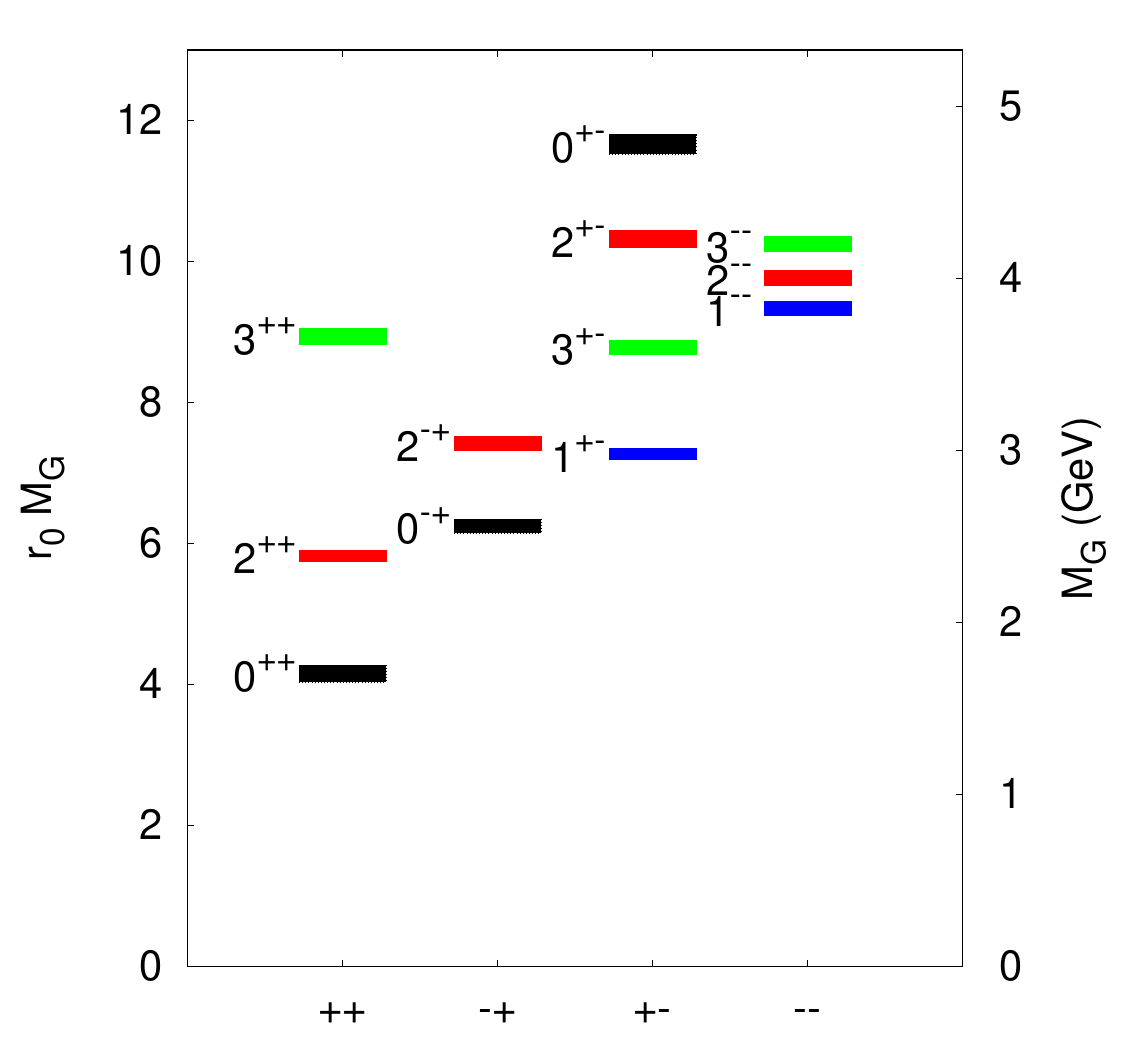}
\caption[]{\label{fig:glueballspectrum}
(color online) The mass spectrum of glueballs. The height of each box indicates the 
statistical uncertainty. Figure used with permission from reference~\cite{Chen:2005mg}.}
\end{figure}

To date, most lattice calculations have been carried out in the quenched approximation,
where the fluctuation of a gluon into a quark-antiquark pair is left out. As computer
power continues to increase, and more efficient ways of carrying out calculations 
evolve, this is starting to change. 

There is also a lattice artifact that can affect the mass calculations of the scalar
glueball~\cite{Heller:1995bz}. A singularity not related to QCD can cause the mass
of the scalar glueball to be artificially small. This effect is particularly apparent
when Wilson fermions are used with too-large a lattice spacing. Other choices are
less sensitive to this, and when the lattice spacing is small enough, the effect does
go away. However, for Wilson fermions, the critical value of $\beta$ is $5.7$, which 
is very close to the values used in many glueball calculations.

Some of the earliest lattice calculations of the glueball spectrum were carried out
in the quenched approximation on relatively small lattices~\cite{Michael:1987wf,Michael:1988jr}. 
These calculations indicated that the mass of the lightest glueball spectrum started 
at about $1.5$~GeV/$c^{2}$. As both computational resources increased and the lattice 
actions and methods improved, calculations on a larger lattice were carried out, and 
the spectrum of the states began to emerge~\cite{Bali:1993fb}. After extrapolating to 
the continuum limit, the lightest three states emerge as the scalar ($J^{PC}=0^{++}$), 
tensor ($J^{PC}=2^{++}$) and the pseudoscalar ($J^{PC}=0^{-+}$), with the scalar around 
$1.55\pm 0.05$~GeV/$c^{2}$, the tensor at $2.27\pm 0.1$~GeV/$c^{2}$ and the pseudoscalar 
at about the same mass. It was also possible to identify a number of other states with 
the first exotic (non-$q\bar{q}$) quantum number state above $3$~GeV/$c^{2}$.

\begin{table}[b!]\centering
\addtolength{\extrarowheight}{3pt}
\begin{tabular}{cc}\hline\hline
$J^{PC}$&  $M_{G}$~(GeV/c$^{2}$)      \\
\hline
$0^{++}$ &    $1.710(.050)(.080)$     \\
$2^{++}$ &    $2.390(.030)(.120)$     \\
$0^{-+}$ &    $2.560(.035)(.120)$     \\
$1^{+-}$ &    $2.980(.030)(.140)$     \\
$2^{-+}$ &    $3.040(.040)(.150)$     \\
$3^{+-}$ &    $3.600(.040)(.170)$     \\
$3^{++}$ &    $3.670(.050)(.180)$     \\
$1^{--}$ &    $3.830(.040)(.190)$     \\
$2^{--}$ &    $4.010(.045)(.200)$     \\
$3^{--}$ &    $4.200(.045)(.200)$     \\
$2^{+-}$ &    $4.230(.050)(.200)$     \\
$0^{+-}$ &    $4.780(.060)(.230)$     \\
\hline\hline
\end{tabular}
\vspace{2mm}
\caption{\label{tab:glueballmasses}  The glueball mass spectrum in physical
units. For the mass of the glueballs ($M_{G}$), the first error comes from 
the combined uncertainty of $r_0 M_G$, the second from the uncertainty of 
$r_0^{-1}=410(20)\,{\rm MeV}$. Data are taken from~\cite{Chen:2005mg}.} 
\end{table}

A later calculation using a larger lattice and smaller lattice parameters yielded 
a mass for the scalar glueball of $1.625\pm 0.094$~GeV/$c^{2}$~\cite{Sexton:1995kd,
Weingarten:1996pp}. The authors also calculated the decay of the scalar glueball 
to pairs of pseudoscalar mesons and estimated that the total width of the glueball 
would be under $0.2$~GeV/$c^{2}$. They also found that the decay width of the scalar 
glueball depended on the mass of the daughter mesons, with coupling increasing with 
mass. This was in contradiction to the lore that glueballs should decay in a 
flavor-blind fashion with the coupling to pairs of pseudoscalar mesons being independent 
of flavor or mass. Other work has followed this in discussions of violations of 
flavor-blind decays~\cite{Close:2001ga,Giacosa:2004ug}. This breaking is (effectively) 
accomplished by introducing a parameter, $r$, in the matrix that mixes quarkonium with 
glueballs. For flavor blind decays, $r=1$. Values that are close to $1$ are typically 
found. On the lattice~\cite{Sexton:1995kd,Weingarten:1996pp}, it is found that $r=1.2\pm 
0.07$, while a fit to data~\cite{Close:2001ga} finds $r=1\pm 0.3$. Finally, in a 
microscopic quark/gluon model~\cite{Giacosa:2004ug}, $r=1.1-1.2$ is determined. Taken 
together, one should probably expect small violations of flavor-blind decays for glueballs, 
but not large. 

Using an improved action, detailed calculations for the spectrum of glueballs were 
carried out by Morningstar and colleagues~\cite{Morningstar:1999rf,Chen:2005mg}. These
are shown in Figure~\ref{fig:glueballspectrum} and the corresponding masses reported
in Table~\ref{tab:glueballmasses}. These calculations are currently the state of the 
art in lattice glueball mass predictions.

Hart and Tepper~\cite{Hart:2001fp} carried out an unquenched calculation of the 
scalar and tensor glueballs using Wilson fermions. They found that the tensor
mass did not move, while the scalar mass came out at about 85\,\% of the unquenched 
mass. McNeile~\cite{McNeile:2007fu} speculates that this may be a result of the 
lattice artifact mentioned above. A later study by Hart and colleagues~\cite{Hart:2006ps}
seems to confirm that the mass of the glueball may be pushed down significantly 
for unquenched calculations. However, the authors point out that the effect may be
due to the lattice artifact, but feel that this is not the case. Another concern arises
due to the small pion mass which opens up the two-pion threshold. No explicit two-pion
operators were included in the calculation and it is possible that the mass reduction
is related to missing operators. In another unquenched calculation, Gregory~\cite{Gregory:2005yr} 
found that unquenching the lattice calculations for glueballs appears not to significantly 
alter the results from unquenched calculations. Given the difficulty of the unquenched
glueball calculation, we note that at this time it is difficult to know what the true 
situation is and look forward to better calculations in the future.

\section{Experimental Methods and Major Experiments}
\begin{figure}[b]\centering
\begin{tabular}{c}
\includegraphics*[angle=-90,viewport=270 200 360 600,width=1.0\textwidth]{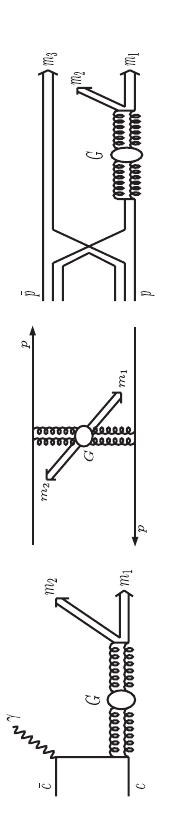}
\end{tabular}
\caption[]{\label{fig:gluerich} Feynman diagrams describing glue-rich production
mechanisms in favor of glueball formation: radiative $J/\psi$ decays, Pomeron-Pomeron
collisions in $pp$ central production, and proton-antiproton annihilation. Picture
taken from~\cite{Klempt:2007cp}.}
\end{figure}

Results on meson spectroscopy have come from a large variety of experiments
using different experimental techniques. Production of glueballs has mainly 
been predicted for glue-rich environments~\cite{Close:1987rj}. The most promising 
examples are proton-antiproton annihilation, central $pp$ collisions through 
double-Pomeron exchange or radiative decays of quarkonia, where one of 
the three gluons arising from the quark-antiquark annihilation is replaced by
a photon leaving two gluons to form bound states. The most prominent example is 
the radiative decay $J/\psi\to\gamma G$~\cite{Cakir:1994jf, Close:1996yc}. 
Fig.~\ref{fig:gluerich} shows some diagrams of gluon-rich processes. The 
following sections describe the main experimental methods and major experiments 
devoted to the study of meson resonances and the search for glueballs.

\subsection{\emph{Proton-Antiproton Annihilation}}
In $p\bar{p}$ annihilations, glueballs may be formed when quark-antiquark
pairs annihilate into gluons. Though not very likely, this may proceed via 
{\it formation} (as opposed to {\it production}) without a recoil particle; 
in this case, exotic quantum numbers are forbidden and the properties of the 
glueball candidate can be determined from the initial states. Instead, production 
of a heavier resonance recoiling against another meson is normally expected, 
but interaction of gluons forming glueballs sounds likely. However, the usually 
observed final states consisting of light $u$- and $d$-quarks can be just as 
effectively produced by quark rearrangements, i.e. without glueballs in the 
intermediate state. In $p\bar{p}$ annihilations at rest, mesons with masses 
up to 1.7 GeV/$c^2$ can be produced.

The $p\bar{p}$ annihilation at rest offers a natural way of limiting the
number of partial waves involved in the process facilitating spin-parity
analyses. At the Low-Energy Antiproton Ring (LEAR) at CERN, slow antiprotons 
of about 200 MeV/$c$ were decelerated in liquid or gaseous hydrogen (or 
deuterium) by ionizing hydrogen molecules. These eventually stop and are captured 
by protons forming hydrogen-like atoms called {\it protonium}. A highly excited 
$p\bar{p}$ state is formed by ejecting an electron via {\it Auger effect\/}:
\begin{center}
$\bar{p}$~+ H$_2$ $\to$~\framebox{$p\bar{p}$} + H + $e^-$
\end{center}
Annihilation takes place from atomic orbits. The capture of the $\bar{p}$
typically occurs at a principal quantum number of $n\approx 30$ and at a high 
angular momentum between the proton and the antiproton of $L\approx n/2$. 
For $n\approx 30$, the radius of the protonium atom matches the size of hydrogen 
atoms in their ground state. For lower $n$~values, the protonium radius becomes 
much smaller; the first Bohr radius of the $p\bar{p}$~atom is 57~fm. Due to 
this small size and due to the fact that protonium carries no charge, it can 
diffuse through hydrogen molecules. The $\bar{p}$ reaches an atomic state with 
angular momentum $L=0$ or $L=1$ when annihilation takes place. In media of 
high density like liquid H$_2$, the protonium is exposed to extremely large 
electromagnetic fields so that rotation invariance is broken. Transitions 
between different nearly mass-degenerate angular momentum states at the same 
high principal quantum number $n$ occur ({\it Stark mixing\/}). The effect is 
proportional to the target density and for liquid targets the rate is very high, 
such that about~90~\%~of all annihilations appear to be $S$-wave annihilations. 
For gaseous hydrogen targets, the $P$-wave contribution is much larger. The 
incoherent superposition of the $L=0$ or $L=1$ angular momentum eigenstates of the 
$p\bar{p}$~atom corresponds to six different partial waves: 
$^1S_0,~^3S_1,~^1P_1,~^3P_0,~^3P_1,~^3P_2$.

Proton and antiproton both carry isospin $|I,I_3\rangle = |\frac{1}{2},\pm\frac{1}{2}
\rangle$ and can couple to either $|I=0, I_3 = 0\rangle$ or $|I=1, I_3 = 0\rangle$. 
Given that initial state interactions due to $p\bar{p}\to n\bar{n}$ are 
small~\cite{Abele:2000xt}, we have
\begin{eqnarray}
\label{reaction:protonium}
p\bar{p} = \sqrt{\frac{1}{2}} (|I=1, I_3 = 0\rangle + |I=0, I_3 = 0\rangle). 
\end{eqnarray}

A large number of meson resonances was studied in $\bar{p}N$ annihilations
at rest and in flight. The most recent experiments were carried out using LEAR 
at CERN; the accelerator was turned off in 1996. The OBELIX collaboration 
at LEAR had a dedicated program on light-meson spectroscopy. The detector
system allowed operation of a variety of hydrogen targets: liquid H$_2$,
gaseous H$_2$ at normal temperature and pressure, and also a target at very 
low pressures. A special feature of the detector was the possibility to study
antineutron interactions, where the $\bar{n}$ beam was produced by charge
exchange in a liquid H$_2$~target. The Open Axial-Field Magnet of the experimental 
setup provided a magnetic field of 0.5~T. Particle detection proceeded via a 
Spiral Projection Chamber acting as vertex detector, a time-of-flight (TOF) 
system, a Jet Drift Chamber (JDC) for tracking and identification of charged 
particles by means of $dE$/$dx$, and a High-Angular Resolution Gamma Detector 
(HARGD), a system of four supermodules for the identification and measurement 
of neutral annihilation products. A detailed description of the experimental 
setup is given in~\cite{Adamo:1992bb}. The resolution of pions in the reaction 
$p\bar{p}\to\pi^+\pi^-$ was determined to 3.5\,\% at 928 MeV/$c$ and the mass 
resolution of $\pi^0$ mesons given by $\sigma = 10$ MeV/$c^2$. Relevant results 
are given in~\cite{Adamo:1992bb,Bertin:1995fx,Bertin:1996nq,Bertin:1998sb,
Cicalo:1999sn,Nichitiu:2002cj,Bargiotti:2003ev,Salvini:2004gz} and discussed
in section~\ref{section:obelix}.

\begin{figure}[tb!]
\begin{center}
\includegraphics*[viewport=0 25 450 265,width=1.0\textwidth]{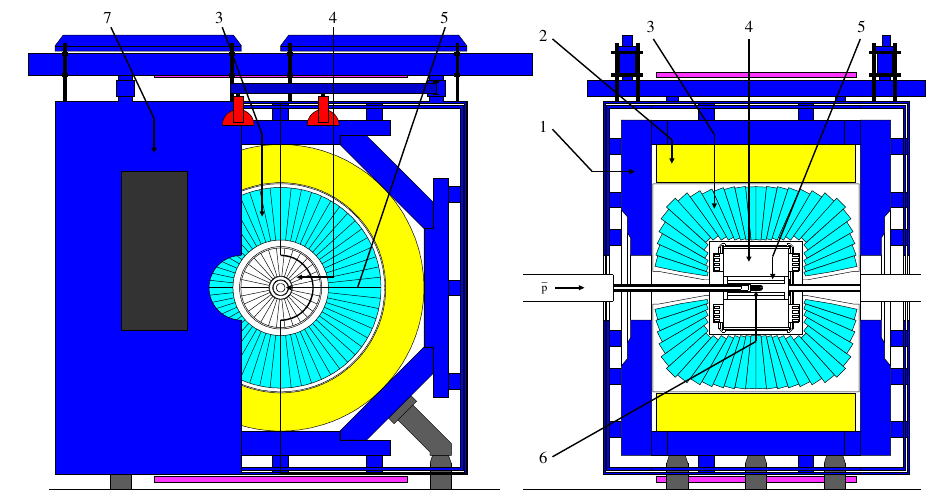}
\end{center}
\caption[]{\label{fig:cb_setup} Cross section of the Crystal-Barrel detector 
  at LEAR. From the outside: (1)~magnet yoke, (2)~magnet coils, (3)~CsI(Tl) 
  barrel calorimeter, (4)~jet drift chamber, (5)~proportional wire chamber, 
  (6)~target, (7)~one half of the endplate}
\end{figure}

Adjacent to OBELIX in the experimental hall, the Crystal Barrel spectrometer was 
operational from 1989-1996. The apparatus is described in~\cite{Aker:1992ny} and 
shown in Fig.~\ref{fig:cb_setup}. It could measure multi-meson final states including 
charged particles and photons from the decay of neutral mesons.
The detector consisted of two concentric cylindrical multi-wire proportional chambers 
(PWC), a jet drift chamber (JDC), and a barrel-shaped, modular electromagnetic CsI(Tl) 
calorimeter giving the detector system its name. The PWC was replaced by a silicon 
vertex detector in September 1995. The JDC had 30 sectors with each sector having 
23~sense wires and allowed the detection and identification of charged particles 
with a momentum resolution for pions of less than 2~\% at 200~MeV/$c$. Separation 
of $\pi/K$ below 500~MeV/$c$ proceeded via ionization sampling. The magnetic field 
of up to 1.5~T with a relative homogeneity of 2~\% in the region of the drift chamber 
was created by a conventional solenoid (B$_r \approx{\rm B}_\phi\approx 0$, B$_z\neq 0$) 
which was encased in a box-shaped flux return yoke. Data were taken on hydrogen and 
deuterium at rest~\cite{Amsler:1992rx,Amsler:1993pr,Amsler:1993xd,Amsler:1994ah,
Amsler:1994rv,Amsler:1995bf,Amsler:1995bz,Amsler:1995gf,Amsler:1995wz,Abele:1996fr,
Abele:1996nn,Abele:1997dz,Abele:1997qy,Abele:1998kv,Abele:1999en,Abele:1999fw,
Abele:2001js,Abele:2001pv,Amsler:2004kn} and at different incident beam 
momenta~\cite{Adomeit:1996nr,Abele:1999pf,Abele:2000qq,Seth:2000tm,Amsler:2002qq,
Amsler:2006du}. Results are presented in section~\ref{section:cb}.

The basic idea of the ASTERIX experiment was to measure $p\bar{p}$ annihilations
with and without X-rays from the $p\bar{p}$ cascade in coincidence. Antiprotons
were stopped in a cylindrical H$_2$ gas target at room temperature and pressure 
(45~cm in length, 14~cm in diameter). The detector consisted of an X-ray drift chamber
and seven multi-wire proportional chambers. Two end-cap detectors with three wire
planes and cathode readout on both sides gave large solid-angle coverage. The
detection volume was situated in a homogeneous axial magnet field of up to 0.82~T.
Pion and kaon separation was possible up to 400~MeV/$c$. The solid angle for charged
particles was 50\,\% of $4\pi$ with a momentum resolution of $\sigma = 4.2$\,\% at 
1~GeV/$c$. The detector is fully described in~\cite{Ahmad:1989gx}. Since the ASTERIX 
experiment was designed to stop antiprotons in a H$_2$ gas target, data were taken 
at the lowest available beam momenta: 300~MeV/$c$ in 1983, 200~MeV/$c$ in 1984, and 
105~MeV/$c$ in 1985 and 1986. Typical intensities used for data taking were $10^4 - 
10^5$ antiprotons per second. The experiment was dismantled and removed from the floor 
in 1986. The physics program on meson spectroscopy was then extended by the Crystal-Barrel 
and OBELIX experiments.
\subsection{\emph{e$^+$e$^-$ Annihilation Experiments and Radiative Decays of Quarkonia}}
\label{subsubsection:flavor-tagging}

The study of radiative decays of quarkonia is considered most suggestive in 
the glueball search. Most of the information in this field has centered on
$J/\psi$ decays; after photon emission, the $c\bar{c}$ annihilation can go 
through $C$-even $gg$ states, and hence may have a strong coupling to the
low-lying glueballs. Study of $J/\psi$ decays facilitates the search because 
the $D\bar{D}$ threshold is above the $J/\psi$ mass of 3097 MeV/$c^2$ and the 
OZI rule\footnote{The OZI rule states that decays corresponding to disconnected 
quark diagrams are forbidden.} suppresses decays of the $c\bar{c}$ system into 
light quarks. Radiative $\Upsilon(1S)$ decays are also supposed to be glue rich
and a corresponding list of two-body decay branching ratios for $\Upsilon(1S)$ 
is desirable. Results have been recently reported by the CLEO collaboration
\cite{Athar:2005nu,Besson:2005ud}. The search for glueballs in $\Upsilon(1S)$
decays is however challenging because the ratio
\begin{eqnarray}
  \label{eq:upsilon}
  \frac{\Gamma\,^{\rm rad}_\Upsilon}{\Gamma\,^{\rm tot}_\Upsilon}\,=\,
  \biggl(\frac{\alpha_s(\Upsilon)}{\alpha_s(J/\psi)} \biggr)^4\,
  \biggl( \frac{\Gamma_{\Upsilon\,\to\, e^+e^-}}{\Gamma\,^{\rm tot}_\Upsilon}
  \cdot\frac{\Gamma\,^{\rm tot}_{J/\psi}}{\Gamma_{J/\psi\,\to\, e^+e^-}}
  \cdot\frac{\Gamma\,^{\rm rad}_{J/\psi}}{\Gamma\,^{\rm tot}_{J/\psi}}\biggr) 
  \approx 0.013
\end{eqnarray}
is much smaller than for $J/\psi$ mesons and a naive suppression factor of 
$\sim 6$ is expected. Moreover, information on any possible $\chi_{cJ}$ hadronic 
decays provides valuable insight into possible glueball dynamics. 

$J/\psi$ decays can also be used to study the flavor content of mesons in the
so-called {\it flavor-tagging} approach. The reaction $J/\psi\to V\,X$ may serve
as an example, where $V$ is one of the light vector mesons $(\omega,\,\phi,\,\rho^0)$
and $X$ is the exclusive final state of interest. Since the $\omega$ and $\phi$
are ideally mixed, i.e. $|\omega\,\rangle = |u\bar{u} + d\bar{d}\,\rangle/\sqrt{2}$ 
and $|\phi\,\rangle = |s\bar{s}\,\rangle$, the flavor of $X$ can now be tagged based 
on the OZI rule as the structure of the light vector is known. If the $\phi(1020)$ is 
produced, for instance, then the reaction $J/\psi\to\phi\,X$ indicates those recoiling 
mesons $X$, which are produced through their $s\bar{s}$ component. For this reason, 
$J/\psi$ decays to $\omega f_2(1270)$ and $\phi f_{2}^{\,\prime}(1525)$ are clearly 
observed, but decays to $\omega f_{2}^{\,\prime}(1525)$ and $\phi f_2(1270)$ are 
missing. It is however pointed out in~\cite{Asner:2008nq} that the role played
by the empirical OZI rule has not been fully understood yet in the charmonium 
energy region and large OZI-violations have been found in some QCD-inspired 
calculations~\cite{Geiger:1992va,Lipkin:1996ny,Li:1996yn}. Nevertheless,
high-statistics flavor-tagging is a promising tool and has helped determine 
mixing angles as discussed in section~\ref{spectroscopy}. Similar decays of the 
$\psi\,^{\prime}$(3770), the first radial excitation of the $J/\psi$, provides 
access to mesons with even higher masses.

A {\it flavor filter} was proposed for $0^{++}$, $0^{-+}$, and $2^{++}$ states
in $J/\psi\to\gamma\,[\gamma\, V]$ decays with $V=\rho,\phi$~\cite{Close:2002sf}. 
Since radiative decays don't change the flavor structure, decays of resonances 
into $\gamma V$ can provide information of the flavor content of these states.

Radiative decays of $c\bar{c}$ states can best be studied in {\it formation}
at $e^+ e^-$ colliders via a virtual photon in the process
\begin{eqnarray}
\label{reaction:radiativeDecays}
e^+e^-\to\gamma^\ast\to c\bar{c}\,.
\end{eqnarray}
Only states with the quantum numbers of the photon ($J^P=1^-$) can be created
and the lowest-mass candidate is the $1^3S_{1}$\,$J/\psi$ state. 

\begin{figure}[t!]
\begin{center}
\includegraphics*[angle=-90,viewport=230 215 380 585,width=1.0\textwidth]{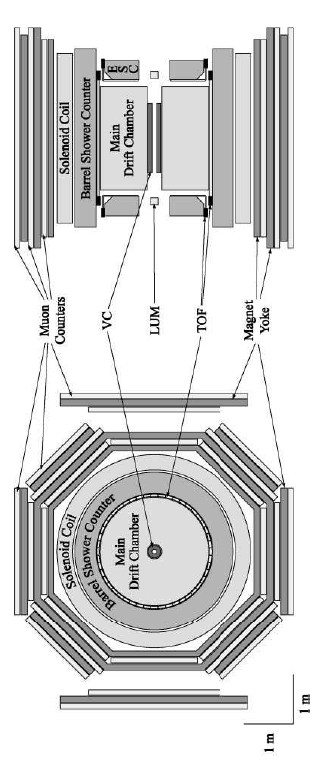}
\end{center}
\caption[]{\label{fig:bes_setup}End view (left) and side view (right) of the BES-II 
detector at IHEP, Beijing.}
\end{figure}

Several reactions have been studied in the BES experiment at the $e^+ e^-$ 
collider BEPC at IHEP, Beijing. Operation started in 1989 with a maximum collider 
energy of $2E = 4.4$~GeV and a luminosity of up to $10^{31}$/cm$^2$/s. A layout of
the BES-II detector is shown in Fig.~\ref{fig:bes_setup}. The detector is a large 
solid-angle magnetic spectrometer based on a conventional 0.4~T solenoidal 
magnet~\cite{Bai:2001dw}. It identifies charged particles using $dE$/$dx$ measurements 
in the drift chambers and time-of-flight measurements in a barrel-like array of 
48~scintillation counters. The barrel shower counter measures the energy of photons 
with a resolution of $\sigma_E/E = 28\,\%/\sqrt{E}$ ($E$ in GeV). More than 
$10^7~J/\psi$~events and more than $10^6~\psi\,^{\prime}(3770)$~events have been 
accumulated. The BES-III detector is a major upgrade aiming at $L = 10^{33}$/cm$^2$/s 
luminosity and recorded its first hadronic event in July 2008. Results relevant to
this review are given in~\cite{Bai:1996wm,Bai:1998tx,Bai:2000,Bai:2003,Ablikim:2004aa,
Ablikim:2004bb,Bai:2004qj,Ablikim:2005aa,Ablikim:2006aw,Ablikim:2006dw} and discussed
further in section~\ref{section:e+e-}.

\begin{figure}[tb!]
\begin{center}
\includegraphics*[viewport=0 10 600 500,width=0.7\textwidth]{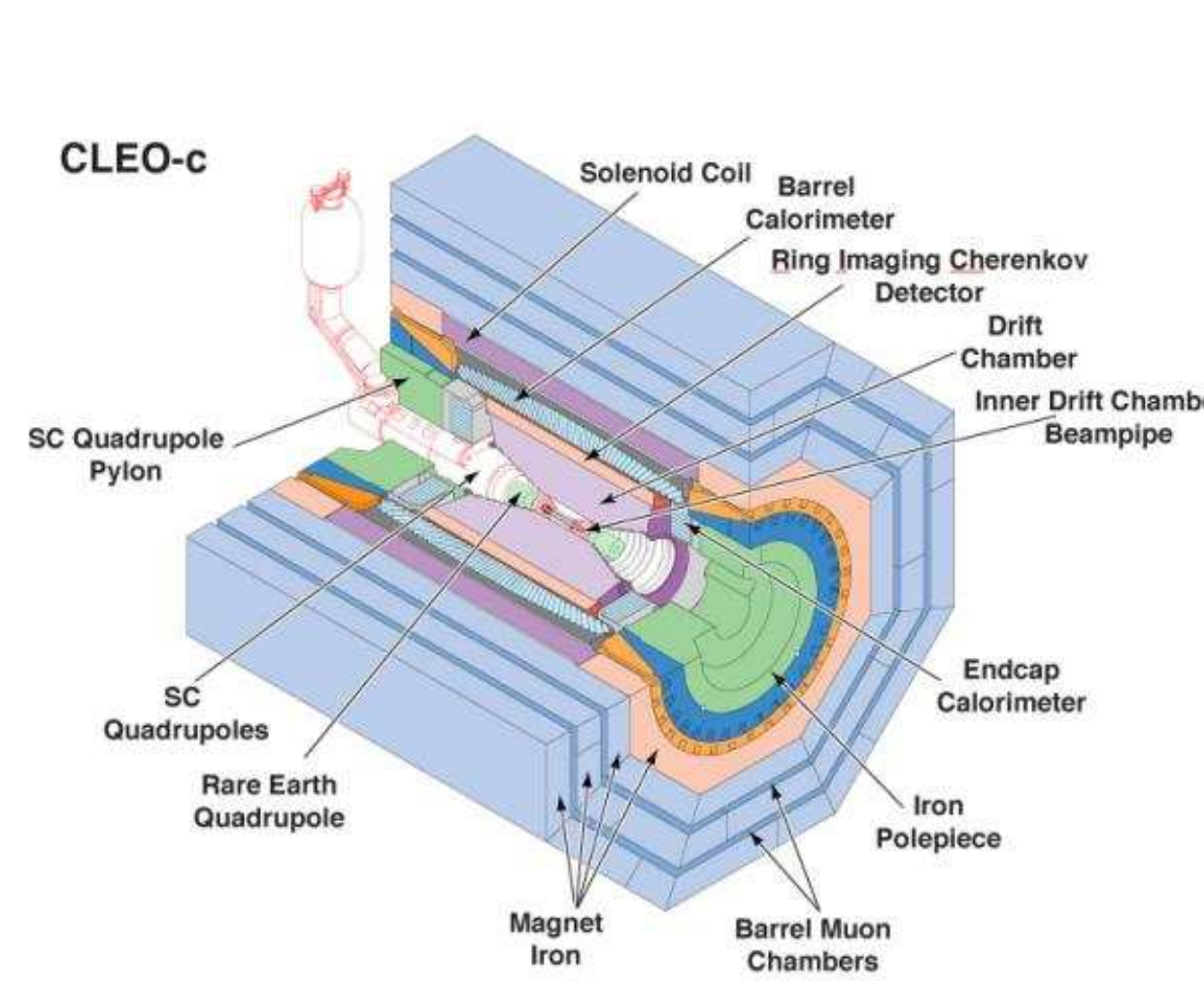}
\end{center}
\vspace{-1mm}
\caption[]{\label{fig:cleo_setup}The CLEO-c detector at the $e^+ e^-$ collider CESR, 
Cornell University.}
\end{figure}

Radiative decays of $c\bar{c}$ and $\Upsilon\,(b\bar{b})$ states have been studied 
with the CLEO detector at the $e^+ e^-$ collider CESR at Cornell University. 
Based on the CLEO-II detector, the CLEO-III detector started operation in
1999~\cite{Kopp:1996kg}. CLEO consisted of drift chambers for tracking and 
$dE$/$dx$ measurements and a CsI electromagnetic calorimeter based on
7800~modules inside a 1.5~T magnetic field. For CLEO-III, a silicon-strip 
vertex detector and a ring-imaging $\rm\check{C}$erenkov detector for particle 
identification were added. The integrated luminosity accumulated by the CLEO-III 
detector in 1999-2003 was 16~fb$^{-1}$. In 2003, CLEO was upgraded to CLEO-c in 
order to study charm physics at high luminosities. Fig.~\ref{fig:cleo_setup} shows
the detector setup. The CLEO-c operations ended in Spring 2008. The anticipated program 
of collecting data at $\sqrt{s}\sim 3.10$~GeV for the $J/\psi$ was given up due to 
technical difficulties in favor of a total of 572~pb$^{-1}$ on the $\psi\,^{\prime}(3770)$. 
Selected results of the CLEO collaboration can be found in~\cite{Kopp:2000gv,
Muramatsu:2002jp,Benslama:2002pa,Rubin:2004cq,Athar:2005nu,CroninHennessy:2005sy,
Besson:2005ud,Cawlfield:2006hm,Bonvicini:2007tc,:2008zi} and section~\ref{section:e+e-}.

The KLOE collaboration has studied radiative $\phi(1020)$ decays to $f_0(980)$
and $a_0(980)$ at the Frascati $\phi$~factory DAPHNE. These decays play an 
important role in the study of the controversial structure of the light scalar 
mesons. In particular, the ratio $\cal{B}$($\phi\to f_0(980)\gamma$) /
$\cal{B}$($\phi\to a_0(980)\gamma$) depends strongly on the structure of the 
scalars~\cite{Close:2001ay}. The KLOE detector consists of a cylindrical drift 
chamber, which is surrounded by an electromagnetic calorimeter and a 
superconducting solenoid providing a 0.52~T magnetic field. The energy 
resolution of the calorimeter is $\sigma_E/E = 5.7\,\%/\sqrt{E}$ ($E$ in GeV).
The detector accumulated $1.4\times 10^9$ $\phi$~decays in 2 years (2001-2002)
with a maximum luminosity of up to $7.5\times 10^{32}$/cm$^2$/s. KLOE resumed
data taking in 2004 with an upgraded machine. Relevant results are given
in~\cite{AloIsio2002,Aloisio2002b,Ambrosino:2005wk} and section~\ref{section:kloe}.

\subsection{\emph{Central Production}}
In contrast to {\it formation} experiments like $e^+e^-$ and $p\bar{p}$ 
annihilation discussed in the previous sections, the total energy in 
{\it production} experiments is shared among the recoiling particle(s) and 
the multi-meson final state. The creation of new particles is limited by
the center-of-mass energy of the reaction. The mass and quantum numbers of the 
final state cannot be determined from the initial state and thus, many resonant 
waves with different angular momenta can contribute to the reaction; resonances 
are not simply observed in the cross section.

In central production, it was suggested that glueballs would be produced 
copiously in the process~\cite{Robson:1977pm}
\begin{eqnarray}
\label{reaction:central}
{\rm hadron\,_{\rm beam}}\,p\to {\rm hadron\,_f}\,X\,{p_s}\,,
\end{eqnarray}
where the final-state hadrons carry large fractions of the initial-state
hadron momenta and are scattered diffractively into the forward direction.
To fulfill this requirement in a (proton) fixed-target experiment, a slow 
proton and a fast hadron needs to be observed in the final state. Mostly 
proton beams were used for these kinds of experiments, but some also involved 
pions or even kaons. The triggers in these experiments enhanced double-exchange 
processes -- Reggeon-Reggeon, Reggeon-Pomeron, or Pomeron-Pomeron -- relative 
to single-exchange and elastic processes. Early theoretical predictions 
suggested that the cross section for double-Pomeron exchange is constant with
center-of-mass energy $\sqrt{s}$, whereas a falling cross section is expected
for the other exchange mechanisms~\cite{Close:1987rj}:
\begin{eqnarray}
\label{reaction:double-exchange}
\sigma~({\rm Reggeon-Reggeon}) & \sim & 1/s\\
\sigma~({\rm Reggeon-Pomeron}) & \sim & 1/\sqrt{s}\\
\sigma~({\rm Pomeron-Pomeron}) & \sim & {\rm constant}\,.
\end{eqnarray}
At sufficiently high center-of-mass energies, reaction 
(\ref{reaction:central}) is expected to be dominated by double-Pomeron exchange.
The pomeron carries no charges -- neither electric nor color charges -- and
is expected to have positive parity and charge conjugation. Thus, double-Pomeron 
exchange should favor production of isoscalar particles with positive $G$-parity
in a glue-rich environment as no valence quarks are exchanged. 

\begin{figure}[tb!]
\begin{center}
\includegraphics*[viewport=20 55 800 600,width=0.9\textwidth]{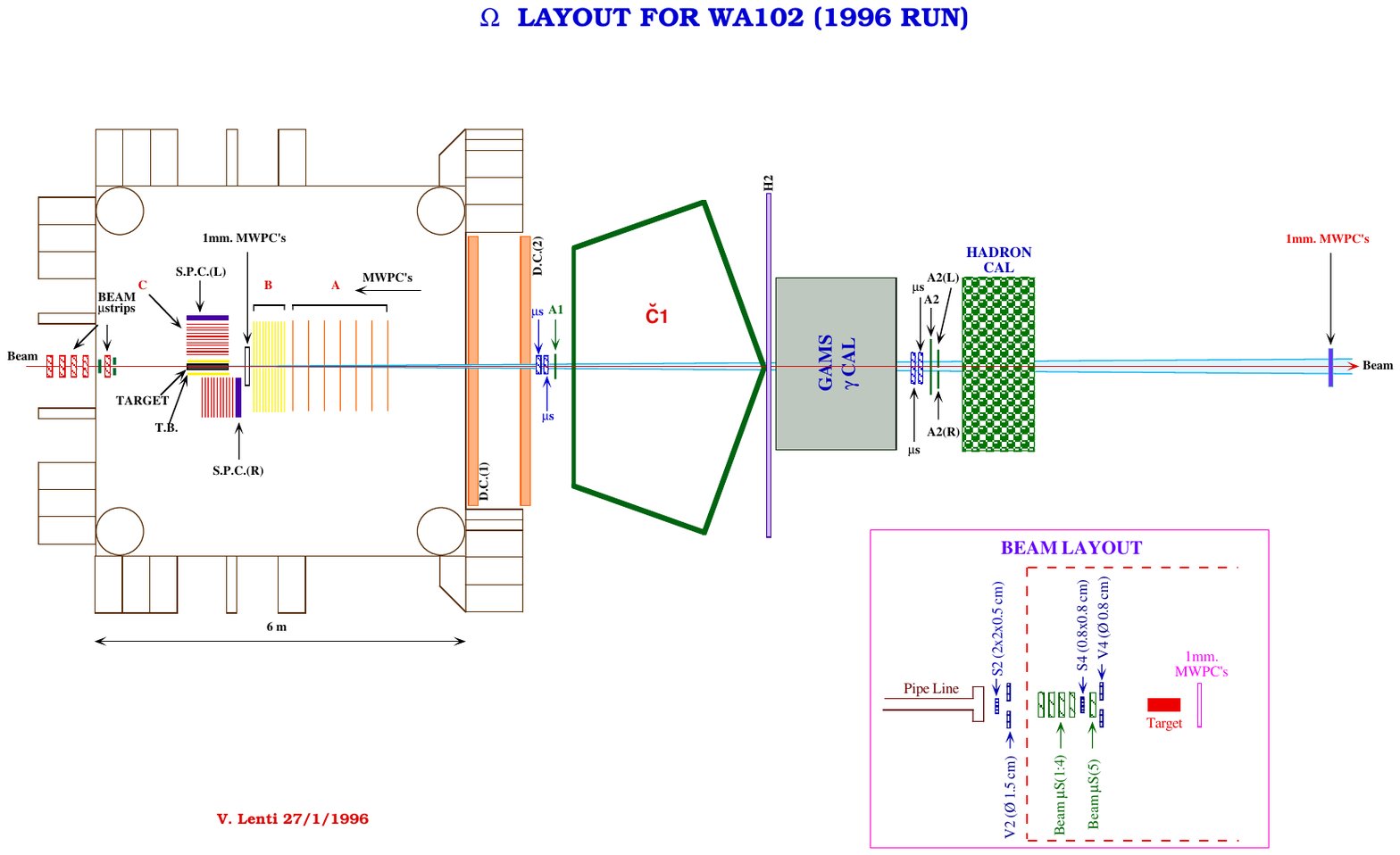}
\end{center}
\vspace{-2mm}
\caption[]{\label{fig:wa102_setup}WA102 experiment using the CERN
  $\Omega$ spectrometer and GAMS-4000 (1996 run)}
\end{figure}

The observation in central production of a significant enhancement of glueball 
candidates over the production of conventional $q\bar{q}$ mesons at small transverse 
momenta led to the idea of a {\it glueball filter}~\cite{Kirk:1999av}. No dynamical 
explanation has been given for this empirical finding, yet. In fact, the data 
collected by the WA102 collaboration show a strong kinematical dependence at small 
transverse momenta for all mesons. Some authors have argued that the glueball filter 
may just be a momentum filter which supresses angular momentum and hence enhances
scalar production~\cite{Klempt:1998wr}.

Several experiments, especially at CERN, made remarkable contributions. The 
WA76 collaboration recorded data, mainly at 300 GeV/$c$, using a 60 cm long 
H$_2$ target. Multi-wire proportional chambers triggered on exactly one ``fast'' 
particle in the forward direction. Some results from WA76 can be found 
in~\cite{Armstrong:1991ch}.

The electromagnetic multi-photon spectrometer GAMS-2000 originally took data
at the IHEP proton synchrotron, Protvino, in a 38 GeV/$c$ negative-pion 
beam~\cite{Binon:1985ag}. The detector was later upgraded for experiments at 
the CERN SPS~\cite{Alde:1985rr}. GAMS-4000 used a 50 cm long liquid H$_2$ 
target and comprised a matrix of $64\times 64$ lead glass cells covering almost 
6 m$^2$.

Both the WA91 and WA102 collaborations reported strong kinematical dependences 
of central meson production. The layout of the WA102 experiment is shown in 
Fig.~\ref{fig:wa102_setup}. The experiment was a continuation of the WA76, WA91 
and NA12/2 experiments at CERN aiming at a more complete study of the mass region 
from 1.2 to 2.5~GeV/$c^2$. The experimental setup was a combination of the WA76 setup 
serving as a charged-particle tracker and the GAMS-4000 detector in the forward 
region providing the opportunity to detect and study events with charged and neutral 
particles. Results from WA102 are given in~\cite{Abatzis:1995xx,Barberis:1997ve,
Barberis:1997vf,Barberis:1998bq,Barberis:1998in,Barberis:1998tv,Barberis:1999wn,
Barberis:1999ap,Barberis:1999an,Barberis:1999am,Barberis:1999cq,Barberis:1999id,
Barberis:1999be,Barberis:2000em,Barberis:2000cd,Barberis:2000kc,Barberis:2000cx,
Barberis:2001bs} and discussed in more detail in section~\ref{section:wa102}.

\subsection{\emph{Two-Photon Fusion at \boldmath{$e^+e^-$} Colliders}}
In contrast to direct glueball signals, the corresponding absence of states or 
glueball candidates in certain reactions can be as informative. Models predict 
for example that the $\gamma\gamma$~coupling of non-$q\bar{q}$ mesons is small.
In particular, glueball production should be strongly suppressed in two-photon 
fusion because there is no valence charge to couple to photons. However, a small 
branching fraction for $X\to\gamma\gamma$ does not necessarily provide compelling 
evidence that $X$ is a glueball because of possible interference effects with mixed 
components. In~\cite{Close:1996yc} for example, the branching fractions for $f_0\to
\gamma\gamma$ were found to be $f_0(1710)\,:\,f_0(1500)\,:\,f_0(1370)\,\approx\, 
3\,:\,1\,:\,12$.

The collision of two photons can best be studied in {\it inelastic Bhabha scattering} 
at $e^+ e^-$~colliders via the the reaction $e^+e^-\to e^+e^-\,\gamma\gamma\to e^+e^-
\,X$~\cite{Cooper:1988km}. This is analogous to the central production process in 
which the photons replace the less-well-understood Pomeron. In $e^+e^-$~annihilations, 
only final states with $J^{PC} = 1^{--}$ can be formed, whereas two-photon collisions 
provide access to most of the $C = +1$ mesons. 

In the two-photon process, each photon can be almost massless, though both can
combine to form a massive state $X$. The square of the mass of each photon is
determined by the scattering angle $\theta$ and the initial and final energies
$E$ and $E\,^\prime$ of its scattered lepton: 
\begin{eqnarray}
\label{eq:ggfusion1}
m^2_\gamma\equiv q^2\approx -2EE\,^\prime (1 - cos\,\theta)\,.
\end{eqnarray}
Spectroscopic data from two-photon collisions are generally separated into 
{\it tagged} and {\it untagged} event samples. If both leptons scatter at very 
small angles, $\theta\approx 0$, then they remain within the beam pipe and most 
likely escape undetected. In such a {\it no-tag} experiment, the two-photon invariant 
mass $W$ comes from the electron-positron energy losses:
\begin{eqnarray}
\label{eq:ggfusion2}
W\approx\sqrt{(E_1 - E_1\,^\prime)(E_2 - E_2\,^\prime)}\,,
\end{eqnarray}
and one can safely assume that the exchanged photons are essentially ``real'' with
$q^2\approx 0$. The allowed quantum numbers for a system formed by two quasi-real 
photons can be found by considering gauge invariance and symmetry principles. Since 
photons are identical bosons, two of them must be in a $C = +1$ state. Moreover, Yang's 
theorem forbids any spin-1 state and odd-spin states with negative parity~\cite{Yang:1950rg}, 
leaving $J^{PC} = 0^{\pm +}, 2^{\pm +}, 3^{++}$, etc. for {\it untagged} events. However,
if the incident $e^+e^-$ pair is scattered through a large angle sufficient to be
{\it tagged} by the experimental setup, the exchanged photons will be virtual and 
will also have a significant component of longitudinal polarization. A factor $1/M^3$
in the cross section indicates that the formation of lighter mesons is favored over 
heavier ones with the same $\Gamma_{\gamma\gamma}$.

In addition to the {\it stickiness} (as defined in Eq.~\ref{eq:sticky}), a further 
quantitative test was proposed whether a meson state is a glueball or a conventional 
$q\bar{q}$ meson. In \cite{Close:1996yc}, the normalized quantity {\it gluiness} 
denotes the ratio of the two-gluon to the two-photon coupling of a particle and is 
expected to be near unity for a $q\bar{q}$ meson within the accuracy of the 
approximations made in~\cite{Close:1996yc}:
\begin{eqnarray}
\label{eq:gluiness}
G & = & \frac{9e_Q^4}{2}\,\biggl( \frac{\alpha}{\alpha_s} \biggr)^2\,
\frac{\Gamma_{R\to gg}}{\Gamma_{R\to\gamma\gamma}}\,.
\end{eqnarray}
Two-photon fusion has been studied in {\it production} with the CLEO~II and
upgraded CLEO~II.V detectors at CESR using 13.3~fb$^{-1}$ of $e^+e^-$ data. 
The hadron is produced in the fusion of two space-like photons emitted by the 
beam electron and positron. Results of two-photon fusion studies from CLEO can 
be found in~\cite{Benslama:2002pa,Ahohe:2005ug} and also in 
sections~\ref{section:pseudoscalars}.

Further results on the $\gamma\gamma$-width of mesons have been reported by the 
LEP program at CERN. Though mainly focussing on electroweak physics, significant
results on meson spectroscopy were achieved. The $4\pi$ detector ALEPH was designed 
to give as much detailed information as possible about complex events in high-energy 
$e^+ e^-$ collisions~\cite{Decamp:1990jra}. A superconducting coil produced a uniform 
1.5-T field in the beam direction. Inside the coil, in order of increasing radius, 
there was a microstrip solid-state device, an Inner Tracking Chamber (ITC) using drift 
wires, a $3.6\times 4.4$~m Time Projection Chamber (TPC), and an electromagnetic 
calorimeter of 2~mm lead sheets with proportional wire sampling. A hadron calorimeter 
and a double layer of drift tubes aiding in good electron/muon identification were 
located outside the magnet coil. ALEPH data-taking ended on November 2000. The L3 detector 
was designed to measure the energy and position of leptons with the highest obtainable 
precision allowing a mass resolution of $\Delta m/m$ smaller than 2\,\% in dilepton final 
states~\cite{:1989kxa}. Hadronic energy flux was detected by a fine-grained calorimeter,
which also served as a muon filter and a tracking device. The outer boundary of the 
detector was given by the iron return-yoke of a conventional magnet. The field was 
0.5~T over a length of 12~m. Radially inwards was a combined hadron calorimeter and 
muon absorber. The electromagnetic energy flow was determined by approximately 11,000 
BGO crystals. Full electromagnetic shower containment over nearly $4\pi$ solid angle 
coverage was achieved. L3 data-taking also ended on November 2000. Important ALEPH and 
L3 results on the $\gamma\gamma$-width of mesons are published in~\cite{Barate:1999ze,
Acciarri:2000ex,Acciarri:2000ev} and presented in section~\ref{section:gg_fusion}.

\subsection{\emph{Other Experiments}}
Many more experiments have significantly contributed to meson spectroscopy 
though these were not particularly devoted to the glueball search. Pion and 
kaon beams were exploited in charge-exchange reactions:
\begin{eqnarray}
\label{reaction:exchange}
\pi^- (K^-) + {\rm proton} \to {\rm neutron} + {\rm meson}\,. 
\end{eqnarray}
In case of a kaon-induced reaction, the baryon in the final state can also
be a $\Lambda$. Reactions with positively-charged beams on proton targets
are also possible. These peripheral processes are viewed as exciting the ``beam''
particles by means of an exchange with the ``target'' particle, leaving the target
essentially unchanged. This is in contrast to ``central production'', which refers
to the case in which there is a collision between exchange particles. Peripheral
reactions are characterized by the square of the exchanged 4-momentum, $t^2\equiv
(p_{\rm beam} - p_X)^2 < 0$. These meson-nucleon reactions are strongly forward
peaked at high energies since the cross section is approximately exponentially
falling with $t$. The fairly-well understood $\pi^- p$ charge-exchange reaction at
small values of $-t$ provides access only to states with $J^{PC} = {\rm even}^{++}$
and ${\rm odd}^{--}$ states, so-called ``natural-parity'' states. Other states
such as $J^{PC} = 0^{-+}$ can for example be produced by neutral $J^{PC} = 0^{++}$ 
Pomeron exchange, which is not as well understood.

The LASS facility at SLAC was developed for strangeonium spectroscopy~\cite{Aston:1986jb}. 
This spectrometer was designed for charged-particle final states and based on a 
superconducting solenoid producing a 2.24~T field along the beam axis and a subsequent 
3~Tm dipole magnet with a vertical field. Particles under large scattering angles at 
low momenta and high-energy secondaries could be detected, respectively, with very 
good angular and momentum resolution. The LASS collaboration recorded over $135\cdot 
10^{6}$ kaon-induced events.

The VErtex Spectrometer (VES) setup at Protvino was a large-aperture magnetic 
spectrometer including systems of proportional and drift chambers, a multi-channel 
threshold \v{C}erenkov counter, beam-line \v{C}erenkov counters, a lead-glass 
$\gamma$-detector~(LGD) and a trigger hodoscope. This arrangement permits full 
identification of multi-particle final states. A negative particle beam $(\pi^-,\,K^-)$ 
with momenta between 20-40 GeV/$c$ was provided by the 70~GeV/$c$ proton synchrotron. 
A description of the setup can be found in~\cite{Bityukov:1991ri}.

Experiment E852 at Brookhaven National Laboratory was an experiment in meson 
spectroscopy configured to detect both neutral and charged final states of 
$18\,\mathrm{GeV/c}$ $\pi^{-}p$ collisions in a search for meson states incompatible 
with the constituent quark model. The E852 apparatus was located at the Multi-Particle 
Spectrometer facility (MPS) of Brookhaven's Alternating Gradient Synchrotron (AGS). 
The apparatus consisted of a fully instrumented beamline, a hydrogen target, a recoil
particle spectro\-meter, forward charged particle tracking, a large, segmented lead glass 
calorimeter and a nearly hermetic photon detection system. A flexible, programmable 
trigger allowed multiple final-state topologies to be collected simultaneously. A 
detailed description of the E852 apparatus is given in Reference~\cite{Teige:1996fi}.
Relevant results on glueballs are discussed in section \ref{section:e852}.

\subsubsection*{Light-Meson Spectroscopy at Heavy-Flavor Experiments}
Decays of $B$ mesons offer a wide phase space at the expense of small event numbers
per MeV/$c^2$. Among many other things, they provide interesting and surprising 
insight into the nature of lighter mesons. In 1999, two so-called $B$-factories 
started data-taking with the main goal to study time-dependent $CP$ asymmetries in 
the decay of these mesons: BaBar at the Stanford Linear Accelerator Center (SLAC) 
and Belle at the KEKB $e^+e^-$ collider in Tsukuba, Japan. The very high luminosities 
of the electron-positron colliders and the general-purpose character of the detectors 
make them suitable places also for the study of lighter mesons. Results on light 
scalar mesons from $B$-factories are discussed in section~\ref{section:b_factories}.

BaBar is a cylindrically-shaped detector~\cite{Aubert:2001tu} with the interaction 
region at its center. The 9~GeV electron beam of the PEP-II facility collides with a 
3.1~GeV positron beam to produce a center-of-mass energy of 10.58~GeV, corresponding to 
the $\Upsilon(4S)$ resonance. This highly unstable state decays almost instantly into 
two $B$~mesons; about $10^9$ $B$ mesons have been recorded. The momenta of charged particles 
are measured with a combination of a five-layer silicon vertex tracker and a 40-layer 
drift chamber in a solenoidal magnetic field of 1.5~T. The momentum resolution is about 
$\sigma_{p_t}\approx 0.5\,\%$ at $p_t = 1.0$~GeV/$c$. A detector of internally reflected 
\v{C}erenkov radiation is used for charged particle identification. The electromagnetic
calorimeter is a finely segmented array of CsI(Tl) crystals with energy resolution of 
$\sigma_E/E\approx 2.3\,\%\times E^{-1/4} + 1.9\,\%$, where the energy is in GeV. The 
instrumented flux return contains resistive plate chambers for muon and long-lived neutral 
hadron identification. Some selected results are given in~\cite{Aubert:2005wb}.

The Belle experiment operates at the KEKB $e^+e^-$ accelerator, the world's highest 
luminosity machine with a world record in luminosity of $1.7\times 10^{34}~{\rm cm}
^{-2}$s$^{-1}$ and with an integrated luminosity exceeding $700~{\rm fb}^{-1}$. The Belle
detector~\cite{Iijima:2000cq} is a large-solid-angle magnetic spectrometer based on 
a 1.5~T superconducting solenoid. Charged particle tracking is provided by a three-layer 
silicon vertex detector and a 50-layer central drift chamber (CDC) surrounding the 
interaction point. The charged particle acceptance covers laboratory polar angles between 
$\theta = 17^\circ$ and $150^\circ$, corresponding to about 92\,\% of the total solid 
angle in the center-of-mass system. The momentum resolution is determined from cosmic 
rays and $e^+e^-\to\mu^+\mu^-$ events to be $\sigma_{p_t}/p_t = (0.30/\beta\oplus 
0.1\,p_t)\,\%$, where $p_t$ is the transverse momentum in GeV/$c$. Charged hadron 
identification and pion/kaon separation is provided by $dE/dx$ measurements in the CDC, 
an array of 1188 aerogel \v{C}erenkov counters (ACC), and a barrel-like array of 128 
time-of-flight scintillation counters with rms resolution of 0.95 ps. Electromagnetic 
showering particles are detected in an array of 8736 CsI(Tl) crystals of projective 
geometry that covers the same solid angle as the charged particle system. The energy 
resolution of electromagnetic showers is $\sigma_E/E = (1.3\oplus 0.07/E\oplus 
0.8/E^{1/4})\,\%$ with $E$ in GeV. Results on light-meson spectroscopy relevant to this
review can be found in~\cite{Garmash:2004wa,Garmash:2006fh}.

\section{The Known Mesons}
\subsection{\emph{The Scalar, Pseudoscalar and Tensor Mesons}}

\begin{table}[bt!]\centering
\addtolength{\extrarowheight}{3pt}
\begin{tabular}{|l|ll|r|}\hline
Name & Mass [\,MeV/$c^2$\,] & Width [\,MeV/$c^2$\,] & Decays \\ \hline
$f_{0}(600)~\ast$  & $400-1200$  & $600-1000$ & $\pi\pi$, $\gamma\gamma$ \\
$f_{0}(980)~\ast$  & $980\pm 10$ & $40-100$ & $\pi\pi$, $K\bar{K}$, $\gamma\gamma$ \\
$f_{0}(1370)~\ast$ & $1200-1500$ & $200-500$  & 
                     $\pi\pi$, $\rho\rho$, $\sigma\sigma$, $\pi(1300)\pi$, $a_{1}\pi$,
                     $\eta\eta$, $K\bar{K}$ \\
$f_{0}(1500)~\ast$ & $1507\pm 5$ & $109\pm 7$ & 
                     $\pi\pi$, $\sigma\sigma$, $\rho\rho$, $\pi(1300)\pi$, $a_{1}\pi$,
                     $\eta\eta$, $\eta\eta\,^\prime$ \\
                   & & & $K\bar{K}$, $\gamma\gamma$ \\
$f_{0}(1710)~\ast$ & $1718\pm 6$ & $137\pm 8$ &
                     $\pi\pi$, $K\bar{K}$, $\eta\eta$, $\omega\omega$, $\gamma\gamma$ \\
$f_{0}(1790)$      & $$ & $$ & \\
$f_{0}(2020)$      & $1992\pm 16$ & $442\pm 60$ & 
                     $\rho\pi\pi$, $\pi\pi$, $\rho\rho$, $\omega\omega$, $\eta\eta$ \\
$f_{0}(2100)$      & $2103\pm 7$  & $206\pm 15$ & 
                     $\eta\pi\pi$, $\pi\pi$, $\pi\pi\pi\pi$, $\eta\eta$, $\eta\eta\,^\prime$ \\
$f_{0}(2200)$      & $2189\pm 13$ & $238\pm 50$ & 
                     $\pi\pi$, $K\bar{K}$, $\eta\eta$ \\
\hline
\end{tabular}
\caption[]{\label{tab:scalarmesons}The $I=0$, $J^{PC}=0^{++}$ mesons as listed by 
  the Particle Data Group~\cite{Amsler:2008zz}. Resonances marked with $\ast$ are 
  listed in the Meson Summary Table. The well-established $f_0(1500)$ is often 
  considered a candidate for the scalar glueball.}
\end{table}

If we focus on the expected three lightest-mass glueballs, $J^{PC}=0^{++}$, $2^{++}$, 
and $0^{-+}$, we note that these are all quantum numbers of normal $q\bar{q}$ mesons. 
As such, it is important to understand what the known spectra and multiplet assignments 
of these states are. As a starting point, we take the point of view of the Particle 
Data Group (PDG)~\cite{Amsler:2008zz}. The $I=0$ mesons are listed for $J^{PC} = 0^{++}$ 
in Table~\ref{tab:scalarmesons}, $J^{PC} = 2^{++}$ in Table~\ref{tab:tensormesons} and
$J^{PC} = 0^{-+}$ in Table~\ref{tab:pseudoscalarmesons}. In addition to the listings, 
the Particle Data Group also has mini-reviews on the three sectors: the scalar 
mesons~\cite{pdg_scalar1,pdg_scalar2,pdg_scalar3}, the pseudoscalars~\cite{pdg_pseudoscalar} 
and the tensors~\cite{pdg_tensors2}. The following sections describe the main experimental 
findings in the search for the lightest-mass glueballs. Results are not always consistent
among the different experiments, but we have to bear in mind that analyses are often
model-dependent. For low-statistics data sets, Breit-Wigner parametrizations are the
only option, whereas large data samples can be analyzed using a K-matrix approach. 
Interference effects are also neglected in many cases.

\begin{table}[tb!]\centering
\addtolength{\extrarowheight}{3pt}
\begin{tabular}{|l|ll|r|}\hline
Name & Mass [\,MeV/$c^2$\,] & Width [\,MeV/$c^2$\,] & Decays \\ \hline
$f_{2}(1270)~\ast$      & $1275.4\pm 1.1$ & $185.2^{+3.1}_{-2.5}$ &  
                          $\pi\pi$, $\pi\pi\pi\pi$, $K\bar{K}$, $\eta\eta$, $\gamma\gamma$ \\
$f_{2}(1430)$           & $~1430$ & $13-150$ & $K\bar{K}$, $\pi\pi$ \\
$f_{2}^{\,\prime}(1525)~\ast$ & $1525\pm 5$ & $73^{+6}_{-5}$ & 
                          $K\bar{K}$, $\eta\eta$, $\pi\pi$, $K\bar{K}^{*}+cc$\\
$f_{2}(1565)$           & $1546\pm 12$ & $126\pm 12$ & 
                          $\pi\pi$, $\rho\rho$, $\eta\eta$, $a_{2}\pi$, $\omega\omega$,
                          $K\bar{K}$, $\gamma\gamma$ \\
$f_{2}(1640)$           & $1638\pm 6$ & $99^{+28}_{-24}$ &
                          $\omega\omega$, $4\pi$,$K\bar{K}$ \\ 
$f_{2}(1810)$           & $1815\pm 12$ & $197\pm 22$ & 
                          $\pi\pi$, $\eta\eta$, $4\pi$, $K\bar{K}$ \\
$f_{2}(1910)$           & $1915\pm 7$ & $163\pm 50$ &
                          $\pi\pi$,$K\bar{K}$, $\eta\eta$, $\omega\omega$, $\eta\eta\,^{\prime}$,
                          $\eta\,^{\prime}\eta\,^{\prime}$, $\rho\rho$ \\ 
$f_{2}(1950)~\ast$      & $1944\pm 12$ & $472\pm 18$ & 
                          $K^{*}\bar{K}^{*}$, $\pi\pi$, $4\pi$, $a_{2}\pi$, $f_{2}\pi\pi$,
                          $\eta\eta$,$K\bar{K}$, $\gamma\gamma$ \\ 
$f_{2}(2010)~\ast$      & $2011^{+60}_{-80}$ & $202\pm 60$ &   
                          $\phi\phi$,$K\bar{K}$ \\
$f_{2}(2150)$           & $2156\pm 11$ & $167\pm 30$ &  
                          $\pi\pi$, $\eta\eta$, $K\bar{K}$, $f_{2}\eta$, $a_{2}\pi$ \\
$f_{2}(2300)~\ast$      & $2297\pm 28$ & $149\pm 40$ &   
                          $\phi\phi$, $K\bar{K}$, $\gamma\gamma$ \\
$f_{2}(2340)~\ast$      & $2339\pm 60$ & $319^{+80}_{-70}$ &
                          $\phi\phi$, $\eta\eta$ \\
\hline
\end{tabular}
\caption[]{\label{tab:tensormesons}The $I=0$, $J^{PC}=2^{++}$ mesons as listed by the 
Particle Data Group~\cite{Amsler:2008zz}. Resonances marked with $\ast$ are listed in 
the Meson Summary Table. A mini-review in the 2004 edition of the PDG discusses more
solid evidence for the $f_{2}(1565)$~\cite{pdg_tensors2}.}
\end{table}

\begin{table}[tb!]\centering
\addtolength{\extrarowheight}{3pt}
\begin{tabular}{|l|ll|r|}\hline
Name & Mass [\,MeV/$c^2$\,] & Width [\,MeV/$c^2$\,] & Decays \\ \hline
$\eta(548)~\ast$     & $547.51\pm 0.18$ & $1.30\pm .07$~keV & $\gamma\gamma$, $3\pi$ \\
$\eta\,^\prime(958)~\ast$ & $957.78\pm 0.14$  & $0.203\pm 0.016$ & 
                       $\eta\pi\pi$, $\rho\gamma$, $\omega\gamma$, $\gamma\gamma$ \\
$\eta(1295)~\ast$    & $1294\pm 4$ & $55\pm 5$ & 
                       $\eta\pi\pi$, $a_{0}\pi$, $\gamma\gamma$, $\eta\sigma$, $K\bar{K}\pi$ \\
$\eta(1405)~\ast$    & $1409.8\pm 2.5$ & $51.1\pm 3.4$ & 
                       $K\bar{K}\pi$, $\eta\pi\pi$, $a_{0}\pi$, $f_{0}\eta$, $4\pi$ \\
$\eta(1475)~\ast$    & $1476\pm 4$ & $87\pm 9$ &
                       $K\bar{K}\pi$, $K\bar{K}^{*}+cc$, $a_{0}\pi$, $\gamma\gamma$ \\
$\eta(1760)$         & $1760\pm 11$ & $60\pm 16$ & $\omega\omega$, $4\pi$ \\
$\eta(2225)$         & $2220\pm 18$ & $150^{+300}_{-60}\pm 60$ & $K\bar{K}K\bar{K}$ \\
\hline
\end{tabular}
\caption[]{\label{tab:pseudoscalarmesons}The $I=0$, $J^{PC}=0^{-+}$ mesons as listed by 
the Particle Data Group~\cite{Amsler:2008zz}. Resonances marked with $\ast$ are listed 
in the Meson Summary Table. The $\eta(1295)$ and the $\eta(1475)$ are often considered
radial excitations of the $\eta$ and $\eta\,^\prime$, respectively, leaving the $\eta(1405)$ 
as a potential glueball candidate.}
\end{table}
\subsection{\emph{Results from \boldmath{$p\bar{p}$} Annihilation: The Crystal Barrel Experiment}}
\label{section:cb}
The Crystal Barrel experiment~\cite{Aker:1992ny} studied $p\bar{p}$ annihilation both 
at rest and in flight and observed final states with multiple charged particles and
photons. In particular, many all-neutral final states were observed for the first time. 
The experiment accumulated about $10^8~p\bar{p}$ annihilations at rest in liquid hydrogen 
and thus, exceeded statistics collected in bubble-chamber experiments by about three 
orders of magnitude. At the beginning of the Crystal-Barrel data taking in late 1989, 
only three scalar states were well established: $a_0(980)$, $f_0(980)$, and the 
$K^\ast_0(1430)$. The high-statistics data sets collected at rest provided firm evidence 
for new states, among others the $f_0(1500)$ scalar state.

For annihilations at rest into three-pseudoscalar final states, Crystal Barrel studied 
the following reactions:
$p\bar{p}\rightarrow \pi^{0}\pi^{0}\pi^{0}$~\cite{Amsler:1995bf,Amsler:1995gf},
$p\bar{p}\rightarrow \pi^{0}\pi^{0}\eta$~\cite{Amsler:1995bf},
$p\bar{p}\rightarrow \pi^{0}\eta\eta$~\cite{Amsler:1992rx,Amsler:1995bf,Amsler:1995bz},
$p\bar{p}\rightarrow \pi^{0}\eta\eta\,^{\prime}$~\cite{Amsler:1994ah},
$p\bar{p}\rightarrow K^{0}_{L}K^{0}_{S}\,\pi^{0}$~\cite{Amsler:1993xd},
$p\bar{p}\rightarrow K^{0}_{L}K^{0}_{S}\eta$~\cite{Amsler:1993xd},
$p\bar{p}\rightarrow K^{+}K^{-}\pi^{0}$~\cite{Abele:1999en},
$p\bar{p}\rightarrow K_{L}K_{L}\pi^{0}$~\cite{Abele:1996nn},
$p\bar{p}\rightarrow \pi^{0}\pi^{0}\eta\,^{\prime}$~\cite{Abele:1997dz},
$p\bar{p}\rightarrow \eta\pi\pi$~\cite{Amsler:1995wz},
$n\bar{p}\rightarrow \pi^{-}\pi^{0}\pi^{0}$~\cite{Abele:1997qy},
$n\bar{p}\rightarrow \pi^{+}\pi^{-}\pi^{-}$~\cite{Abele:1999fw}. Studies were also
carried out for several four-pseudoscalar-meson final states, where 
$p\bar{p}\rightarrow \eta\pi^{0}\pi^{0}\pi^{0}$~\cite{Abele:1998kv} was studied, for instance. 
In addition, a number of five-pion final states were analyzed,
$p\bar{p}\rightarrow 5\pi$~\cite{Amsler:1994rv,Abele:1996fr,Abele:2001js,Abele:2001pv}, 
as well as the reactions 
$p\bar{p}\rightarrow \omega\pi^{0}\pi^{0}$~\cite{Amsler:1993pr} and
$n\bar{p}\rightarrow \omega\pi^{-}\pi^{0}$~\cite{Amsler:2004kn}.
Measurements were also made in flight at several different 
incident $\bar{p}$ momenta:
$p\bar{p}\rightarrow K\bar{K}\pi^{\circ}$~\cite{Amsler:2006du},
$p\bar{p}\rightarrow \eta\pi^{0}\pi^{0}$~\cite{Abele:1999pf},
$p\bar{p}\rightarrow \eta\pi^{0}\pi^{0}\pi^{0}$~\cite{Adomeit:1996nr},
$p\bar{p}\rightarrow \omega\pi^{\circ}$~\cite{Abele:2000qq},
$p\bar{p}\rightarrow \omega\eta$~\cite{Abele:2000qq},
$p\bar{p}\rightarrow \omega\eta\,^{\prime}$~\cite{Abele:2000qq},
$p\bar{p}\rightarrow \pi^{0}\pi^{0}\pi^{0}$~\cite{Amsler:2002qq},
$p\bar{p}\rightarrow \pi^{0}\pi^{0}\eta$~\cite{Amsler:2002qq},
$p\bar{p}\rightarrow \pi^{0}\eta\eta$~\cite{Amsler:2002qq},
$p\bar{p}\rightarrow K^{+}K^{-}\pi^{0}$~\cite{Amsler:2002qq}.

\subsubsection*{\emph{Results on Scalar States}}
The first experimental hint for an isoscalar state around 1500 MeV/$c^2$ came in 1973
from a low-statistics analysis of $p\bar{p}$ annihilations at rest into three pions
\cite{Devons:1973rs}. The state was later confirmed with a mass of $1527$~MeV/$c^2$
suggesting a spin-0 assignment and also reporting a missing $K\bar{K}$ decay mode
\cite{Gray:1983cw}. A very broad, somewhat higher-mass $S$-wave state called $G(1590)$
was reported in 38~GeV/$c$ pion-induced reactions by the GAMS-2000 collaboration 
decaying to $\eta\eta$~\cite{Binon:1983ny} and $\eta\eta\,^{\prime}$~\cite{Binon:1984ip}. 
The group reported that the decay rate into two neutral pions is at least three times 
lower than the rate into $\eta\eta$.
\begin{figure}[tb!]
\begin{center}
\includegraphics*[viewport=80 130 550 660,
                  width=0.9\textwidth,height=0.86\textwidth]{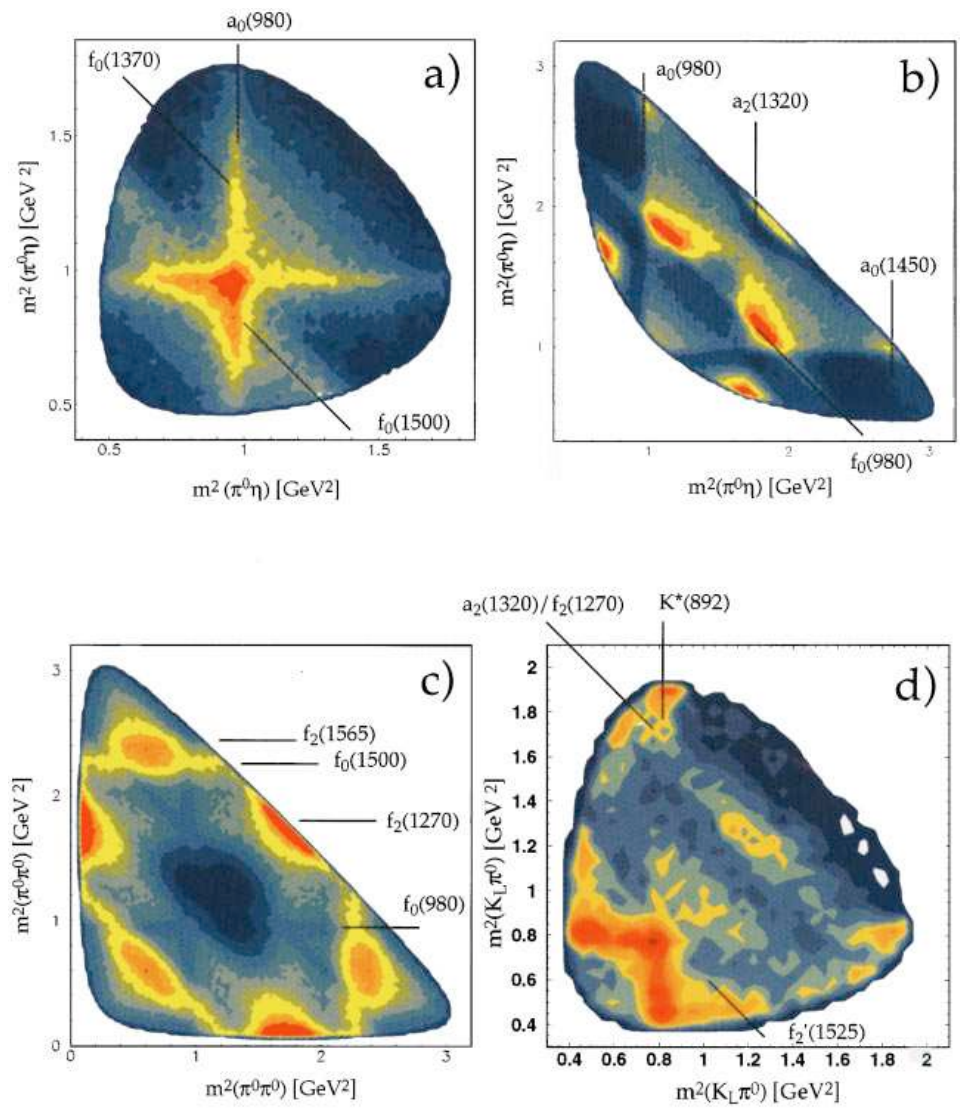}
\end{center}
\caption[]{\label{fig:cb} Dalitz plots for $p\bar{p}$ annihilation at rest from 
  Crystal Barrel into (a) $\pi^0\eta\eta$ ($\sim 2\times 10^5$ events), (b) $\pi^0
  \pi^0\eta$ ($\sim 2.8\times 10^5$ events), (c) $3\pi^0$ ($\sim 700,000$ events),
  (d) $\pi^0 K_L K_L$ ($\sim 37,000$ events). All events are entered more than once 
  for symmetry reasons. Figure taken from~\cite{Amsler:1997up}.}
\end{figure}

Crystal Barrel provided high-statistics data samples for final states with three light 
pseudoscalar mesons. In summary, a consistent description of all these data was achieved in a 
coupled-channel analysis by using four (isoscalar) scalar $\pi^0\pi^0$ waves: $f_0(980)$, 
$f_0(1370)$, $f_0(1500)$, and a broad structure $f_0(400-1200)$ -- listed as $f_0(600)$ 
by the PDG. In particular, the $3\pi^0$ and $\pi^0\eta\eta$ channels needed the two scalar 
states, $f_0(1370)$ and $f_0(1500)$, decaying to $\pi^0\pi^0$ and $\eta\eta$. Consistency 
in the description of the data sets further required two poles for the $\eta\pi^0$ $S$-wave
in annihilation into $\pi^0\pi^0\eta$, $a_0(980)$ and $a_0(1450)$, in addition to a 
tensor meson in the $\pi\pi$ $P$-wave, $f_2(1565)$~\cite{Amsler:1995bf}.  

Dalitz plots from various Crystal Barrel analyses are shown in Fig.~\ref{fig:cb} for 
$p\bar{p}$ annihilation into $\pi^0\eta\eta$~(a), $\pi^0\pi^0\eta$ (b), $3\pi^0$ (c),
and $\pi^0 K_L K_L$ (d). The most prominent features are labeled in the figure. The 
Dalitz plot for proton-antiproton annihilation into $\pi^0\eta\eta$ is dominated by the 
crossing vertical and horizontal bands for the isovector state $a_0(980)$ decaying into 
$\pi\eta$~(Fig.~\ref{fig:cb}~(a)). A good description of these data requires two isoscalar 
states, labeled $f_0(1370)$ and $f_0(1500)$, decaying to $\eta\eta$ in the $6\gamma$
final state~\cite{Amsler:1995bz}. In fact, an earlier analysis based on a reduced 
$\pi^0\eta\eta$ data set provided the first evidence for the $f_0(1370)$~\cite{Amsler:1992rx}. 
The observation of two necessary scalar states decaying to $\eta\eta$ is confirmed in an 
analysis of the $\pi^0\eta\eta\to 10\gamma$ final state, which exhibits entirely different 
systematics. It is now widely accepted that the $f_0(1500)$ observed by Crystal Barrel is 
identical to the $G(1590)$ observed by the GAMS collaboration~\cite{Binon:1983ny}. In the 
Crystal Barrel coupled-channel analysis, the  K-matrix mass and width of the $f_{0}(1500)$ 
come out to be $M\sim 1569$~MeV/$c^2$ and $\Gamma\sim 191$~MeV/$c^2$ which are quite 
similar to the reported $G(1590)$.

A narrow band of about constant intensity is observed in the $3\pi^0$ Dalitz plot 
(Fig.~\ref{fig:cb}~(c)) indicating the presence of the $f_0(1500)$. Further visible 
features include an increased population at the edges of the Dalitz plot along the 
$\pi\pi$ band marked $f_2(1270)$. This indicates that one decay $\pi^0$ is preferentially 
emitted along the flight direction of the resonance, which is typical of a spin-2 resonance 
decaying with an angular distribution of $(3\,{\rm cos}^2\,\theta - 1)^2$ from the 
$^1S_0$ initial state. Striking are the corner blobs, which follow a sin$^2\,\theta$ 
angular distribution and correspond to the $f_{2}^{\,\prime}(1525)$ interfering constructively 
with the two $\pi\pi$ $S$-waves. The fit also requires a small contribution from the 
$f_2(1565)$.
\begin{table}[t!]\centering
\addtolength{\extrarowheight}{3pt}
\begin{tabular}{|c|lcl|} \hline
Ratio & \multicolumn{1}{c}{$f_{0}(1370)$} & & \multicolumn{1}{c|}{$f_{0}(1500)$}\\\hline
$\cal{B}$$(K\bar{K})\,/\,$$\cal{B}$$(\pi\pi)$ & $(0.37\pm 0.16)$ to $(0.98\pm 0.42)$~\cite{Abele:2001pv} & & $^a\,0.186\pm 0.066$~\cite{Amsler:1995gf,Abele:1996nn}\\
                                      & & & $^b\,0.119\pm 0.032$~\cite{Amsler:1995bf}\\
$\cal{B}$$(\eta\eta)\,/\,$$\cal{B}$$(\pi\pi)$ & $0.020\pm 0.010$~\cite{Abele:2001pv} & & $^a\,0.226\pm 0.095$~\cite{Amsler:1995gf,Amsler:1995bz}\\
                                      & & & $^b\,0.157\pm 0.062$~\cite{Amsler:1995bf}\\
$\cal{B}$$(\eta\eta\,^{\prime})\,/\,$$\cal{B}$$(\pi\pi)$ & & & $^a\,0.066\pm 0.028$~\cite{Amsler:1995gf,Amsler:1994ah}\\
                                               & & & $^b\,0.042\pm 0.015$~\cite{Amsler:1995bf}\\\hline
$\cal{B}$$(\rho\rho)\,/\,$$\cal{B}$$(4\pi)$ & $0.260\pm 0.070$~\cite{Abele:2001js} & & $0.130\pm 0.080$~\cite{Abele:2001pv,Abele:2001js}\\
$\cal{B}$$(\sigma\sigma)\,/\,$$\cal{B}$$(4\pi)$ & $0.510\pm 0.090$~\cite{Abele:2001js} & & $0.260\pm 0.070$~\cite{Abele:2001js}\\
$\cal{B}$$(\rho\rho)\,/\,$$\cal{B}$$(2[\pi\pi]_S)$ & & & $0.500\pm 0.340$~\cite{Abele:2001js}\\
$\cal{B}$$(4\pi)\,/\,$$\cal{B}$$_{\rm tot}$ & $0.800\pm 0.050$~\cite{Gaspero:1992gu} & & $0.760\pm 0.080$~\cite{Abele:2001js}\\ \hline
\end{tabular}
\caption[]{\label{tab:cb_scalar-decays}A summary of Crystal-Barrel results on the decay
of scalar mesons. Branching ratios for decays into $4\pi$ are determined from $\bar{p}n$
annihilation. Results labeled $^a$ are from single channel analyses and $^b$ from a 
coupled channel analysis including $3\pi^0,~2\pi^0\eta$, and $\pi^0\eta\eta$.}
\end{table}
An important piece of information to clarify the internal structure of the $f_0(1500)$
is to study its $K\bar{K}$ decay mode. In fact, no strange decay was reported by a 
previous bubble-chamber experiment~\cite{Gray:1983cw}, which had however very limited statistics
and no partial wave analysis was performed. The Dalitz plot for the $\pi^0 K_L K_L$ channel 
at rest from Crystal Barrel is shown in Fig.~\ref{fig:cb}~(d). One $K_L$ was missing in 
the analysis and the other $K_L$ interacted hadronically in the CsI calorimeter. Events
with three clusters in the barrel were used for the analysis~\cite{Abele:1996nn}. The
contributions from the $f_0(1370)$ and $f_0(1500)$ were found to be small. The precise
determination is however challenging because the isovector state $a_0(1450)$ also decays 
to $K_L K_L$, which can form both an $I=0$ and $I=1$ system. Contributions from $a_0(1450)$
were thus determined from the $K_L K^\pm \pi^\mp$ final state by using isospin conservation
and the fact that no isoscalar $S$-wave contributes. 

The $4\pi$ decay modes of scalar mesons were studied at rest in proton-antiproton 
annihilation into $3\pi^0\pi^+\pi^-$~\cite{Amsler:1994rv} and $5\pi^0$~\cite{Abele:1996fr} 
as well as in antiproton-neutron annihilation into $4\pi^0\pi^-$~\cite{Abele:2001js,Abele:2001pv}
and also into $2\pi^0 2\pi^-\pi^+$~\cite{Abele:2001pv}. All data sets are dominated by $4\pi$ 
scalar isoscalar interactions and at least the two states, $f_0(1370)$ and $f_0(1500)$, are 
required in the analysis. It is observed that the $4\pi$-decay width of the $f_0(1370)$ is 
more than 6~times larger than the sum of all observed partial decay widths to two pseudoscalar 
mesons. This may indicate a dominant $n\bar{n}$ component over a $s\bar{s}$~structure. The 
$4\pi$-decays of the $f_0(1500)$ represent about half of its total width. The analyses also 
yield important couplings to $(\pi\pi)_S(\pi\pi)_S$ and to $\rho\rho$ 
(Table~\ref{tab:cb_scalar-decays}). It was pointed out in~\cite{StrohmeierPresicek:1998fi} 
that the $\rho\rho$ decay should dominate $2[\pi\pi]_S$ if the $f_0(1500)$ was a mixture of 
the ground state glueball with nearby $q\bar{q}$ states, at least in the framework of the 
$^3P_0$~$q\bar{q}$-pair creation model. In leading order of this scheme, the decay mechanism
of the $f_0(1500)$ proceeds dominantly via its quarkonia components. Unfortunately, results 
from Crystal Barrel and WA102 on $4\pi$~decays entirely disagree leaving the experimental 
situation unsettled~(Tables~\ref{tab:cb_scalar-decays}~and~\ref{tab:wa102_scalar-decays}). 
A possible source of this disagreement may be due to the different ways of treating the decay 
to a pair of broad resonances, $\rho\rho$ or $(\pi\pi)_{S}(\pi\pi)_{S}$, where the mass of the 
scalars are close to the ``nominal'' thresholds for $\rho\rho$. Additional work is required 
to develop a consistent way to handle this problem. 

\begin{figure}[t!]
\begin{center}
\includegraphics*[angle=-90,viewport=220 200 385 600,width=1.0\textwidth]{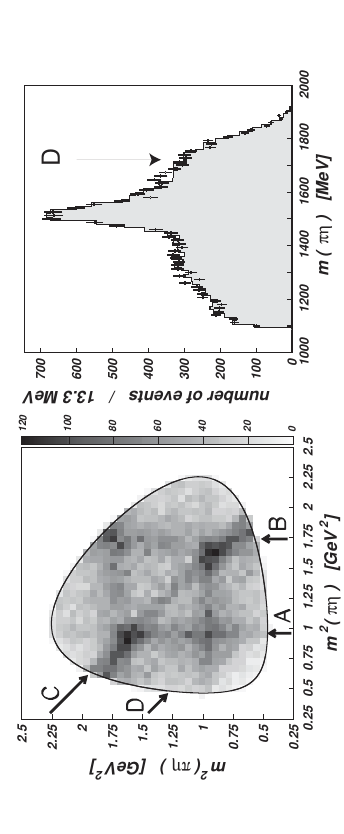}
\end{center}
\vspace{-2mm}
\caption[]{\label{fig:cb_1710}Dalitz plot from Crystal Barrel for $p\bar{p}\to\pi^0\eta\eta$
  in flight at 900~MeV/$c$ (left). The arrows indicate $a_0(980)$~(A), $a_2(1320)$~(B), $f_0(1500)/
  f_{2}^{\,\prime}(1525)$~(C), whereas (D) shows the expected location of the $f_0(1710)$.
  The $\eta\eta$ mass projection (right) is dominated by the $f_0(1500)/f_{2}^{\,\prime}(1525)$
  peak; the shaded area represents the fit~\cite{Amsler:2002qq}.
  Figures are taken from~\cite{Amsler:2004ps}.}
\end{figure}

In the limit of no $s$-quark admixture in the proton wave function, the OZI rule does not 
support production of pure $s\bar{s}$ states in $p\bar{p}$ annihilation. For this reason, 
observation of the $f_0(1710)$ scalar state should be strongly suppressed, which is 
assumed to have a dominant $s\bar{s}$ component. However, deviations from the OZI rule are 
observed, e.g. in low-energy proton-antiproton $S$-wave annihilation to $\phi\pi$ or $\phi
\gamma$, but also for some pion-induced reactions on the nucleon~\cite{Amsler:1997up}. 
Violations of the OZI rule can possibly be understood by the presence of an $s\bar{s}$ 
component in the nucleon. In~\cite{Ellis:1994ww} for example, the strong excess of $\phi$ 
production in $S$-wave $p\bar{p}$ annihilation is interpreted in terms of the polarization 
of the nucleon's $s\bar{s}$ component indicated by deep inelastic scattering experiments.
Alternatively, rescattering can also explain the effect~\cite{Locher:2001zc}.

The $f_0(1710)$ state was discovered by the Crystal-Ball collaboration in radiative 
$J/\psi$~decays into $\eta\eta$~\cite{Edwards:1981ex}, but the spin ($J=0$~or~2) remained 
controversial for a long time. The WA102 collaboration later determined the spin in favor 
of $0^{++}$ in central production at 450~GeV/$c$. Crystal Barrel data for the reactions 
$p\bar{p}\to\pi^0\pi^0\pi^0$, $p\bar{p}\to\pi^0\pi^0\eta$, and $p\bar{p}\to\pi^0\eta\eta$ 
in flight at 900~MeV/$c$ were used to search for isoscalar $0^{++}$ and $2^{++}$ states in 
the 1000-2000~MeV/$c^2$ mass range, in particular for the $f_0(1710)$~\cite{Amsler:2002qq}. 
A satisfactory signal around 1700~MeV/$c^2$ was neither observed for a scalar nor for a 
tensor state in the partial wave analyses of both the $\pi^0\pi^0\pi^0$ and $\pi^0\eta\eta$ 
channels. The $\pi^0\eta\eta$ Dalitz plot and the corresponding $\eta\eta$ mass projection 
are shown in Fig.~\ref{fig:cb_1710}. The (D)~arrow indicates the expected location of the 
$f_0(1710)$. None of the fits using the PDG mass and width for the $f_0(1710)$ had a stable 
solution and the log-likelihood improvement was not significant even in the best fit. The 
state does not seem to be produced in proton-antiproton annihilations in flight at 900~MeV/$c$ 
and upper limits at the 90\,\% confidence level were derived using PDG mass and width of 
$M = 1715$~MeV/c$^2$ and $\Gamma = 125$~MeV/c$^2$~\cite{Groom:2000in}:
\begin{alignat}{2}
\frac{{\cal{B}}(p\bar{p}\to\pi^0 f_0(1710)\to \pi^0\pi^0\pi^0)}
     {{\cal{B}}(p\bar{p}\to\pi^0 f_0(1500)\to \pi^0\pi^0\pi^0)}\: <\: 0.31\\[1ex] 
\frac{{\cal{B}}(p\bar{p}\to\pi^0 f_0(1710)\to \pi^0\eta\eta)}
     {{\cal{B}}(p\bar{p}\to\pi^0 f_0(1500)\to \pi^0\eta\eta)}\: <\: 0.25\,.
\end{alignat}
\noindent
For this reason, the non-observation of the $f_0(1710)$ scalar state in $p\bar{p}$ 
reactions is consistent with a dominant $s\bar{s}$ assignment to this state assuming 
it has a $q\bar{q}$ structure. Though the WA102 collaboration supported this conclusion 
by reporting a much stronger $K\bar{K}$ coupling of the $f_0(1710)$ than $\pi\pi$ 
coupling, it was not directly observed in the amplitude analysis of the reaction 
$K^- p\to K_S K_S\Lambda$~\cite{Aston:1987am}. As mentioned before, the spin assignment 
was controversial for a long time and much later settled in favor of $0^{++}$. The 
assumption in the analysis of $K^-$-induced data was $J=2$ and may explain the absence.

\subsubsection*{\emph{Results on Pseudoscalar States}}
%
In addition to the familiar $\eta$ and $\eta\,^{\prime}$ mesons, Crystal Barrel observed
a pseudoscalar state identified with the $\eta(1405)$ in the reactions $\bar{p}{p}\to
\eta\pi^{+}\pi^{-}\pi^{0}\pi^{0}$~\cite{Amsler:1995wz} and $\bar{p}{p}\to\eta\pi^{+}
\pi^{-}\pi^{+}\pi^{-}$~\cite{Reinnarth:2003} with the following masses and widths:
\begin{alignat*}{2}
M_{\eta(1405)} & = 1409\pm 3~{\rm MeV}/c^2 & \qquad \Gamma & = 86\pm 10~{\rm MeV}/c^2\\[1ex]
M_{\eta(1405)} & = 1407\pm 5~{\rm MeV}/c^2 & \qquad \Gamma & = 57\pm 9~{\rm MeV}/c^2\,,
\end{alignat*}
respectively. The state was observed to decay to both $\eta\pi^{0}\pi^{0}$ and $\eta
\pi^{+}\pi^{-}$. Partial wave analysis indicated that the decays were via two reactions, 
$\eta(1405)\rightarrow a_{0}(980)\pi$ and $\eta(\pi\pi)_{S}$ with~\cite{Amsler:1995wz}:
\begin{eqnarray*}
\frac{{\cal{B}}(\eta\rightarrow \eta (\pi\pi)_{S})}{{\cal{B}}(\eta\rightarrow a_{0}\pi)} 
& = & 0.78\pm 0.12\pm 0.10\, .
\end{eqnarray*}

\begin{figure}[tb!]
\begin{center}
\includegraphics*[viewport=150 190 640 380,width=1.0\textwidth]{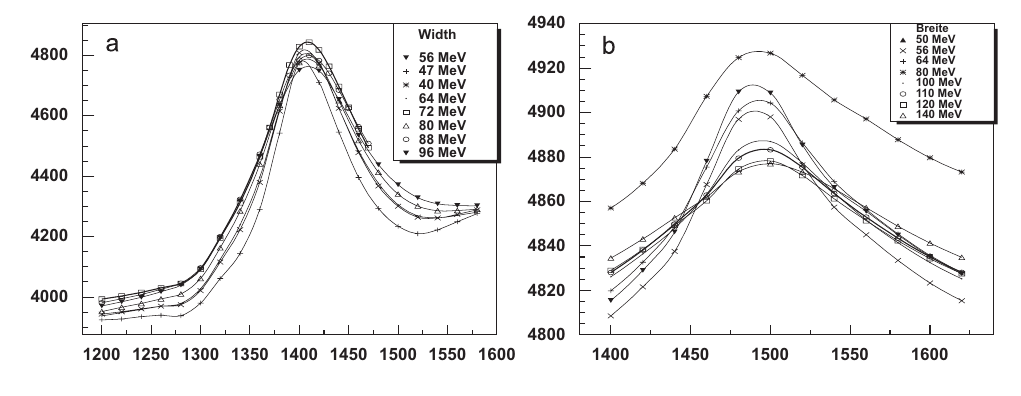}
\end{center}
\vspace{-7mm}
\caption[]{\label{fig:cb_1440}Analysis of the annihilation reaction $p\bar{p}\to
\pi^+\pi^-\pi^+\pi^-\eta$~\cite{Reinnarth:2003}: mass scans for $I^G\,J^P = 0^+\,0^\pm$ 
resonances assuming different widths. (a) The left side shows a clear peak identified as 
the $\eta(1405)$. (b) The right side shows the search for a second pseudoscalar state
indicating the presence of the $\eta(1475)$ with ($M=1490\pm 15,~\Gamma=74\pm 10$)~MeV/$c^2$.}
\end{figure}

It was reported by the Mark-III collaboration in 1990~\cite{Bai:1990hs} that in radiative $J/\psi$ 
decays in the $1400$~MeV/$c^2$ mass region, there were actually two pseudoscalar states. The 
lighter decaying via $a_{0}\pi$ and the heavier via $K^{*}K$. This is not inconsistent with the 
Crystal Barrel observation of a single state~\cite{Amsler:1995wz} which is likely the lighter 
of the two Mark-III states. The later Crystal Barrel analysis~\cite{Reinnarth:2003} of the 
$\eta\pi^{+}\pi^{-}\pi^{+}\pi^{-}$ final state confirmed the lighter pseudoscalar state (see Fig.
\ref{fig:cb_1440}, left side) and also searched for an additional $0^+\,0^\pm$ resonance. 
A scan shown in Fig.~\ref{fig:cb_1440}~(b) provides evidence for the heavier $\eta(1475)$ 
state with a mass of $M=1490\pm 15$~MeV/$c^2$.

Oddly, the expected lighter pseudoscalar state, the $\eta(1295)$, was not observed by 
Crystal Barrel~\cite{Amsler:2004rd}, even though its dominant decays are expected to lead
to the same final states as for the $\eta(1405)$. In~\cite{Amsler:2004rd}, a peak at 1285~MeV/$c^2$ 
could be described best with the $f_1(1285)$.

\subsection{\emph{Results from the OBELIX Experiment}}
\label{section:obelix}
The OBELIX detector system was operated at LEAR with a liquid H$_2$~(D$_2$) target, a 
gaseous H$_2$ target at room temperature and pressure, and a target at low pressures 
(down to 30 mbar). Among other things, the wide range of target densities provided 
detailed information about the influence of the atomic cascade on the annihilation 
process.

\subsubsection*{\emph{Results on Scalar States}}
The OBELIX collaboration recently studied the $\pi^+\pi^-\pi^0$, $K^+ K^- \pi^0$, and $K^\pm 
K^0 \pi^\mp$ final states in proton-antiproton annihilation at rest at three different 
hydrogen target densities in the framework of a coupled-channel analysis together with
$\pi\pi$, $\pi K$, and $K\bar{K}$ scattering data~\cite{Bargiotti:2003ev}. One of the 
main goals of the analysis was to determine branching ratios as well as $\pi\pi$ and 
$K\bar{K}$ partial widths of all the involved $(J^P = 0^+,~1^-,~2^+)$ resonances. Dalitz-plot 
projections of the three annihilation reactions are shown in Fig.~\ref{fig:obelix}.
The scattering data, in particular the $\theta_0(\pi\pi\to\pi\pi)$ phase shift, clearly
require contributions from the $f_0(980)$ pole. The authors further report on two 
additional poles required by annihilation data, a broad $f_0(1370)$ and a relatively 
narrow $f_0(1500)$. The introduction of a fourth scalar state improves the data and 
splits the initially broad $f_0(1370)$ into a broad $f_0(400-1200)$ and a relatively 
narrow $f_0(1370)$. The $f_0(1710)$ does not seem to be needed by the data. In addition,
a good description requires the $f_2(1270)$ and $f_{2}^{\,\prime}(1525)$ tensor states. The
$f_2(1565)$ pole is needed for the $\pi^+\pi^-\pi^0$ and $K^+ K^-\pi^0$ data at low
pressure and in hydrogen gas at normal temperature and pressure. The following 
$\cal{B}$$_{K\bar{K}}/$$\cal{B}$$_{\pi\pi}$ ratios of branching fractions for $f_0(1370)$,
$f_0(1500)$, and $f_2(1270)$ were determined:
\begin{figure}[tb!]
\begin{center}
\includegraphics*[viewport=90 80 525 730,width=0.75\textwidth]{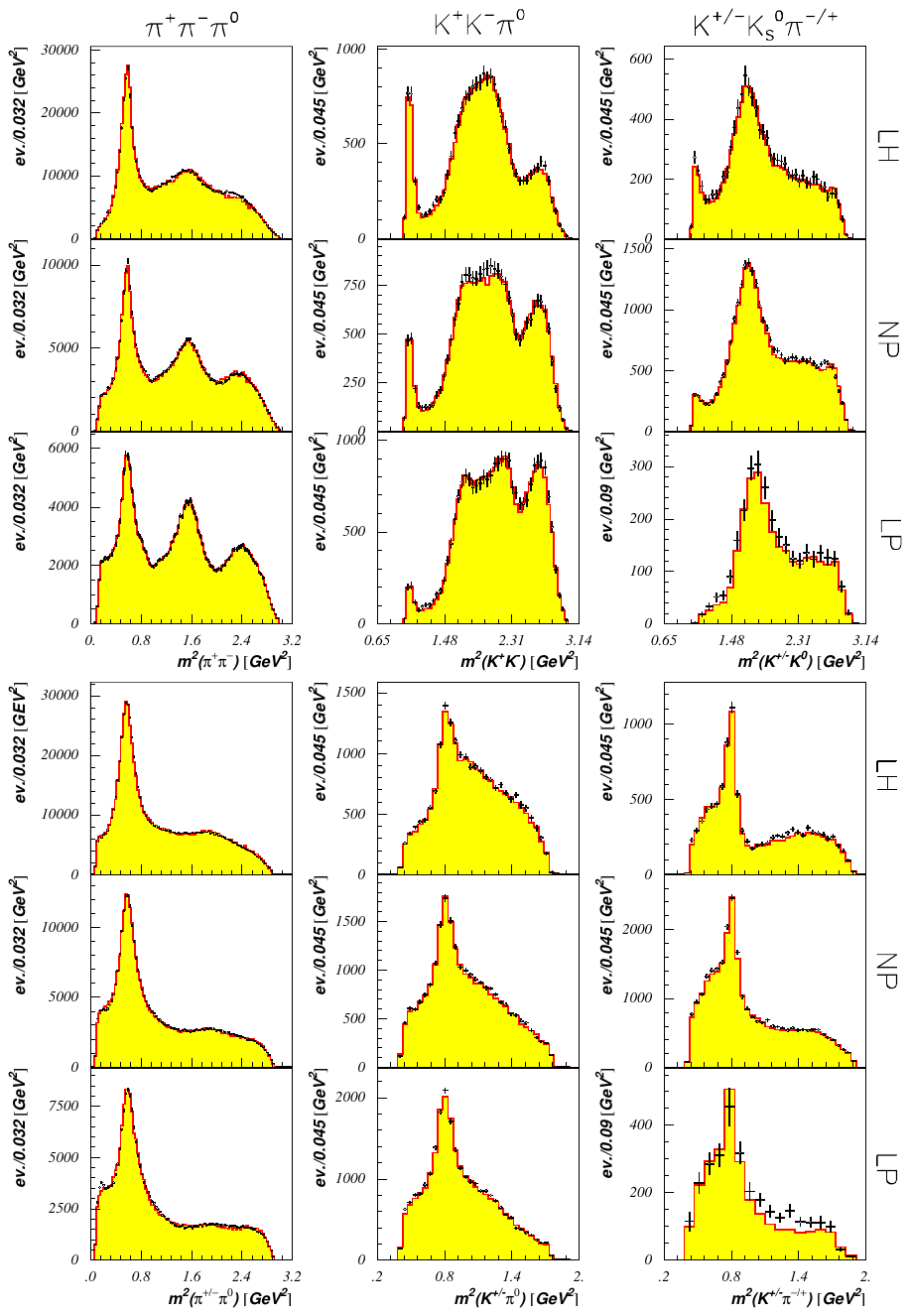}
\end{center}
\vspace{-0.2cm}
\caption[]{\label{fig:obelix}Theoretical (shaded histograms) and experimental (background
subtracted) Dalitz-plot projections from the OBELIX experiment of the three annihilation 
reactions $p\bar{p}\to\pi^+\pi^-\pi^0$, $p\bar{p}\to K^+ K^-\pi^0$, and $p\bar{p}\to K^\pm 
K_S^0 \pi^\mp$ in liquid (LH) hydrogen, H$_2$ gas at normal pressure and temperature (NP), 
and low pressure (LP) hydrogen gas. Theoretical and experimental errors are summed. Figure 
is taken from~\cite{Bargiotti:2003ev}.} 
\end{figure}

\begin{align}
\frac{{\cal{B}}(p\bar{p}\to f_0(1370)\pi^0,~f_0\to K\bar{K})}{{\cal{B}}(p\bar{p}\to f_0(1370)\pi^0,~f_0\to \pi\pi)} &\;=\;\begin{cases} \; 1.000\pm 0.200\quad ^1S_0\\ \; 0.940\pm 0.200\quad ^3P_1\end{cases}\\[1ex]
\frac{{\cal{B}}(p\bar{p}\to f_0(1500)\pi^0,~f_0\to K\bar{K})}{{\cal{B}}(p\bar{p}\to f_0(1500)\pi^0,~f_0\to \pi\pi)} &\;=\;\begin{cases} \; 0.240\pm 0.040\quad ^1S_0\\ \; 0.300\pm 0.040\quad ^3P_1\end{cases}\\[1ex]
\frac{{\cal{B}}(p\bar{p}\to f_2(1270)\pi^0,~f_2\to K\bar{K})}{{\cal{B}}(p\bar{p}\to f_2(1270)\pi^0,~f_2\to \pi\pi)} &\;=\;\begin{cases} \; 0.043\pm 0.010\quad ^1S_0\\ \; 0.045\pm 0.010\quad ^3P_1\\ \; 0.048\pm 0.010\quad ^3P_2\end{cases}
\end{align}

\noindent
The values for $f_0(1500)$ from OBELIX obtained in a coupled-channel framework are 
somewhat greater than earlier coupled-channel results from Crystal-Barrel
(Table~\ref{tab:cb_scalar-decays}) and agree well with results from WA102
\cite{Barberis:1999cq}~(Table~\ref{tab:wa102_scalar-decays}). The $f_2(1270)$ ratios
agree with the PDG values within the experimental errors.

\subsubsection*{\emph{Results on Pseudoscalar States}}
The collaboration has performed several studies looking at the $\eta(1405)$ and 
$\eta(1460)$ in the $K\bar{K}\pi$ final states. The first study looked at $\bar{p}p\to 
K^{\pm}K^{0}_{miss}\pi^{\mp}\pi^{+}\pi^{-}$ at rest~\cite{Bertin:1995fx} where they 
confirmed two pseudoscalar states previously announced by the Mark-III collaboration. 
The lighter decayed mainly to $K\bar{K}\pi$ (via $a_{0}(980)\pi$) while the heavier 
decayed to $K^\ast K$. Further evidence for the two states is provided in references
\cite{Bertin:1996nq} and~\cite{Cicalo:1999sn} based on the analyses of the reactions 
$\bar{p}p\to\eta(1440)(\pi\pi)\to K^\pm K_L^0\pi^\mp(\pi\pi)$ and $\bar{p}p\to K^\pm 
K_S^0\pi^\mp \pi^+\pi^-$, respectively. Production rates of $\eta(1440)\to K^\pm
K_L^0\pi^\mp$ were determined for the first time with the same detector setup for three 
different hydrogen target densities: $f_{\eta(1440)}({\rm liquid~H}_2) = (6.0\pm 0.5)
\cdot 10^{-4}$, $f_{\eta(1440)}(NTP) = (2.9\pm 0.4)\cdot 10^{-4}$, and $f_{\eta(1440)}
({\rm 5~mbar}) = (1.0\pm 0.2)\cdot 10^{-4}$. 

A spin-parity analysis of $p\bar{p}\rightarrow K^{+}K^{-}\pi^{+}\pi^{-}\pi^{0}$ at rest
in a gaseous hydrogen target, which included the observation of the axial vector meson 
$f_1(1420)$ decaying into $K^\ast \bar{K}$ (produced from $^3P_1$ protonium), provided 
more information on the $\eta(1405)$ and $\eta(1460)$~\cite{Nichitiu:2002cj}. Also hints 
of the $f_{0}(1710)$ decaying to $f_{0}(1370)(\pi\pi)_{S}$ were found in this analysis. 
Masses and widths of the two pseudoscalar states from OBELIX are summarized in Table
\ref{tab:obelix}.

\begin{table}[t!]\centering
\addtolength{\extrarowheight}{3pt}
\begin{tabular}{llllc}
             & Reaction & Mass & Width & Reference\\\hline
$\eta(1405)$ & $\bar{p}p\to K^\pm K_S^0 \pi^\mp \pi^+\pi^-$ & $1405\pm 5$  & $50\pm 4$ & \cite{Cicalo:1999sn}\\
             & $\bar{p}p\to K^+ K^- \pi^+\pi^-\pi^0$ & $1413\pm 2$  & $51\pm 4$ & \cite{Nichitiu:2002cj}\\\hline
$\eta(1460)$ & $\bar{p}p\to K^\pm K_S^0 \pi^\mp \pi^+\pi^-$ & $1500\pm 10$ & $100\pm 20$ & \cite{Cicalo:1999sn}\\
             & $\bar{p}p\to K^+ K^- \pi^+\pi^-\pi^0$ & $1460\pm 12$ & $120\pm 15$ & \cite{Nichitiu:2002cj}\\\hline
\end{tabular}
\caption{\label{tab:obelix} Masses and widths of $\eta(1405)$ and $\eta(1460)$ from OBELIX.}
\end{table}

OBELIX also observed an isovector scalar state with a mass of about $1.3$~GeV/$c^{2}$ in 
its $K\bar{K}$ decay mode~\cite{Bertin:1998sb,Bargiotti:2003ev}; the state is relatively 
narrow, with a width of $80$~MeV. In formation, the collaboration reported on the $3\pi$ 
decays of the $\pi(1300)$ as well as suggested a $3\pi$ decay of the hybrid candidate 
$\pi_{1}(1400)$ from $p\bar{p}\rightarrow 2\pi^{+}2\pi^{-}$~\cite{Salvini:2004gz} at rest 
and in flight.

\subsection{\emph{Results from Central Production: The WA102 Experiment}}
\label{section:wa102}
The WA102 experiment looked at $450$~GeV/$c$ protons incident on a proton target
to study the reaction $pp\rightarrow p\,_{(f)ast}\,X\,p\,_{(s)low}$ -- so-called 
\emph{central production}. Such reactions are believed to have a significant 
contribution from double-Pomeron exchange -- a reaction that is supposed to be 
glue rich. Relevant to the search for scalar glueballs, the collaboration carried 
out partial wave analysis on a large number of final states:
$pp\rightarrow pp 4\pi$~\cite{Abatzis:1995xx,Barberis:1997ve,Barberis:1999wn,Barberis:2000em},
$pp\rightarrow pp\pi^{0}\pi^{0}$~\cite{Barberis:1999ap},
$pp\rightarrow pp\pi^{+}\pi^{-}$~\cite{Abatzis:1995xx,Barberis:1999an,Barberis:1999cq},
$pp\rightarrow pp K^{+}K^{-}$~\cite{Barberis:1999am,Barberis:1999cq},
$pp\rightarrow pp K^{0}_{S}K^{0}_{S}$~\cite{Barberis:1999am},
$pp\rightarrow pp\eta\eta$~\cite{Barberis:2000cd},
$pp\rightarrow pp\eta\eta\,^\prime$~\cite{Barberis:1999id},
$pp\rightarrow pp\eta\,^\prime\eta\,^\prime$~\cite{Barberis:1999id},
$pp\rightarrow pp\phi\phi$~\cite{Barberis:1998bq},
$pp\rightarrow pp\omega\omega$~\cite{Barberis:2000kc},
$pp\rightarrow pp\phi\omega$~\cite{Barberis:1998tv} and
$pp\rightarrow pp K^{*}(892)\bar{K}^{*}(892)$~\cite{Barberis:1998tv}.

\begin{table}[b!]\centering
\addtolength{\extrarowheight}{3pt}
\begin{tabular}{|c|ccc|ccc|} \hline
Scalar        & $\pi\pi/K\bar{K}$ 
              & $\pi\pi/\eta\eta$ 
              & $\eta\eta/K\bar{K}$  
              & $\rho\rho/2[\pi\pi]_S$
              & $\rho\rho/4\pi$
              & $\sigma\sigma/4\pi$\\ \hline
$f_{0}(1370)$ & $2.17\pm 0.90$ & & $0.35\pm 0.21$ & & $\sim 0.9$ & $\sim 0$\\
%
%
$f_{0}(1500)$ & $3.13\pm 0.68$ & $5.5\pm 0.84$ & & $2.6\pm 0.4\,^1$ & $0.74\pm 0.03$ & $0.26\pm 0.03$\\
              &                &               & & $3.3\pm 0.5\,^2$ & &\\
$f_{0}(1710)$ & $0.20\pm 0.03$ & & $0.48\pm 0.14$ & & &\\
\hline
\end{tabular}
\caption[]{\label{tab:wa102_scalar-decays}A summary of WA102 results on the decay
of scalar mesons~\cite{Abatzis:1995xx,Barberis:1997ve,Barberis:1997vf,
Barberis:1999am,Barberis:1999an,Barberis:1999ap,Barberis:1999id,
Barberis:1999wn,Barberis:2000em}. The result for the decay of the $f_{0}(1500)$ into
$4\pi$ is derived $^1$\,from $2\pi^+2\pi^-$ and $^2$\,from $\pi^+\pi^- 2\pi^0$.}
\end{table}

In addition, a number of studies that bear on the search for pseudoscalar states 
were also performed: $pp\rightarrow pp K \bar{K}\pi$~\cite{Barberis:1997vf}, 
$pp\rightarrow pp\pi^{+}\pi^{-}\pi^{0}$~\cite{Barberis:1998in},
$pp\rightarrow pp\eta\pi^{+}\pi^{-}$~\cite{Barberis:1999be},
$pp\rightarrow pp\eta\pi^{0}$~\cite{Barberis:2000cx}, and
$pp\rightarrow pp\pi^{0}\pi^{0}\pi^{0}$~\cite{Barberis:2001bs}.

\begin{figure}[tb!]
\begin{center}
\includegraphics*[angle=-90,viewport=195 155 420 645,width=1.0\textwidth]{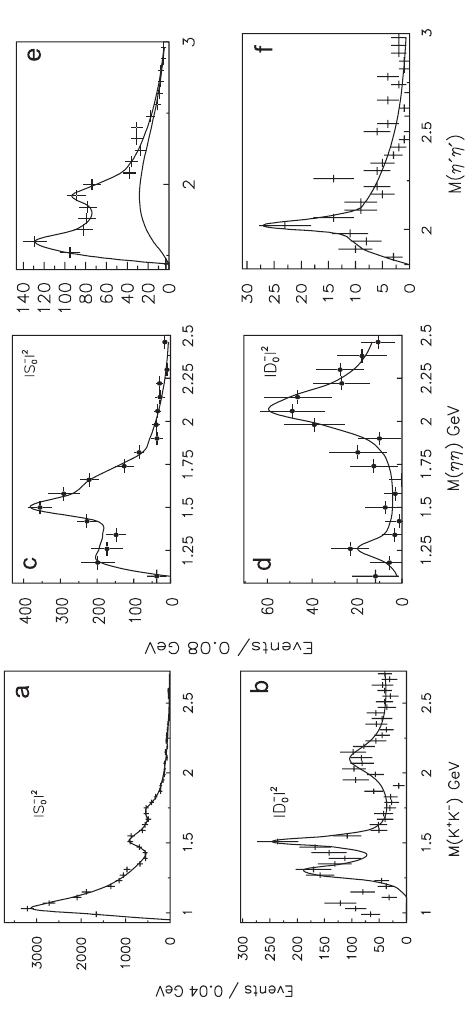}
\end{center}
\vspace{-2mm}
\caption[]{\label{fig:wa102}$K^+K^-$ $S$- (a) and $D$-wave (b) from a coupled-channel
analysis of WA102 $\pi^+\pi^-$ and $K^+K^-$ data~\cite{Barberis:1999cq}. The
T-matrix formalism was used for overlapping resonances to account for unitarity. 
Signals for $f_{0}(1500)$ and $f_{0}(1710)$ are clearly observed. Mass distributions
for $\eta\eta$ $S$- and $D$-wave are shown in (c) and (d). The right panels show 
$\eta\eta\,^{\prime}$~(e) and $\eta\,^{\prime}\eta\,^{\prime}$~(f) mass distributions.}
\end{figure}

Results of measured decay branching ratios into two pseudoscalar mesons are listed
in Table~\ref{tab:wa102_scalar-decays}~\cite{Abatzis:1995xx,Barberis:1997ve,Barberis:1997vf,
Barberis:1999am,Barberis:1999an,Barberis:1999ap,Barberis:1999id,Barberis:1999wn,Barberis:2000em}. 
Further selected PWA results from the WA102 experiment on central production are presented 
in Fig.~\ref{fig:wa102}~\cite{Klempt:2007cp,Barberis:1999cq,Barberis:1999id,Barberis:2000cd}. 
The $K^+K^-$ $S$-wave from a coupled-channel analysis of $\pi^+\pi^-$ and $K^+K^-$ data is shown 
in (a). The long tail beyond the dominant threshold enhancement for the $f_{0}(980)$ includes 
signals for the three scalar resonances of Table~\ref{tab:wa102_scalar-decays}. Only the 
two higher-mass states are observed as peaks~\cite{Barberis:1999cq}. The corresponding 
$K^+K^-$~$D$-wave (b) shows resonant peaks for the two established tensor mesons 
$f_{2}(1270)$ and $f_{2}^{\,\prime}(1525)$. A structure at 2.15~GeV/$c^2$ is also observed 
in the mass distribution. The pole positions for the $f_{0}(1370)$ and $f_{0}(1500)$ are 
in excellent agreement with results from the Crystal-Barrel experiment. 
Fig.~\ref{fig:wa102}~(c) and (d) present the $\eta\eta$ $S$- and 
$D$-wave~\cite{Barberis:2000cd}. The $f_{0}(1500)$ is clearly seen in the $S$-wave 
$\eta\eta$ mass distribution (c). In addition to a weak $f_{2}(1270)$ signal, the resonant 
structure at 2.15~GeV/$c^2$ is also observed in the $\eta\eta$ $D$-wave (d). The $f_{0}(1500)$ 
was also observed in studies of the $\eta\eta\,^\prime$ decay mode~\cite{Barberis:1999id}, 
shown in Fig.~\ref{fig:wa102}~(e), and in the $4\pi$ final state~\cite{Barberis:1999wn}.

The small $\cal{B}$$(\pi\pi)/$$\cal{B}$$(K\bar{K})$ value for the $f_{0}(1710)$ in 
Table~\ref{tab:wa102_scalar-decays} clearly indicates that this resonance must have 
a large $s\bar{s}$ component. By contrast, the same ratio is much greater than one for 
the $f_{0}(1500)$. If interpreted as $q\bar{q}$ state, the $f_{0}(1500)$ cannot have a 
large $s\bar{s}$ component since pure $s\bar{s}$ mesons do not decay to pions. Moreover, 
we recall that an enhancement of gluonic states is expected in Pomeron-Pomeron fusion 
(Close-Kirk glueball filter)~\cite{Kirk:1999av}. Though the $f_{0}(1710)$ couples more 
strongly to $K\bar{K}$, the $K^+K^-$ $S$-wave signal for the $f_{0}(1500)$ shown in 
Fig.~\ref{fig:wa102}~(a) is larger in agreement with predictions of the glueball filter.
As mentioned earlier, there are significant discrepancies with Crystal Barrel results on
the decay of these scalars to $\rho\rho$ and $2[\pi\pi]_S$ final states. This may be
due to how the opening of broad thresholds is treated in the individual analyses. 

\begin{table}[tb!]\centering\small
\addtolength{\extrarowheight}{3pt}
\begin{tabular}{|c|c|c|c|c|c|} \hline
$J^{PC}$ & Res. & $\sigma$ [nb] & $dP_T \leq 0.2$ GeV & $0.2\leq dP_T \leq 0.5$ GeV & $dP_T \geq 0.5$ GeV\\\hline
$0^{-+}$ & $\pi^0$         & $22\,011 \pm 3\,267$ & $12\pm 2$ & $45\pm 2$ & $43\pm 2$\\
         & $\eta$          & $3\,859 \pm 368$ & $6\pm 2$ & $34\pm 2$ & $60\pm 3$\\
         & $\eta\,^{\prime}$ & $1\,717\pm 184$ & $3\pm 2$ & $32\pm 2$ & $64\pm 3$\\\hline
$0^{++}$ & $a_0(980)$      & $638\pm 60$ & $25\pm 4$ & $33\pm 5$ & $42\pm 6$\\
         & $f_0(980)$      & $5\,711\pm 450$ & $23\pm 2$ & $51\pm 3$ & $26\pm 3$\\
         & $f_0(1370)$     & $1\,753\pm 580$ & $18\pm 4$ & $32\pm 2$ & $50\pm 3$\\
         & $f_0(1500)$     & $2\,914\pm 301$ & $24\pm 2$ & $54\pm 3$ & $22\pm 4$\\
         & $f_0(1710)$     & $245\pm 65$ & $26\pm 2$ & $46\pm 2$ & $28\pm 2$\\
         & $f_0(2000)$     & $3\,139\pm 480$ & $12\pm 2$ & $38\pm 3$ & $50\pm 4$\\\hline
$1^{++}$ & $a_1(1260)$     & $10\,011\pm 900$ & $13\pm 3$ & $51\pm 4$ & $36\pm 3$\\
         & $f_1(1285)$     & $6\,857\pm 1\,306$ & $3\pm 1$ & $35\pm 2$ & $61\pm 4$\\
         & $f_1(1420)$     & $1\,080\pm 385$ & $2\pm 2$ & $38\pm 2$ & $60\pm 4$\\\hline
$2^{++}$ & $a_2(1320)$     & $1\,684\pm 134$ & $10\pm 2$ & $38\pm 5$ & $52\pm 6$\\
         & $f_2(1270)$     & $3\,275\pm 422$ & $8\pm 1$ & $29\pm 1$ & $63\pm 2$\\
         & $f_{2}^{\,\prime}(1520)$ & $68\pm 9$ & $4\pm 3$ & $36\pm 3$ & $60\pm 4$\\
         & $f_2(1910)$     & $528\pm 40$ & $20\pm 4$ & $62\pm 7$ & $18\pm 4$\\
         & $f_2(1950)$     & $2\,788\pm 175$ & $27\pm 2$ & $46\pm 5$ & $27\pm 2$\\
         & $f_2(2150)$     & $121\pm 12$ & $3\pm3$ & $53\pm 4$ & $44\pm 3$\\
\hline
\end{tabular}
\caption[]{\label{tab:wa102}A summary of WA102 results on resonance production 
at $\sqrt{s} = 29.1$ GeV~\cite{Kirk:2000ws}. The quoted errors are statistical 
and systematic errors summed in quadrature. Numbers given for resonance
production as a function of $dP_T$ are percentages of the total contribution.}
\end{table}

Table~\ref{tab:wa102} shows further results on resonance production from WA102. 
The cross section is given at 450 GeV/$c$ for the reaction
\begin{eqnarray*}
\label{reaction:central_a}
p\,p\to {p_f}\,X\,{p_s}\,,
\end{eqnarray*}
and the dependence of the production of $X$ on the parameter $dP_T$, denoting
the difference in transverse momentum between the particles exchanged from
the fast and slow vertices. Production of $X=\rho^0$ with $I=1$ cannot proceed 
via double-Pomeron exchange (DPE). The $\pi^+\pi^-$ mass spectra from the WA76 
collaboration at $\sqrt{s} = 12.7$~GeV and $\sqrt{s} = 23.8$~GeV~\cite{Armstrong:1991ch}
show a reduction of the $\rho^0$ yield at higher center-of-mass energies. The
collaboration reports a cross section ratio of $0.44\pm 0.07$ indicating that
DPE becomes more important with increasing energy.

In particular, Table~\ref{tab:wa102} shows that production of isovector states 
is strong, e.g. the $a_1(1260)$, which requires Reggeon exchange. On the other
hand, production of scalar resonances is fully consistent with Pomeron-Pomeron 
fusion. The $a_0(980)$ is suppressed by almost a factor of 10 with respect to 
the $f_0(980)$. The $a_0(1450)$ is not even seen, whereas a large $f_0(1500)$ 
yield is observed. Interestingly, production of the $f_0(1710)$ is very weak
compared to all other scalar resonances with $I=0$.

While strong signals were seen for axial states in the same decay modes, beyond 
the $\eta(548)$ and $\eta\,^{\prime}(958)$, no pseudoscalar states were reported 
in central production by WA102. However, this is believed to be related to the 
production mechanism, and WA102 likely did not run at high-enough energy to 
produce pseudoscalar states more massive than the $\eta\,^{\prime}$. Thus, the 
lack of these states should not be interpreted as anything more than a kinematic
effect.

Though scalar meson production from WA102 indicates dominant Pomeron-Pomeron exchange, 
the situation for other resonances is less clear. It may be questioned that sufficiently 
high energies were already reached in the experiment to form a glue-rich environment.

\subsection{\emph{Light-Meson Spectroscopy in \boldmath{$e^+e^-$} Experiments}}
\label{section:e+e-}

\subsubsection*{\emph{Scalar Mesons from the BES Experiment}}
Several reactions have been studied in the BES experiment. Partial wave analysis has
been performed on several radiative decays of the $J/\psi$ and are reported for the 
following reactions:
$J/\psi\rightarrow\gamma\pi^{+}\pi^{-}\pi^{+}\pi^{-}$~\cite{Bai:2000},
$J/\psi\rightarrow\gamma K^{+} K^{-}$~\cite{Bai:2003},
$J/\psi\rightarrow\gamma K^{0}_{S} K^{0}_{S}$~\cite{Bai:2003},
$J/\psi\rightarrow\gamma\pi^{+}\pi^{-}$~\cite{Ablikim:2006aw},
$J/\psi\rightarrow\gamma\pi^{0}\pi^{0}$~\cite{Bai:1998tx,Ablikim:2006aw}, and
$J/\psi\rightarrow\gamma\phi\omega$~\cite{Ablikim:2006dw}.
In addition to the radiative decays, BES has also examined decays to associated vector
meson decays. Here, analyses have been performed on the reactions:
$J/\psi\rightarrow\omega\pi^{+}\pi^{-}$~\cite{Ablikim:2004aa},
$J/\psi\rightarrow\omega K^{+} K^{-}$~\cite{Ablikim:2004bb},
$J/\psi\rightarrow\phi\pi^{+}\pi^{-}$~\cite{Ablikim:2005aa}, and
$J/\psi\rightarrow\phi K^{+} K^{-}$~\cite{Ablikim:2005aa}.

\begin{figure}[tb!]
\begin{center}
\includegraphics*[viewport=300 200 500 398,width=0.45\textwidth]{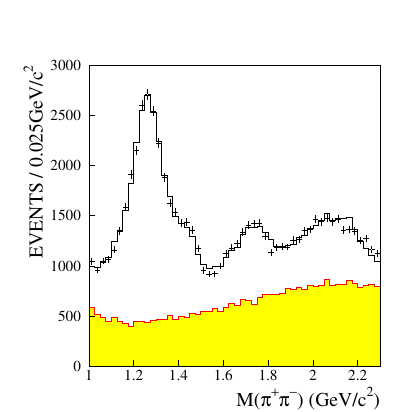}
\includegraphics*[viewport=300 202 500 400,width=0.45\textwidth]{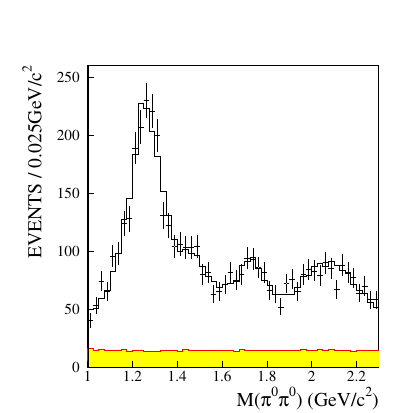}
\end{center}
\vspace{-0.2cm}
\caption[]{\label{fig:bes_twopi} The $\pi^+\pi^-$ invariant mass distribution (a) and
  the $\pi^0\pi^0$ mass distribution (b) from the reaction $J/\psi\to\gamma\pi\pi$ 
  from BES~\cite{Ablikim:2006aw}. The crosses are data, the full histogram shows 
  the maximum likelihood fit, and the shaded area corresponds to the background.}
\end{figure}
\begin{figure}[tb!]
\begin{center}
\includegraphics*[viewport=160 190 450 600,width=0.8\textwidth,height=0.9\textwidth]{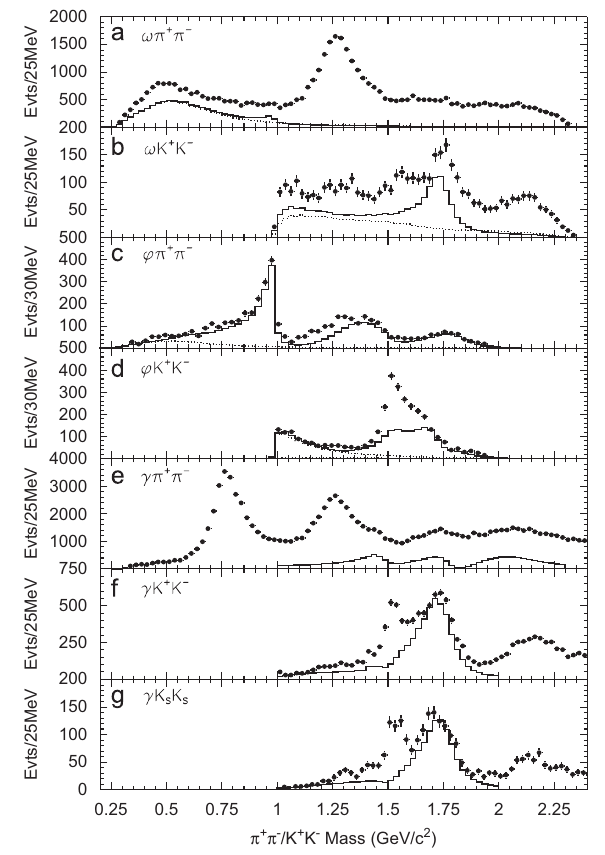}
\end{center}
\vspace{-0.3cm}
\caption[]{\label{fig:bes_summary} Invariant mass distributions of pseudoscalar meson
pairs recoiling against $\omega,~\phi$ or $\gamma$ in $J/\psi$~decays measured at BES II.
The dots with error bars are data, the solid histograms are the scalar contributions from
PWA, and the dashed lines in (a) through (c) are contributions of $\sigma(485)$ from the 
fits, while the dashed line in (d) is the $f_0(980)$. Notice that not the full mass spectra
are analyzed in (e), (f), and (g). The figure is taken from~\cite{Yuan:2005es}; see text
for details.}
\end{figure}

BES II results on $J/\psi$ radiative decays to $\pi^+\pi^-$ and $\pi^0\pi^0$ are
shown in Fig.~\ref{fig:bes_twopi}. A sample of 58\,M~$J/\psi$~events was used for the 
PWA~\cite{Ablikim:2006aw}. Similar structures are visible in both mass spectra. 
Three clear peaks are observed in both distributions in the 1.0 to 2.3 GeV/$c^2$ 
mass range: a strong $f_2(1270)$ signal exhibiting a shoulder on the high-mass side, 
an enhancement at $\sim\,1.7$ GeV/$c^2$ associated with the $f_0(1710)$, and a peak at 
$\sim\,2.1$~GeV/$c^2$. The shaded histogram in Fig.~\ref{fig:bes_twopi} (a) corresponds 
to dominant background from $J/\psi\to\pi^+\pi^-\pi^0$. The estimated background in 
(b) stems from various reactions; PDG branching ratios have been used in the studies. 
Three scalar mesons are observed with approximately consistent results from both fits. 
The lowest $0^{++}$ state is consistent with the $f_0(1500)$ and associated with the 
shoulder in Fig.~\ref{fig:bes_twopi}. The collaboration reports that spin~0 is strongly 
preferred over spin~2 in the analysis. Though not favored in the PWA, the presence of 
the $f_0(1370)$ is not excluded. The fitted masses and widths from 
$J/\psi\to\gamma\pi^+\pi^-$ for the two lowest-mass states are given by:
\begin{alignat*}{2}
\label{eq:bes}
M_{f_0(1500)} & = 1466\pm 6\pm 20~{\rm MeV}/c^2 & \qquad \Gamma & = 108^{+14}_{-11}
                   \pm 25~{\rm MeV}/c^2\\[1ex]
M_{f_0(1710)} & = 1765^{+4}_{-3}\pm 13~{\rm MeV}/c^2 & \qquad \Gamma & = 145\pm 8
                   \pm 69~{\rm MeV}/c^2\,,
\end{alignat*}
whereas PDG values for the $f_0(2020)$ are used for the structure at $\sim\,2.1$~GeV/$c^2$.

The two established scalar mesons, $f_0(1500)$ and $f_0(1710)$, are also significantly 
produced in $J/\psi\to2\pi^+\,2\pi^-$ with masses of $M_{f_0(1500)} = 1505^{+15}_{-20}$ 
and $M_{f_0(1710)} = 1740^{+30}_{-25}$, respectively~\cite{Bai:2000}. In addition, the 
likelihood fit requires a tensor state, $f_2(1950)$, around 2~GeV/$c^2$ confirming earlier 
WA91 and WA102 results~\cite{Antinori:1995wz,Barberis:1997ve}. Branching fractions
determined from $J/\psi\to2\pi^+\,2\pi^-$ are listed in Table~\ref{tab:bes}. The $f_0(1710)$ 
scalar state also dominates the reaction $J/\psi\to\gamma K\bar{K}$. Evidence for the 
$f_0(1500)\to K\bar{K}$ is however insignificant, but included in the partial wave analysis 
interfering with the $f_0(1710)$. For a description of the 1500 MeV/$c^2$ mass range, the 
$f_{2}^{\,\prime}(1525)$ tensor state is required in the analysis~\cite{Bai:2003}.

\begin{table}[bt!]\centering
\addtolength{\extrarowheight}{3pt}
\begin{tabular}{|c|cccc|} \hline
Scalar        & $\pi^{+}\pi^{-}$~\cite{Ablikim:2006aw} 
              & $\pi^{\circ}\pi^{\circ}$~\cite{Ablikim:2006aw}
              & $K\bar{K}$~\cite{Bai:2003}
              & $\pi^{+}\pi^{-}\pi^{+}\pi^{-}$~\cite{Bai:2000} \\ \hline
$f_{0}(1500)$ & $0.67\pm 0.30$ & $0.34\pm 0.15$ & & $3.1\pm 1.12$ \\
$f_{0}(1710)$ & $2.64\pm 0.75$ & $1.33\pm 0.88$ & $9.62\pm 0.29\,_{\rm stat}$ & $3.1\pm 1.12$ \\
$f_{0}(2100)$ &                &        &       & $5.1\pm 1.82$ \\ \hline 
$f_{2}(1270)$ & $9.14\pm 1.48$ & $4.00\pm 0.59$ &        & $1.8\pm 0.63$ \\
$f_{2}^{\,\prime}(1525)$ &                &        & $3.42\pm 0.15\,_{\rm stat}$ & \\
$f_{2}(1565)$ &                &        &       & $3.2\pm 1.12$ \\
$f_{2}(1950)$ &                &        &       & $5.5\pm 1.92$ \\ \hline 
\end{tabular}
\caption[]{\label{tab:bes}BES results on radiative $J/\psi$ decays.
The rates are shown for $J/\psi\rightarrow \gamma f_{0}$ with the 
subsequent decay of the $f_{0}$ to the listed final state. All rates
are multiplied by $10^{-4}$. The quoted errors are statistical and 
systematic errors summed in quadrature.}
\end{table}

In section~\ref{subsubsection:flavor-tagging}, we have discussed the {\it flavor-tagging} 
approach to study the flavor content of mesons in $J/\psi$ decays. Due to the OZI rule,
$J/\psi\to\omega X$ couples to the $n\bar{n}$ component of $X$, while $J/\psi\to\phi X$ 
couples to $s\bar{s}$. Fig.~\ref{fig:bes_summary} shows invariant mass distributions of
pseudoscalar meson pairs recoiling against $\omega,~\phi$ or $\gamma$~\cite{Yuan:2005es}. 
The $K^+K^-$ mass distribution from $\omega K^+K^-$ (b) shows a clear scalar peak at 
1710~MeV/$c^2$ which is not observed in the corresponding spectrum recoiling against the 
$\phi$ meson (d). By contrast, the $\pi^+\pi^-$ mass distribution from $\phi\pi^+\pi^-$ 
(c) indicates an enhancement at $\sim\,1790$~MeV/$c^2$, which is absent in $\phi K^+K^-$ 
(d). This observation is puzzling and does not seem to be compatible with a single $f_0(1710)$ 
state, which is known to decay dominantly to $K\bar{K}$. The BES collaboration suggested 
two distinct scalar states around 1.75~GeV/$c^2$: the known $f_0(1710)$ with ($M\sim 
1740,~\Gamma\sim 150$)~MeV/$c^2$ decaying strongly to $K\bar{K}$ and a broad $f_0(1790)$ 
with ($M\sim 1790,~\Gamma\sim 270$)~MeV/$c^2$ which couples more strongly to 
$\pi\pi$~\cite{Ablikim:2005aa}. This new state is not confirmed by any other experiment 
and not listed in the 2008 edition of the ``Review of Particle Physics'' by the Particle 
Data Group~\cite{Amsler:2008zz}. The BES collaboration emphasizes that the $\phi f_0(1790)$ 
signal is very close to edge of the available phase space, where the reconstruction efficiency 
of the $\phi$ decreases significantly as the momentum of the $\phi$ decreases. Tails of 
broad higher-mass states could also interfere with the $f_0(1710)$ generating a structure 
near the end of the phase space~\cite{Yuan:2005es}. If both states really exist, it remains
a mystery why the $f_0(1710)$ mainly $s\bar{s}$-state is produced recoiling against an 
$\omega$, and the new $f_0(1790)$ mainly $n\bar{n}$-state is observed recoiling against 
a $\phi$. In fact, it's worth noting that many strong signals due to non-strange states
are seen in the $\phi\pi\pi$ data from BES: $f_2(1270)$, $f_0(1370)$, $f_0(1500)$, and
$f_0(1790)$. The collaboration makes a strong argument for the existence of a $f_0(1370)$
resonance, which has been doubted previously by several authors. In the analysis, the
state interferes with the $f_0(1500)$ and $f_2(1270)$ making it more noticeable, but a
determination of its mass and width is challenging for the same reason. 

A recent BES observation has increased the scalar puzzle even more. The group observes a 
state at $M\sim 1812$~MeV/$c^2$ and $\Gamma\sim 105$~MeV/$c^2$ in the doubly OZI suppressed 
process $J/\psi\to\gamma\omega\phi$~\cite{Ablikim:2006dw}. The PWA favors a $0^+$ scalar
assignment, but the signal has also been discussed as a $0^{-+}$ resonance. The state is 
listed as $X(1835)$ by the PDG, but has not been seen by another experiment. The production ratio 
should be suppressed by at least an order of magnitude. A value of ${\cal{B}}(J/\psi\to\gamma 
X)\cdot {\cal{B}}(X\to\omega\phi) = (2.61\pm 0.27~{\rm (stat)}\pm 0.65~{\rm (syst))}\times 
10^{-4}$ is reported. Decay rates for the discussed mesons are listed in Table~\ref{tab:bes}. 
Moreover, BES reports the ratios of branching fractions into $\pi\pi$ and $K\bar{K}$ for 
the $f_0(1370)$~\cite{Ablikim:2005aa} and $f_0(1710)$~\cite{Ablikim:2004bb}:
\begin{alignat}{2}
\frac{{\cal{B}}(f_0(1370)\to K\bar{K})}{{\cal{B}}(f_0(1370)\to\pi\pi)} & \; = \; 
          0.08\,\pm\,0.08\\
\frac{{\cal{B}}(f_0(1710)\to\pi\pi)}{{\cal{B}}(f_0(1710)\to K\bar{K})} & \; < \; 
          0.11~{\rm (@\,95\,\%~C.L.)}\,.
\end{alignat}

\subsubsection*{\emph{Light Scalar Mesons from the CLEO Experiment}}
\label{section:cleo}
Radiative $\Upsilon(1S)$ decays also provide a glue-rich environment for 
producing exotic states. The CLEO collaboration has reported on results from 
$\Upsilon(1S)$ decays to pairs of pseudoscalar mesons~\cite{Athar:2005nu,Besson:2005ud}.
Considering the fact that the quark-photon coupling is proportional to the
electric charge and assuming that the quark propagator is roughly proportional
to $1/m$ for low-momentum quarks, radiative decays from $\Upsilon(1S)$ should 
be suppressed by a factor of
\begin{eqnarray}
\label{eq:scaling}
(q_b/q_c)^2\cdot (m_c/m_b)^2\cdot \Gamma_\Upsilon/\Gamma_\psi \approx 0.04
\end{eqnarray}
relative to the corresponding $J/\psi$ decay. Table~\ref{tab:cleo} summarizes
the results. In particular, the decay rates of $f_{0}(1500)$ and $f_{0}(1710)$ 
to $\pi^{\circ}\pi^{\circ}$ are much smaller -- by more than an order of 
magnitude -- than predicted based on the scalar-glueball mixing matrix 
in~\cite{Close:2001ga}. 

The two listed tensor states dominate the di-gluon spectrum in $\Upsilon(1S)$
decays and fair agreement with the naive scaling argument is observed (to the 
same order of magnitude) for the suppression factor of these states relative 
to $J/\psi$ decays.

\begin{table}[t!]\centering
\addtolength{\extrarowheight}{3pt}
\begin{tabular}{|c|c|ccccc|} \hline
Scalar        & $\Upsilon$(1S)$\to\gamma$ + meson
              & $\pi^{+}\pi^{-}$ 
              & $\pi^{\circ}\pi^{\circ}$
              & $K^+K^-$ 
              & $\pi^{+}\pi^{-}\pi^{+}\pi^{-}$ 
              & $\eta\eta$\\ \hline
$f_{0}(980)$ & & $<3$ &  & & & \\
$f_{0}(1500)$ & $<1.5$ & $ <1.5 $  &  & & & $<0.3$\\
$f_{0}(1710)$ & & & $<0.14$ & $<0.7$ & & $<0.18$\\ \hline
$f_{2}(1270)$ & $10.2\pm 0.8$~\cite{Athar:2005nu} & & $3.0\pm 0.5$ & & & \\
$f_{2}^{\,\prime}(1525)$ & $3.7^{\,+\,0.9}_{\,-\,0.7}$~\cite{Athar:2005nu} & & & & & \\ \hline
\end{tabular}
\vspace{1mm}
\caption[]{\label{tab:cleo}CLEO results on radiative $\Upsilon$(1S) decays.
The rates for the scalar mesons are upper-limit branching fractions at the 
90\,\% confidence level for $\Upsilon\rightarrow \gamma f_{0}$ with the 
subsequent decay of the $f_{0}$ to the listed final state. All rates are 
multiplied by $10^{-5}$.}
\end{table}

The three-body decays of $D$-mesons can provide an excellent test of the 
microscopic structure of scalar mesons. The final state consists of only $u$ and 
$d$ quarks and antiquarks providing enough energy to cover most of the range of 
interest to light quark binding. The initial state is relatively simple with little 
impact on the final state. CLEO published the results of Dalitz-plot analyses for  
$D^0\to\pi^+\pi^-\pi^0$~\cite{CroninHennessy:2005sy}, 
$D^+\to\pi^+\pi^+\pi^-$~\cite{Bonvicini:2007tc},
$D^0\to K^-\pi^+\pi^0$~\cite{Kopp:2000gv}, 
$D^0\to K_S^0\pi^+\pi^-$~\cite{Muramatsu:2002jp}, 
$D^0\to K^0_S\eta\pi^0$~\cite{Rubin:2004cq}, 
$D^\pm\to K^+K^-\pi^\pm$~\cite{:2008zi}, and
$D^0\to K^+K^- \pi^0$~\cite{Cawlfield:2006hm}.

The weak decays of $D$ mesons are expected to be dominated by resonant two-body 
decays, e.g.~\cite{Buccella:1996uy}. In $D^0\to \pi^+\pi^-\pi^0$ decays, 
there is no evidence observed for any $\pi\pi$ $S$-wave contribution, only the 
3~$\rho(770)\pi$ modes are seen. The $f_0(980)$ meson is observed to be 
highly-suppressed in the reaction $D^0\to f_0(980)\pi^0$~\cite{CroninHennessy:2005sy}. 
Though the suppression itself is in good agreement with predictions inserting 
rescattering effects and considering the $f_0(980)$ as $s\bar{s}$ + a light 
$q\bar{q}$~pair~\cite{Buccella:1996uy}, the calculated ratio ${\cal{B}}(D^+\to 
f_0(980)\pi^+)/{\cal{B}}(D^0\to f_0(980)\pi^0)$ of 46.7 is almost an order of 
magnitude smaller than the experimentally determined lower limit of $>340$ at 95\,\% 
confidence level. The Dalitz plot study of $D^+\to\pi^+\pi^+\pi^-$~\cite{Bonvicini:2007tc} 
finds small contributions from $f_0(1370)$ and $f_0(1500)$ with $2.6\pm 1.9$\,\% 
and $3.4\pm 1.3$\,\%, respectively. The existence of $f_0(1370)$ is not questioned 
in the analysis. However, the mass introduced as a free fit parameter is somewhat 
low: $M = 1260$~MeV/$c^2$.

The most precise results on $D^0\to K_S^0\pi^+\pi^-$ come from the CLEO collaboration.
They find a much smaller non-resonant contribution to this reaction than did earlier 
experiments and attribute the source of this non-resonant component to the broad 
scalar resonances $K^\ast_0(1430)$ and $f_0(1370)$~\cite{Muramatsu:2002jp}. Some 
confusion arises from the small $f_0(980)$ contribution to this reaction, which is
inconsistent with a large contribution of $f_0(980)$ in $D^0\to K_S^0 K^+ K^-$ 
reported by the ARGUS and BaBar collaborations. A solution is discussed in a 
mini-review on charm Dalitz plot analyses in the 2008 edition of the PDG~\cite{Amsler:2008zz}. 
The explanation is a large $a_0(980)\to K^+K^-$ contribution to the reaction 
$D^0\to K_S^0 K^+ K^-$, which is also observed by CLEO in the reaction $D^0\to 
K^0_S\eta\pi^0$~\cite{Rubin:2004cq}.

Recently, the existence of the two scalar resonances, $f_0(1370)$ and $f_0(1500)$,
is also assumed in the CLEO analysis of the reaction $D^0\to K^+K^- \pi^0$
\cite{Cawlfield:2006hm}, but not further investigated. The main goal of the analysis 
is to determine the strong phase difference $\delta_D$ between $D^0\to K^{\ast -} 
K^+$ and $D^0\to K^{\ast +} K^-$, which is required to extract the 
Cabibbo-Kobayashi-Maskava (CKM) angle $\gamma$.

In summary, results on how $f_0(1370)$ and $f_0(1500)$ populate various charm Dalitz
plots are still confusing~\cite{pdg_charm}. In the analysis of $D^+\to \pi^+\pi^+\pi^-$, 
some $f_0(1370)$ was seen by E791~\cite{Aitala:2000xu}, but only very little was 
observed by FOCUS~\cite{Malvezzi:2003jp}. The state was also neither seen in 
$D_s^+\to \pi^+\pi^+\pi^-$ by E687 and FOCUS~\cite{Malvezzi:2003jp}, nor by BaBar 
in $D^0\to \bar{K}^0K^+K^-$, while CLEO has observed the $f_0(1370)$ in $D^0\to 
K^0_S\pi^+\pi^-$~\cite{Muramatsu:2002jp}. Evidence for the $f_0(1500)$ is also weak.
Small contributions have been observed only in $D^+\to\pi^+\pi^+\pi^-$ by 
CLEO~\cite{Bonvicini:2007tc}; E687 and FOCUS found a resonance in $D_s^+\to 
\pi^+\pi^+\pi^-$ with parameters similar to the $f_0(1500)$~\cite{Malvezzi:2003jp}.

\subsubsection*{\emph{Scalar Mesons from the \boldmath{$B$} Factories: Belle and BaBar}}
\label{section:b_factories}
The $B$-factories also studied decays of $D$ mesons, but more interesting for this review 
are the results on the three-body charmless decays $B^0\to K^0\pi^+\pi^-$~\cite{Garmash:2006fh}, 
$B^+\to K^+K^+K^-$ and $B^+\to K^+\pi^+\pi^-$~\cite{Garmash:2004wa} from the Belle collaboration 
as well as $B^0\to K^+K^- K_S^0$ and $B^0\to\pi^+\pi^- K_S^0$~\cite{Aubert:2005wb} from BaBar. 
Though statistics is very limited in $B$-decays and hence a proper K-matrix analysis cannot be 
performed, the Belle collaboration finds clear signals in the $B^+\to K^\ast(892)^0\pi^+$, 
$B^+\to\rho(770)^0 K^+$, $B^+\to f_0(980)^0 K^+$, and $B^+\to\phi K^+$ decay channels. 
It's pointed out in~\cite{Garmash:2004wa} that all the quasi-two-body branching fractions 
results from Belle and BaBar are in good agreement. The decay mode $B^+\to f_0(980)^0 K^+$ 
is the first observed example of a $B$ decay to a charmless scalar-pseudoscalar final state. 
They determine the mass and width of the $f_0(980)$ to be $(M = 976\pm 4^{+2}_{-3},~\Gamma = 
61\pm 9^{+14}_{-8})$~MeV/$c^2$.
\begin{figure}[t!]
\begin{center}
\includegraphics*[viewport=0 160 790 430,width=1.0\textwidth]{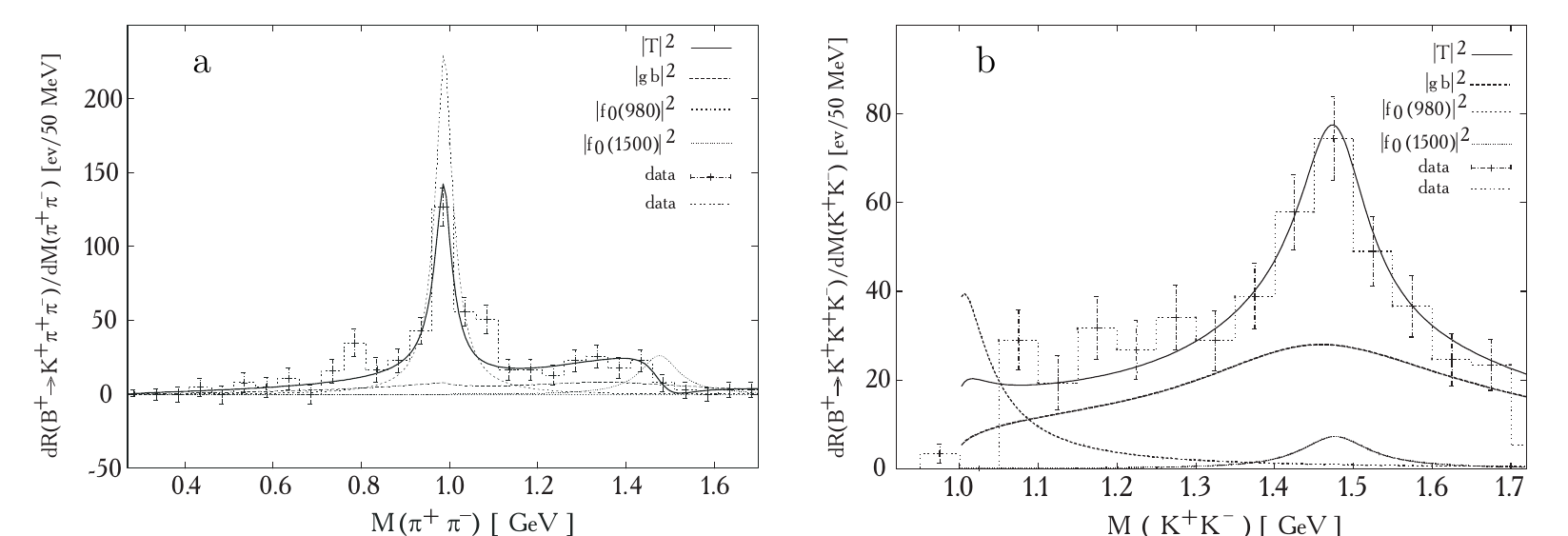}
\end{center}
\vspace{-0.3cm}
\caption[]{\label{fig:belle} Invariant mass spectra in the decays $B^+\to K^+\pi^+\pi^-$
and $B^+\to K^+ K^+ K^-$ measured by the Belle collaboration: (a) $\pi^+\pi^-$ and (b) 
$K^+K^-$. A clear peak at 980~MeV/$c^2$ can be seen in the double-pion spectrum and an 
enhancement around 1400~MeV/$c^2$. The $K^+K^-$ spectrum is dominated by a peak around 
1500~MeV/$c^2$. The solid line represents a fit by Minkowski and Ochs~\cite{Minkowski:2004xf}; 
also shown are the individual resonance terms $|T_R|^2$.}
\end{figure}
Fig.~\ref{fig:belle} shows invariant masses from Belle. A clear signal is visible in
the $\pi^+\pi^-$~mass at 980~MeV/$c^2$ and a broad enhancement in the 1300~MeV/$c^2$
region denoted as $f_X(1300)$, but no signal is observed for the scalar state, 
$f_0(1500)$. However, the invariant $K^+K^-$~mass shows a peak around 1500~MeV/$c^2$ 
denoted as $f_X(1500)$ by the Belle collaboration. The latter is best described in the
analysis as a scalar resonance with mass and width determined from the fit consistent 
with the standard $f_0(1500)$ state. An observation of $f_0(1500)\to K^+K^-$, but no
signal in the decay to $\pi^+\pi^-$ is inconsistent with the standard $f_0(1500)$, 
which is expected to couple more strongly to the two-pion decay. The $f_X(1300)$ 
structure in the $\pi^+\pi^-$~mass is equally well described by a scalar or vector 
amplitude if approximated by a single resonant state. Higher-statistics data are 
required to resolve this issue. All these observations are essentially confirmed by 
BaBar and their conclusion is that the nature of the $f_0(1500)$ remains unclear. An 
identification as the standard PDG $f_0(1500)$ leads to inconsistencies with the 
measurement of the $B^0\to f_0(1500)K_S^0,~f_0(1500)\to\pi^+\pi^-$ decay~\cite{Aubert:2005wb}. 
A new scalar resonance could be proposed with a strong coupling to $K^+K^-$ and a weak 
coupling to $\pi^+\pi^-$. A very similar situation led to the introduction of the 
$f_0(1790)$ by the BES collaboration. 

Also shown in Fig.~\ref{fig:belle} is a fit to the data by Minkowski and Ochs (solid 
line) providing a possible solution to the puzzle~\cite{Minkowski:2004xf}. In their
analysis, the remarkable phenomenon of the $f_0(1500)$ signal in $K\bar{K}$, but
its apparent absence in $\pi^+\pi^-$, is explained by the constructive and destructive
interference with the broad glueball (``red dragon''). They claim that such a 
behavior would be expected from the near octet flavor composition of the $f_0(1500)$. 
The data can then be described using both $f_0(980)$ and $f_0(1500)$ with standard 
properties. No $f_0(1370)$ state is needed in contrast to the analyses performed by
the Belle and BaBar collaborations.

\subsubsection*{\emph{Further Results on Scalar Mesons from Photon-Photon Fusion}}
\label{section:gg_fusion}

\begin{figure}[t!]
\begin{center}
\includegraphics*[viewport=180 70 700 535,width=0.49\textwidth]{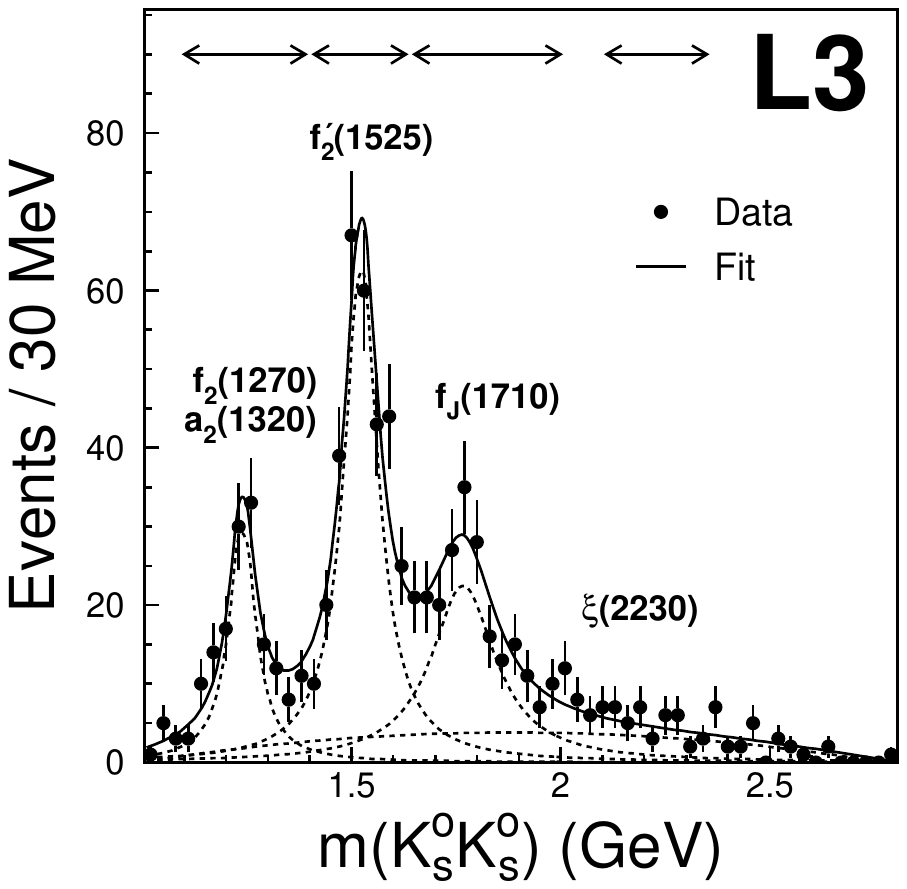}
\includegraphics*[viewport=290 167 540 400,scale=0.85]{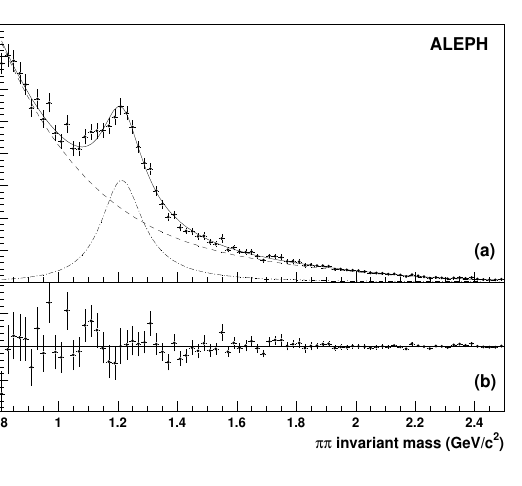}
\end{center}
\vspace{-0.4cm}
\caption[]{\label{fig:lep} Left: invariant $K_S^0 K_S^0$ mass in $\gamma\gamma$ 
  collisions from L3. The solid line corresponds to a maximum-likelihood fit. The 
  arrows represent the $f_2(1270)-a_2^0(1320)$, the $f_{2}^{\,\prime}(1525)$, the 
  $f_J(1710)$, and the $\xi(2230)$ mass regions~\cite{Acciarri:2000ex}. Right: 
  invariant $\pi^+\pi^-$ mass distribution from ALEPH. The top plot (a) shows the 
  fit to the data using a Breit-Wigner function for the $f_2(1270)$ (dot-dashed line), 
  a polynomial for the background (dashed line) and the combination of these functions 
  (solid line). The bottom plot (b) shows the data after subtraction of the fitted 
  curves. Errors are statistical only~\cite{Barate:1999ze}.}
\end{figure}

Apart from proving the existence of particular states, crucial to establishing 
the glueball nature of any glueball candidate is an anti-search in two-photon 
collisions since gluonic states do not couple directly to photons. Results from
$\gamma\gamma$ collisions were reported by the LEP collaborations. Fig.~\ref{fig:lep} 
(left side) shows three peaks below 2~GeV in the invariant $K_S^0 K_S^0$ mass 
distribution observed by the L3 collaboration~\cite{Acciarri:2000ex}. The background 
is fitted by a second-order polynomial and the three peaks by Breit-Wigner functions. 
The mass spectrum is dominated by the formation of tensor mesons, the $f_{2}^{\,\prime}(1525)$
and the $f_2(1270)$ interfering with the $a_2^0(1320)$. A clear signal for the 
$f_J(1710)$ is observed and found to be dominated by the spin-two helicity-two 
state. No resonance is observed in the 2.2 GeV/$c^2$ mass region. The $f_0(1500)$
scalar meson is not seen in its decay to $K_S^0 K_S^0$ in agreement with 
central-production data indicating a small $s\bar{s}$ component if this state is 
interpreted as $q\bar{q}$ meson. Fig.~\ref{fig:lep} (right side) shows the fitted
$\pi^+\pi^-$ spectrum measured by the ALEPH collaboration in $\gamma\gamma$ 
collisions. Only the $f_2(1270)$ is observed and no signals for the $f_0(1500)$
and $f_J(1710)$. Upper limits for the decay into $\pi^+\pi^-$ have been determined
at the 95\,\% confidence level:
\begin{eqnarray}
\label{eqn:lep}
\Gamma(\gamma\gamma\to f_0(1500))\cdot{\cal{B}}(f_0(1500)\to\pi^+\pi^-) <0.31~{\rm keV}\\[1ex]
\Gamma(\gamma\gamma\to f_J(1710))\cdot{\cal{B}}(f_J(1710)\to\pi^+\pi^-) <0.55~{\rm keV}
\end{eqnarray}
Though the $f_0(1500)$ is well established in $\bar{p}p$~annihilation at rest and 
in $pp$ central collisions at 450~Gev/$c$, it has not been observed in $\gamma\gamma$
collisions, yet. Amsler has concluded in~\cite{Amsler:2002ey} that the experimental 
decay rates of the $f_0(1500)$ into two pseudoscalar mesons -- suggesting a dominant 
$n\bar{n}$ structure -- in addition to its missing decay into two photons -- indicating 
a large $s\bar{s}$ component -- are incompatible with a quark-antiquark state. The 
natural explanation for this contradiction would be a mostly gluonic structure of 
the $f_0(1500)$~\cite{Amsler:2002ey}. Further details are discussed in 
section~\ref{section:interpretation_scalars}. Some people, e.g. Close {\it et al.}, 
have also argued that scalar mesons are never observed in $\gamma\gamma$ collisions. 
It would therefore be crucial to study the reaction $\gamma\gamma\to\eta\pi^0\pi^0$ 
because only isovector states should be observed, whereas the discussed $K_S^0 K_S^0$ 
mode has both an isovector and isoscalar component. In particular, the $a_0(1450)$ 
should appear clearly in $a_0(1450)\pi^0\to(\eta\pi^0)\pi^0$.


\subsubsection*{\emph{Radiative \boldmath{$\phi$} Decays into $f_0(980)$ and 
$a_0(980)$ from the KLOE Experiment}}
\label{section:kloe}
\begin{table}[b!]\centering
\addtolength{\extrarowheight}{3pt}
\begin{tabular}{|l|r|c|} \hline
Scalar        & \multicolumn{1}{c|}{${\cal{B}}(\phi\to\gamma$ + meson)} & Number of Events\\ \hline
$f_{0}(980)\to\pi^0\pi^0$ & $(1.49\pm 0.07)\times 10^{-4}$~\cite{Aloisio2002b} & $2438\pm 61$~\cite{AloIsio2002}\\
$f_{0}(980)\to\pi^+\pi^-$ & $(2.1 - 2.40)\times 10^{-4}$~\cite{Ambrosino:2005wk} &\\
$a_{0}(980)\to\pi^0\eta$ & $(7.4\pm 0.70)\times 10^{-5}$~\cite{Aloisio2002b} & 802~\cite{Aloisio2002b}\\ \hline
\end{tabular}
\caption[]{\label{tab:kloe}KLOE results on radiative $\phi$ decays.}
\end{table}
The KLOE collaboration reported on radiative $\phi$ decays into $f_0(980)$ and 
$a_0(980)$~\cite{Aloisio2002b,AloIsio2002,Ambrosino:2005wk}. As pointed out 
in~\cite{Close:2001ay}, these decays have long been recognized as a potential route 
towards disentangling the nature of these states. The magnitudes of the decay widths 
are rather sensitive to the fundamental structures of the $f_0(980)$ and $a_0(980)$, 
and can possibly discriminate amongst models. According to different interpretations, 
the $\phi\to a_0\gamma$ branching fraction can range from $10^{-5}$ for a mostly 
$q\bar{q}$ and $K\bar{K}$ structure to $10^{-4}$ for $q\bar{q}q\bar{q}$. The ratio 
${\cal{B}}(\phi\to f_0(980)\gamma)\,/\,{\cal{B}}(\phi\to a_0(980)\gamma)$ is also 
highly dependent on the structure of the scalars~\cite{Close:2001ay}. The $K\bar{K}$ 
and $\eta\eta$ thresholds produce sharp cusps in the energy dependence of the resonant 
amplitude and pose further challenges in the models. Table~\ref{tab:kloe} shows the 
branching fractions for the decays into two pions and $\pi^0\eta$, respectively. They 
determined the ratio of the two branching fractions and the ratio of the two couplings 
to the $KK$ system to be:
\begin{alignat}{2}
\frac{{\cal{B}}(\phi\to f_0(980)\gamma)}{{\cal{B}}(\phi\to a_0(980)\gamma)}\; & = \; 
          6.1\,\pm\,0.6\\[1ex]
\frac{g^2_{f_0 K K}}{g^2_{a_0 K K}}\; & = \; 7.0\,\pm\,0.7
\end{alignat}
More recently, a new value for the branching ratio of the $\phi\to\pi^0\pi^0\gamma$
process has been obtained~\cite{Ambrosino:2006hb}:
\begin{eqnarray}
{\cal{B}}(\phi\to S\gamma\to\pi^0\pi^0\gamma) = 1.07\,^{+0.01}_{{-0.03}^{\rm 
\,fit}}\,^{+0.04}_{{-0.02}^{\rm \,syst}}\,^{+0.05}_{{-0.06}^{\rm \,mod}}\times 10^{-4}\,
\end{eqnarray}
where the last error reflects the maximum variation observed when changing the fit 
model. The collaboration concludes that the couplings extracted in the Kaon Loop 
model~\cite{Achasov:2005hm} provide a stable description of the data with a large 
coupling of the $f_0(980)$ to kaons, as also indicated by the study of the $\pi^+
\pi^-\gamma$ final state indicating a 4-quark structure of the $f_0(980)$ meson. 
These results are also in good agreement with further predictions for the $f_0(980)$
and $a_0(980)$ to be four-quark states in~\cite{Boglione:2003xh,Oller:2002na}; 
however, this model may be oversimplified~\cite{pdg_scalar3}. Starting from 
unitarized chiral Lagrangians, the scalar mesons can also be generated 
dynamically~\cite{Marco:1999df,Palomar:2003rb,Caprini:2005zr}. In particular,
no direct $\phi\to f_0\gamma,~a_0\gamma$ coupling has been used 
in~\cite{Palomar:2003rb} in accordance with the philosophy that these two
resonances are dynamically generated by meson-meson interaction.

\begin{figure}[t!]
\begin{center}
\includegraphics*[viewport=140 270 470 540,width=0.97\textwidth]{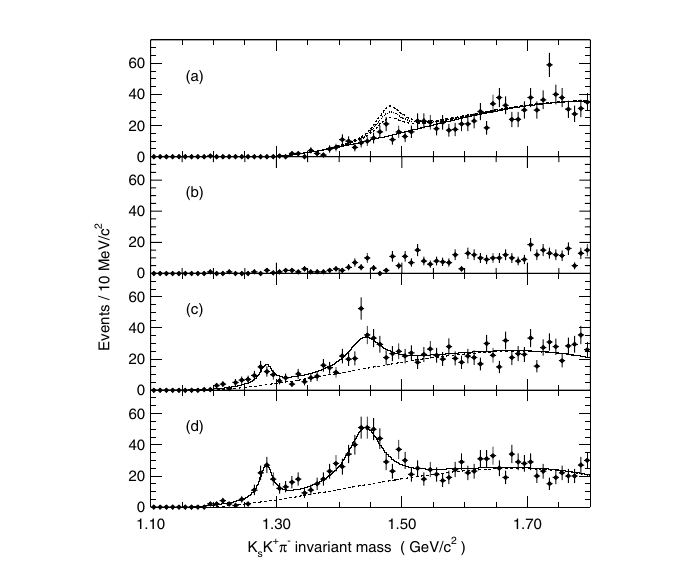}
\end{center}
\vspace{-0.5cm}
\caption[]{\label{fig:cleo}Distributions of the $K_S^0 K^\pm \pi^\mp$ invariant mass
from CLEO for data events detected with (a) $p_\bot\leq 100$~MeV/$c$, (b) $100~{\rm 
MeV}/c\leq p_\bot\leq 200$~MeV/$c$, and (c) $200~{\rm MeV}/c\leq p_\bot\leq 600$~MeV/$c$ 
in the untagged mode, and (d) for all $p_\bot$ in the tagged mode~\cite{Ahohe:2005ug}. 
The dashed curves in (a) show the strength of the expected $\eta(1475)$ signal according 
to the L3 results~\cite{Acciarri:2000ev}. The solid curves in (a), (c) and (d) are the 
results of binned maximum likelihood fits for resonances with a polynomial approximation 
to the non-interfering combinatorial background. See text for details.}
\end{figure}

It remains a strong possibility that the two states $f_0(980)$ and $a_0(980)$, together
with the $f_0(600)$ (or $\sigma$ meson) and the ``missing'' $K_0^\ast(800)$ (or $\kappa$ 
meson) form a low-mass nonet of predominantly four-quark states, where at larger distances
the quarks recombine into a pair of pseudoscalar mesons. The low-lying scalar mesons
can then also be considered meson-meson molecules or what is generally referred to as 
dynamically generated resonances forming by a meson cloud~\cite{Oller:1997ti,pdg_scalar3}.

\subsubsection*{\emph{Pseudoscalar States}}
\label{section:pseudoscalars}

\begin{table}[t!]\centering
\addtolength{\extrarowheight}{3pt}
\begin{tabular}{cccc|c}
 CLEO & $\eta(1295)$ & $\eta(1405)$ & $\eta(1475)$ & L3 on $\eta(1475)$\\\hline
Mass in [MeV/$c^2$] & 1294 & 1410 & 1481 & -\\
Upper limit at 90\,\% CL [eV] & 14 & 24 & 89 & $212\pm 50\pm 23$\\\hline
\end{tabular}
\caption{\label{tab:cleo_pseudoscalars} Upper limits on the two-photon partial
widths of pseudoscalar resonances from CLEO: $\Gamma_{\gamma\gamma}(\eta){\cal{B}}
(K\bar{K}\pi)$. For comparison, the L3 result is also given.}
\end{table}

Two pseudoscalar mesons are reported in the 1400-1500 MeV/$c^2$ mass region and
are listed as two separate states by the Particle Data Group, the
$\eta(1405)$ and $\eta(1475)$. Long considered only one resonance, the 
$\eta(1475)$ was first observed in two-photon collisions in 2001 decaying to $K\bar{K}
\pi$ by the L3 collaboration~\cite{Acciarri:2000ev}. The reported two-photon
partial width is $212\pm 50~({\rm state.})\,\pm 23$ ({\rm sys.}) eV. The second 
$\eta$ state is neither observed in $K\bar{K}\pi$ nor in $\eta\pi\pi$ by L3, 
suggesting a large gluonic content of the $\eta(1405)$. In 2005, CLEO published 
negative results on the search for $\eta(1475)\to K_S^0 K^\pm \pi^\mp$. The
non-observation of any pseudoscalar meson below 1700 MeV/$c^2$ was based on 5~times 
more statistics and the results are more than 2 standard deviations inconsistent  
with the L3 findings. Fig.~\ref{fig:cleo} shows the distributions of the $K_S^0 
K^\pm \pi^\mp$ invariant mass for different ranges of $p_\bot$. Signals due to 
pseudoscalar mesons are expected in (a). The authors observe two statistically 
significant enhancements in the $\eta(1475)$ mass region (Fig.~\ref{fig:cleo}~(c)). 
However, the enhancements have large transverse momentum which rules them out as 
being due to pseudoscalar resonances~\cite{Ahohe:2005ug}. The observation is consistent 
with the production of axial-vector mesons. Table~\ref{tab:cleo_pseudoscalars} summarizes 
CLEO upper limits on the two-photon partial widths of pseudoscalar states. All these 
are consistent with the glueball and the radial excitation hypotheses of the 
$\eta(1295)$, $\eta(1405)$, and $\eta(1475)$. Unfortunately with no state observed, 
it is difficult to come to a conclusion about the nature of the pseudoscalar resonances. 
Only two ground-state axial vector mesons were reported in this mass range consistent 
with quark model expectations~\cite{Ahohe:2005ug}.

\subsubsection*{\emph{The Flavor Filter in the Decay \boldmath{$J/\psi\to\gamma\,[\gamma\,V]$}}}

\begin{figure}[t!]
\begin{center}
\includegraphics*[angle=-90,viewport=190 270 410 520,width=0.49\textwidth]{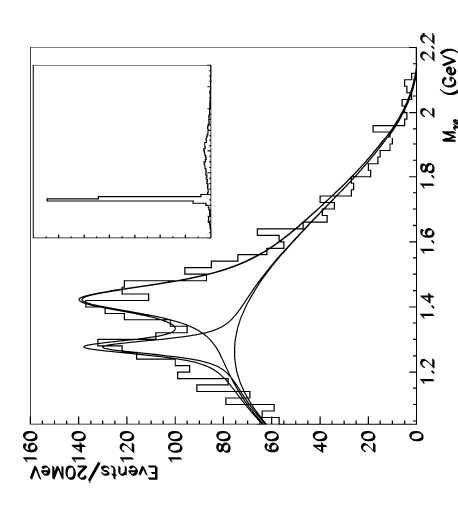}
\includegraphics*[angle=-90,viewport=190 270 410 520,width=0.49\textwidth]{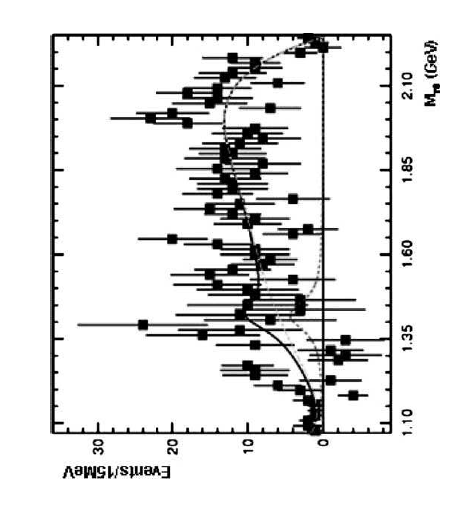}
\end{center}
\vspace{-0.2cm}
\caption[]{\label{fig:bes_gV}Invariant $\gamma V$ mass distributions for $V=\rho$ 
(left) and $V=\phi$ (right) from~\cite{Bai:2004qj}. The insert shows the full mass 
scale with the $\eta\,^\prime$ clearly visible. The $\gamma\phi$ distribution is 
side-band background subtracted and no structure is observed around 1400~MeV/$c^2$.}
\end{figure}

BES also studied $J/\psi\to\gamma\gamma V(\rho,\,\phi)$ decays with the BES-II 
detector~\cite{Bai:2004qj}. A resonance around 1420~MeV/$c^2$ was observed in the 
$\gamma\rho$ mass spectrum. Fig.~\ref{fig:bes_gV} (left) shows the invariant $\gamma
\rho$ mass. The properties of this state were determined to be $M=(1424\pm 10_{\rm stat}
\pm 11_{\rm sys})$~MeV/$c^2$ and $\Gamma = (101.0\pm 8.8\pm 8.8)$~MeV/$c^2$. A branching 
ratio of ${\cal{B}}(J/\psi\to\gamma X(1424)\to\gamma\gamma\rho) = (1.07\pm 0.17\pm 0.11)
\times 10^{-4}$ was obtained. A corresponding search for the state decaying into $\gamma
\phi$ only resulted in a 95\,\% CL upper limit of ${\cal{B}}(J/\psi\to\gamma X(1424)\to
\gamma\gamma\phi) < 0.82\times 10^{-4}$ (right side of Fig.~\ref{fig:bes_gV}). The 
authors do not draw a definite conclusion and the situation is puzzling. If the $X(1424)$ 
is identified with the $\eta(1405)$ and considered a glueball candidate, it should decay 
into $J/\psi\to\gamma\,(\gamma\phi)$.

Radiative decays of a resonance near 1440~MeV/$c^2$ in $J/\psi\to\gamma V~(V = \rho,\,
\phi$) was studied in earlier experiments by Crystal-Ball~(published as Ph.D. thesis),
Mark-III in 1990~\cite{Coffman:1989nk}, and DM2 in 1990~\cite{Augustin:1989zf}. In 
the summary, DM2 concludes that ``the signal observed in $J/\psi\to\gamma\rho\gamma$ 
is hardly consistent with the $\eta(1430)$'' and Mark-III is unable to distinguish
between between the $f_1(1420)$ and a pseudoscalar state.

\subsubsection*{\emph{Tensor States}}
\label{section:tensors}
\begin{figure}[t!]
\includegraphics*[viewport=30 250 430 540,width=0.68\textwidth]{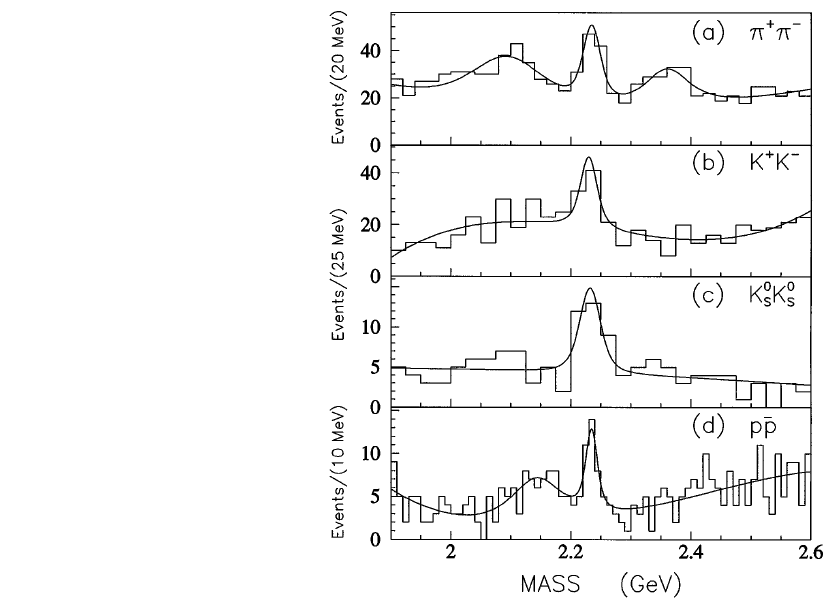}
\caption[]{\label{fig:xi2220}Fitted invariant mass spectra of a) $\pi^+\pi^-$, 
b) $K^+K^-$, c) $K_S^0 K_S^0$, and d) $p\bar{p}$ from BES suggesting the 
existence of the $f_J(2220)$. An unbinned maximum-likelihood method was applied 
using a smooth background plus one or several Breit-Wigners resonances convoluted 
with Gaussian resolution functions~\cite{Bai:1996wm}.}
\end{figure}

Evidence for the $2^{++}$ glueball is weak. The BES collaboration
observed signals in radiative $J/\psi$ decays for the $f_J(2220)$, also known as 
$\xi(2230)$, in a sample of more than $5\times 10^6~J/\psi$ decays in final states 
including $K\bar{K},~\pi^+\pi^-,$ and $p\bar{p}$~\cite{Bai:1996wm,Bai:1998tx}.
Fig.~\ref{fig:xi2220} shows the fitted invariant mass spectra. However, this signal
has not yet been confirmed by the larger data sets from BES II. Visual inspection
of the newer invariant mass spectra seem to show no evidence for this state (see 
for example Fig.~\ref{fig:bes_summary},~e-g). However, it has been pointed out 
in~\cite{Asner:2008nq} that it is difficult to exclude the existence of the 
$\xi(2220)$ based on preliminary results of a partial wave analysis of 
$J/\psi\to\gamma K^+ K^-$ data from BES-II; a signal of $4.5\sigma$ significance 
with mass, width and product branching fraction consistent with the BES-I results
was found. Given that these data can help resolve this puzzle, it is hoped that BES-II 
or BES-III will be able to say more on the issue in the future.

\begin{table}[b!]\centering
\addtolength{\extrarowheight}{3pt}
\begin{tabular}{c|cccr}
\hline
 & Stickiness & Gluiness & & \multicolumn{1}{c}{Reference}\\\hline
$f_0(1500)$ & $>\,1.4$ & & & ALEPH Collaboration~\cite{Jones:2000nq}\\
$f_J(1710)$ & $>\,0.3$ & & & ALEPH Collaboration~\cite{Jones:2000nq}\\
$\eta(1440)$ & $79\pm 26$ & $41\pm 14$ & & L3 Collaboration~\cite{Acciarri:2000ev}\\
$\eta\,^\prime(958)$ & $3.6\pm 0.3$ & $5.2\pm 0.8$ & &\\
$f_2(2220)$ & $>\,109$ & & & CLEO Collaboration~\cite{Benslama:2002pa}\\\hline
\end{tabular}
\caption{\label{tab:stickiness}Stickiness and gluiness for some of the mesons. The 
number for the $f_{2}(2220)$ assumes that the state exists.}
\end{table}

The $\xi(2230)$ signal was first observed by the MARK-III collaboration and published in 
1985 in the reactions $J/\psi\to\gamma K^0_S K^0_S$ and $J/\psi\to\gamma K^+ K^-$ based
on a sample of $5.8 \times 10^6~J/\psi$ decays~\cite{Baltrusaitis:1985pu}. However, 
limits on the product branching fraction, ${\cal{B}}(J/\psi\to\gamma f_J(2220))\cdot 
{\cal{B}}(f_J(2220)\to K^+K^-)$, reported by the DM2 collaboration were in disagreement
with the MARK-III findings~\cite{Augustin:1987da}. The first indication for a spin-2
particle came from the GAMS collaboration in 1986. They observed a signal at 2220 
MeV/$c^2$ decaying to $\eta\eta\,^{\prime}$ in the $\eta\eta\,^{\prime} n$ final state 
from pion-induced reactions on the proton~\cite{Alde:1986nx}. This finding was in 
agreement with the original result from radiative $J/\psi$ decays. 

The CLEO collaboration has also reported on the possible glueball candidate $f_J(2220)$. 
An upper limit of $\Gamma_{\gamma\gamma}{\cal{B}}(f_J(2220)\to K^0_S K^0_S\leq 1.1)$~eV 
at 95\,\% C.L. was derived from two-photon interactions, $\gamma\gamma\to f_J\to 
K^0_S K^0_S$, using the CLEO II detector~\cite{Benslama:2002pa}. The same approach
for the $f_{2}^{\,\prime}(1525)$ leads to consistent results with PDG values. The CLEO 
observation is in agreement with the published result from the L3 collaboration 
(Fig,~\ref{fig:lep}) of $\Gamma_{\gamma\gamma}{\cal{B}}(f_J\to K^0_S K^0_S\leq 1.4)$~eV 
at 95\,\% C.L.~\cite{Acciarri:2000ex}. The non-observation of a signal in the
2.2 GeV/c$^2$ mass region in two-photon fusion is certainly expected for a true 
glueball, but the non-existence of this narrow state is also not excluded. Assuming 
that the BES cross section is correct, the CLEO authors determine a large lower limit 
for the ``stickiness'' of $>109$ at the 95\,\% C.L. Further results on stickiness and 
gluiness are summarized in Table~\ref{tab:stickiness}.

Many other experiments carefully searched for the $f_J(2220)$ in proton-antiproton 
annihilation in-flight, but no evidence was found of this state~\cite{Hasan:1996jy, 
Bardin:1987at,Sculli:1987bg,Evangelista:1997be,Evangelista:1998zg,Buzzo:1997vh}. A 
high-statistics search in 2000 by the Crystal-Barrel collaboration also showed no narrow 
state~\cite{Seth:2000tm}. If the state really exists, it has a very large branching 
fraction in radiative $J/\psi$~decays.

\subsection{\emph{Evidence for \boldmath{$\eta(1295)$} using \boldmath{$\pi^- p$} 
Reactions: The E852 Experiment}}
\label{section:e852}
The $\eta(1295)$ is the lowest-mass pseudoscalar meson above the $\eta\,^\prime$ 
and thus, often interpreted as the first radial excitation of the $\eta$ meson. It
is also degenerate in mass with the $\pi(1300)$ suggesting ideal mixing. The PDG
has listed the $\eta(1295)$ meson in the summary table~\cite{Amsler:2008zz}. Although 
the resonance has not been observed in proton-antiproton annihilation, $\gamma\gamma$
collisions, central production, and $J/\psi$ decays, evidence seems to be solid
in pion-induced reactions. It was first observed in a PWA analysis of the $\eta\pi\pi$
system~\cite{Stanton:1979ya} and later confirmed by other experiments, e.g.~\cite{Ando:1986bn,
Fukui:1991ps,Alde:1997vq,Manak:2000px,Adams:2001sk}. A small $\eta(1295)$ signal at
1255~MeV/$c^2$ was found in the analysis of $\bar{p}p\to\eta\pi^+\pi^-\pi^+\pi^-$
\cite{Anisovich:2001jb}, but the peak was later explained by an insufficient simulation 
of trigger conditions in the Monte Carlo. In particular, the E852 experiment is probably 
the only experiment with clear signals for the $\eta(1295)$, the $\eta(1405)$ and the 
$\eta(1475)$.

\begin{figure}[tb!]
\begin{center}
\includegraphics*[viewport=200 190 600 400,width=1.0\textwidth]{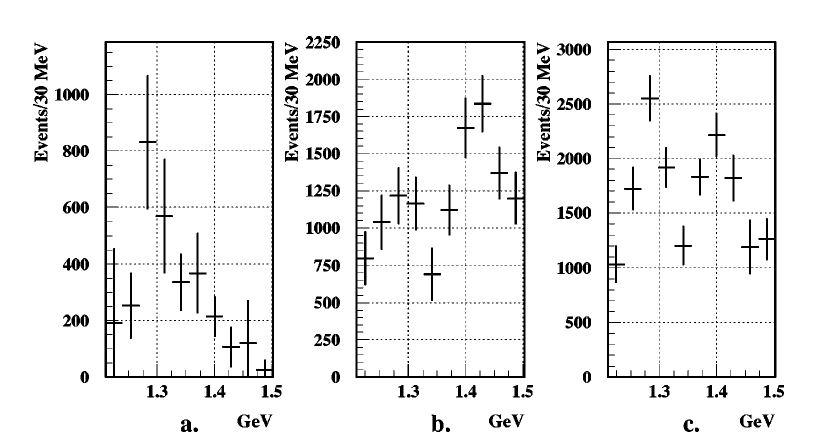}
\end{center}
\vspace{-0.5cm}
\caption[]{\label{fig:e852}Fitted intensity distributions from E852 for the $0^{-+}$
wave~\cite{Manak:2000px}: (a) $a_0\pi$ intensity, (b) $\eta(\pi\pi)_S$ intensity, and
(c) total $0^{-+}$ intensity. A sharp peak at about 1300~MeV/$c^2$ is visible in (a),
consistent with the assumption of $\eta(1295)\to a_0(980)\pi$.}
\end{figure}

\begin{table}[b!]\centering
\addtolength{\extrarowheight}{3pt}
\begin{tabular}{l|ccc}
 & Mass [GeV/$c^2$] & Width [GeV/$c^2$] & $\cal{B}$$(\eta\to a_0\pi)$ /
                                          $\cal{B}$$(\eta\to \sigma\eta)$\\\hline
$\eta(1295)$ & $1.282\pm 0.005$ & $0.066\pm 0.013$ & $0.48\pm 0.22$\\
$\eta(1440)$ & $1.404\pm 0.006$ & $0.080\pm 0.021$ & $0.15\pm 0.04$\\\hline
\end{tabular}
\caption{\label{tab:e852}Properties of the $J^{PC} = 0^{-+}$ states from 
E852~\cite{Manak:2000px}.}
\end{table}

The E852 collaboration performed a partial-wave analysis of the $\eta\pi^+\pi^-$ system 
produced in the reaction $\pi^- p\to \eta\pi^+\pi^-\,n$ at 18~GeV based on 9082 events 
in the $1205\leq M(\eta\pi^+\pi^-)\leq 1535$~MeV/$c^2$~\cite{Manak:2000px}.
Since the $J^{PC} = 0^{-+}$ states often have an $a_0(980)\pi$ decay mode and $a_0(980)$
couples to both $\eta\pi$ and $K \bar{K}$~final states, a comparison of resonances 
produced in $\eta\pi^+\pi^-$ and $K \bar{K}\pi$ can help identify states. 
Fig.~\ref{fig:e852} shows the fitted intensity distributions for the $0^{-+}$ wave.
A clear enhancement is observed around 1300~MeV/$c^2$ in panel (a), which is identified 
with the $\eta(1295)\to a_0(980)\pi$. The distribution in (b) shows a double-peak
structure indicating contributions from $\eta(1295)$ and $\eta(1440)$. It's pointed out
in the publication that the dominant higher-mass peak is somewhat inconsistent with 
previous analyses~\cite{Fukui:1991ps,Ando:1986bn}, which do not see it as dominant.
The result of the incoherent sum of the two modes is shown in Fig.~\ref{fig:e852}~(c).
In summary, the authors find that the lower-mass region is dominated by the $\eta(1295)$,
accounting for about 80\,\% of the signal. The dominance has a significant influence
on the $f_1(1285)$ branching fractions. The properties of the pseudoscalars states are
summarized in Table~\ref{tab:e852}.

\begin{figure}[t!]\centering
\includegraphics[viewport=0 0 600 560,width=0.6\textwidth]{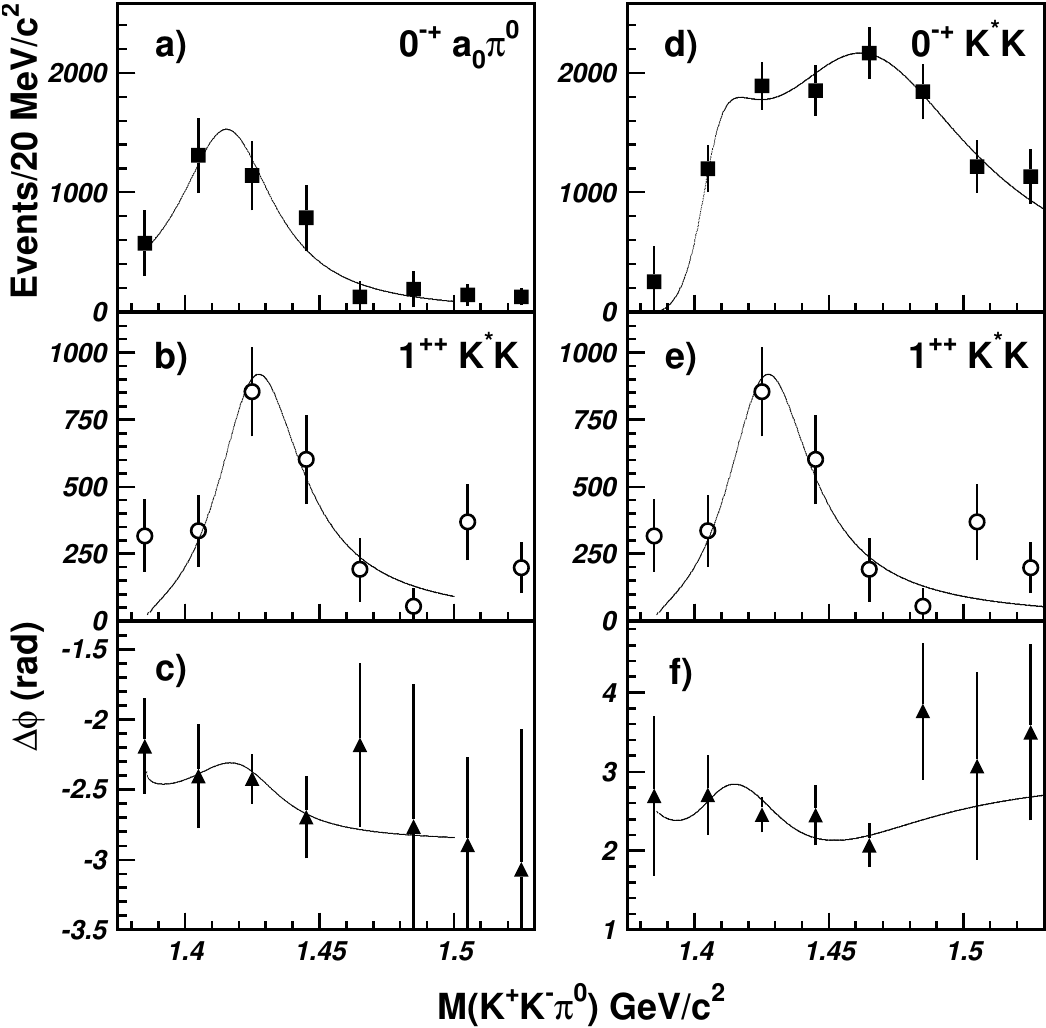}
\vspace{1mm}
\caption[]{\label{fig:e852KKpi}The intensity observed in the $K\bar{K}\pi$ final 
state from E852~\cite{Adams:2001sk}. 
(a) shows the $0^{-+}$ intensity in the $a_{0}(980)\pi$ partial wave,
(b) and (e) show the $1^{++}$ intensity in the $K^{*}K$ partial wave,
(d) shows the $0^{-+}$ intensity in the $K^{*}K$ partial wave, 
and (e) and (f) shows the phase differences between (a) and (b) and between
(d) and (e) respectively.}
\end{figure}

In a later analysis, E852 looked at the reaction $\pi^{-}p\rightarrow K\bar{K}\pi p$
\cite{Adams:2001sk}. In the $K\bar{K}\pi$ final state, they observed two partial
wave, $KK^{*}$ and $a_{0}\pi$, both of which coupled to $0^{-+}$ quantum numbers.
Figure~\ref{fig:e852KKpi} shows the results of the partial wave analysis of this
channel. There is clear evidence for  two pseudoscalar states above $1.4$~GeV as 
well as the $f_{1}(1485)$ state. The lower mass $0^{-+}$ state, $\eta(1405)$, couples
to both $a_{0}\pi$ and $KK^{*}$, while the higher mass state, $\eta(1475)$, couples
mostly to $KK^{*}$. 

\section{Gluonic Excitations}
\subsection{\label{sec:pseudoscalar}\emph{The Pseudoscalar Mesons}}
Within the pseudoscalar sector, the ground states are the well established
$\eta(548)$ and $\eta\,^\prime(958)$. Only radial excitations of these states
are expected in the framework of the quark model. Beyond the simple quark
model, a nonet of hybrid pseudoscalar mesons is expected in the $1.8$ to 
$2.2$~GeV/c$^{2}$ mass region, and a glueball is expected in the $\sim 2$~GeV/c$^{2}$
region. Table~\ref{tab:pseudoscalarmesons} shows the current known pseudoscalar 
states. The $\eta(1295)$ and the $\eta(1475)$ are often considered as the radial
excitations. The $\eta(1760)$ is often taken as the partner of the $\pi(1800)$
which leaves the $\eta(1405)$ as the odd state out. The higher mass $\eta(2225)$
has a mass that is not inconsistent with a glueball interpretation, but the single
measurement of this state needs confirmation, and more of its decay modes need to
be measured. 

In the initial studies of radiative $J/\psi$ decays, a great deal of excitement
was raised over the very large cross section for a pseudoscalar state known as
the $\iota$~\cite{Edwards:1982nc,Augustin:1989zf,Scharre:1980zh}.
The apparently large production of this state, now known as the $\eta(1405)$, led 
to speculation that this was a glueball, or at least a state with sizable glueball 
content. Later analysis by MARK III~\cite{Bai:1990hs} indicated that there are 
actually two pseudoscalar states in this mass region. This splitting was later
confirmed by the OBELIX collaboration~\cite{Bertin:1995fx}. 

While the two pseudoscalar states near $1400$~MeV/$c^2$ are listed by the 
PDG~\cite{Amsler:2008zz}, there has been some speculation by Klempt~\cite{Klempt:2007cp} 
that they are in fact a single state with a node in the wave function. There is also 
some speculation~\cite{Klempt:2007cp} that the $\eta(1295)$ may not exist. In its 
clearest observations, it is always seen in conjunction with the $f_{1}(1285)$ and 
it could possibly be explained as feed through from the $1^{++}$ state. 

Under this latter interpretation, the $\eta(1405)$ and $\eta(1475)$ become the 
radial excitations of the ground state pseudoscalar mesons. This is consistent 
with the observations in $\bar{p}p$ annihilation, where both the $\eta(1405)$ and
$\eta(1475)$ are strongly produced, but the $\eta(1295)$ is not seen. 

A detailed review of the pseudoscalar sector was carried out by Masoni~\cite{Masoni:2006rz} 
and colleagues. They argue that since three states are observed where two are expected
that the extra state is the pseudoscalar glueball. They suggest that the $\eta(1405)$
is in fact the extra state. This arguement relies on the existence of the $\eta(1295)$,
and they argue that while the evidence is weak in all experiments, there are a number
of them in different production mechanisms. From this they conclude that the state 
does likely exist.

Our concern here is the existance of the $\eta(1295)$. The original DM2~\cite{Augustin:1989zf} 
observation has not been confirmed by any of the later $J/\psi$ experiments, even
with higher statistics. It also appears that while the higher mass pseudoscalers
are strongly produced in $p\bar{p}$ annihilation, no evidence of the $\eta(1295)$ 
is observed. The only other observations of the $\eta(1295)$ are always in conjunction
with the nearby $f_{1}(1285)$. Without a confirmation of the $\eta(1295)$, we feel that
one is unable to associate the pseudoscalar glueball with the $\eta(1405)$. A good 
discussion of the history of this and the current state of these two states can be 
found in the PDG mini-review~\cite{Amsler:iota:2006} on this topic. There is hope that 
new data from BES-III and COMPASS will be able to help clarify the situation.


\subsection{\label{sec:tensormesons}\emph{The Tensor Mesons}}
Within the quark model, there are two quark configurations which yield $2^{++}$ 
quantum numbers, the $^3P_{2}$ $(L=1,S=1,J=2)$ and the $^3F_2~(L=3,S=1,J=2)$
nonets. Members of the former are the well established isoscalar states $f_{2}(1270)$ 
and $f_{2}^{\,\prime}(1525)$, while the latter are expected in mass similar to the $4^{++}$ 
states of the same orbital angular momentum (observed near 2~GeV/$c^2$). For 
both of these nonets, radial excitations are also expected. 

Evidence for a tensor glueball is essentially non-existent. Below 2~GeV/$c^2$, only one 
further resonance is listed in the Meson Summary Table, the $f_{2}(1950)$. Four additional 
reported resonances require confirmation. For this reason, none of the reported 
isoscalars above the $f_{2}^{\,\prime}(1525)$ can be definitely assigned to one of the nonets 
2$^3P_{2}$, 3$^3P_{2}$, or 1$^3F_2$, and the identification of a glueball as a 
resonance that has no place in the $q\bar{q}$ nonets is premature. The latest mini-review 
published by the Particle Data Group in 2004 considers evidence for the $f_2(1565)$ 
observed in $p\bar{p}$ annihilation at rest as more solid. It could be a member of the 
2$^3P_{2}$ nonet and is perhaps the same state as the $f_2(1640)$, seen in its decay 
to $\omega\omega$ and $4\pi$. However, there has been speculation from Dover~\cite{Dover:1990kn}
that since this state only appears in $p\bar{p}$ annihilation, it may be associated 
with a quasinuclear $\bar{N}N$ bound state.

Above 2~GeV/$c^2$, the BES collaboration has reported on the $f_J(2220)$, likely $J=2$, 
and has considered it a glueball candidate since it is produced strongly in radiative 
$J/\psi$ decays and seems to be non-existent in $\gamma\gamma$ collisions. However,
careful searches by other experiments could not confirm this resonance. It has neither
been observed in radiative $\Upsilon$ decays, nor in formation in $p\bar{p}$ annihilation
into $K^+ K^-$, $K_S K_S$, $\phi\phi$, $\eta\eta$, or $\pi\pi$. The evidence is thus
very weak and more data is needed to clarify the situation. The data from BES-II shown
in Figure~\ref{fig:bes_twopi} seem to show no evidence for the BES peaks seen in
Figure~\ref{fig:xi2220}. The BES-II data have substantially more statistics than those
of BES-I, so this could presumably set a much lower limit on the production of such a
state. Hopefully, the analysis of the BES-II data will be able to make definitive 
statements about the $\xi(2230)$. Data from BES-III will also help in this,
confirmation in both the much higher statistics data sets as well as from some other 
channel would certainly be desirable. At this point, the existence of this state looks
extremely dubious.

Moving beyond the quark-model picture, a nonet of hybrid states is expected
in the $1.8$ to $2.2$~GeV/c$^{2}$ mass region, and a $2^{++}$ glueball is expected
in the $~2$~GeV/c$^{2}$ mass region. Needless to say, the tensors are expected
to be an extremely busy sector, and this is clearly born out by the large number
of states in the PDG (see Table~\ref{tab:tensormesons}).

\subsection{\emph{The Scalar Sector}}
\label{section:interpretation_scalars}
The $J^{PC} = 0^{++}$ $(L=1,~S=1)$ scalar sector is without doubt the most complex 
one and the interpretation of the states' nature and nonet assignments are still 
very controversial. In particular, the number of observed $I=0$ isosinglet states 
with masses below 1.9~GeV/$c^2$ is under debate. According to the PDG mini-review 
on non-$q\bar{q}$ candidates~\cite{Amsler:2008zz}, five isoscalar resonances are 
well established: the very broad $f_0(600)$ or so-called $\sigma$ state, the 
$f_0(980)$, the broad $f_0(1370)$, and the rather narrow $f_0(1500)$ and $f_0(1710)$ 
resonances. Naive arguments without chiral symmetry constraints and the close
proximity of states in other $J^{PC}$ nonets suggest that the $f_0(600)$, $f_0(980)$, 
and $a_0(980)$ are members of the same nonet. The missing $I=1/2$ state -- usually 
called $K^\ast_0(800)$ or $\kappa$ -- is not listed by the Particle Data Group as
well established state in its latest 2008 edition. The nature of this nonet is not 
necessarily $q\bar{q}$. Very often, $f_0(980)$ and $a_0(980)$ are interpreted as 
multi-quark states or $K\bar{K}$ molecules~\cite{Weinstein:1990gu,Locher:1997gr}.

Using the same naive arguments, the $a_0(1450)$, $K^\ast(1430)$, and two states out 
of the $I=0$ group, $f_0(1370)$, $f_0(1500)$, and $f_0(1710)$, would form an SU(3) 
flavor nonet. These nonet assignments however pose some serious challenges. While 
almost all models agree on the $K^\ast(1430)$ to be the quark model $s\bar{u}$ or 
$s\bar{d}$ state, the situation is very ambiguous for the isoscalar resonances. The 
most striking observation is that one $f_0$ state appears supernumerary, thus leaving 
a non-$q\bar{q}$ (most likely) glueball candidate. Both the $f_0(1370)$ and $f_0(1500)$ 
decay mostly into pions. In fact, all analyses agree that the $4\pi$ decay mode accounts 
for at least half of the $f_0(1500)$ decay width and dominates the $f_0(1370)$ decay
pointing to a mostly $n\bar{n}$ content of these states. On the other hand, the LEP 
experiments indicate that the $f_0(1500)$ is essentially absent in $\gamma\gamma\to 
K\bar{K}$ (L3 collaboration~\cite{Acciarri:2000ex}) and $\gamma\gamma\to\pi^+\pi^-$ 
(ALEPH collaboration~\cite{Barate:1999ze}). If the state were of $q\bar{q}$~nature, 
the extremely small upper limit for the branching fraction into $\pi^+\pi^-$~(\ref{eqn:lep}) 
would suggest a mainly $s\bar{s}$ content.

\begin{figure}[t!]
\begin{center}
\includegraphics*[viewport=190 165 600 440,width=0.69\textwidth]{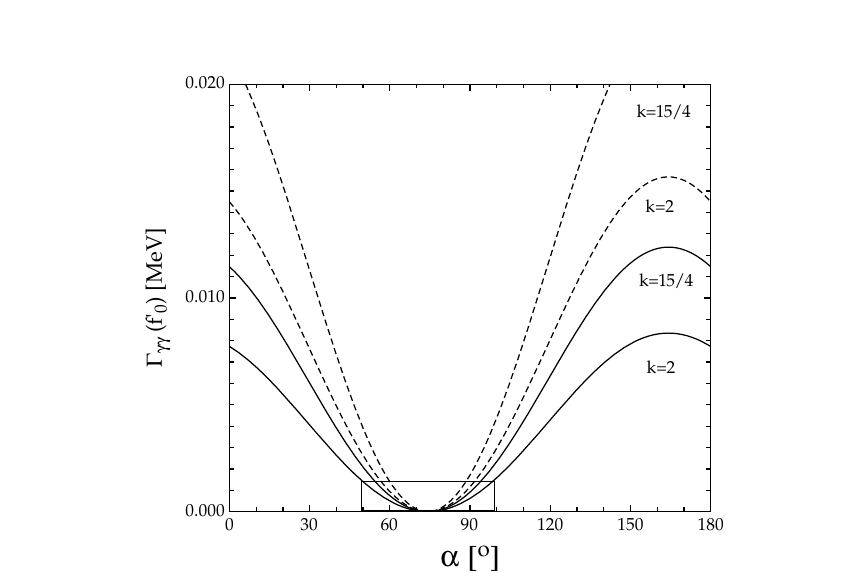}
\end{center}
\vspace{-0.2cm}
\caption[]{\label{fig:two_photon_width}Predicted two-photon width for the $f_0(1500)$
(solid curves) and $f_0(1710)$ (dashed curves) assuming a $q\bar{q}$ structure. The 
box shows the 95\,\% confidence-level upper limit from ALEPH for the $f_0(1500)$. 
The picture is taken from~\cite{Amsler:2002ey}.}
\end{figure}

This large $s\bar{s}$ component can be understood from the following argument. The 
two-photon width of an isoscalar meson with mass $m$ is given within SU(3) by~\cite{Amsler:2002ey}:
\begin{eqnarray}
  \Gamma_{\gamma\gamma}\,=\,c\,(5\,{\rm cos}\,\alpha - \sqrt{2} {\rm sin}\,\alpha)^2\,\,m^3\,,
\end{eqnarray}
where $c$ is a nonet constant, which can be calculated for the tensor nonet from the
measured two-photon width of the $f_{2}^{\,\prime}(1525)$. In a non-relativistic 
calculation, the two-photon width for a scalar meson with mass $m_0$ can be derived from 
the two-photon width for a tensor with mass $m_2$~\cite{Amsler:2002ey}:
\begin{eqnarray}
  \Gamma_{\gamma\gamma}(0^{++})\,=k\,\,\biggl(\frac{m_0}{m_2}\biggr)^3\,\,
  \Gamma_{\gamma\gamma}(2^{++})\,.
\end{eqnarray}
Fig.~\ref{fig:two_photon_width} shows the two-photon width as expected for the $f_0(1500)$
(solid curves) and for the $f_0(1710)$ (dashed curves) assuming a $q\bar{q}$ structure. 
The box shows the 95\,\% confidence-level upper limit from ALEPH for the $f_0(1500)$
indicating that $50^\circ\leq\alpha\leq 100^\circ$ with $\mid f\,\rangle\, = \, {\rm cos}\,
\alpha\,\mid n\bar{n}\,\rangle - {\rm sin}\,\alpha\,\mid s\bar{s}\,\rangle$.

This contradiction emphasizes the non-$q\bar{q}$ nature of the $f_0(1500)$ resonance. 
On the other hand, the observed decays into $\pi\pi$, $\eta\eta$, $\eta\eta\,^\prime$, 
and $K\bar{K}$ are not in agreement with predictions for a pure glueball. For this reason, 
a large variety of mixing scenarios of the pure glueball with the nearby $n\bar{n}$ and 
$s\bar{s}$ isoscalar mesons has been described. These are discussed in section~\ref{sec:mixing}.

Challenges in the interpretation of the scalar sector involve both experimental
and theoretical efforts. The following key questions account for the major differences
in the models on scalar mesons and need to be addressed in the future:
\begin{itemize}
  \item What is the nature of the $f_0(980)$ and $a_0(980)$? Do they have mainly a
        quarkonium structure or are these additional non-$q\bar{q}$ states? Though very 
        often interpreted as a $K\bar{K}$ molecule, the decay $\phi\to\gamma f_0~(a_0)
        \to\gamma K\bar{K}$ is kinematically suppressed and has not been observed, yet.
        Data from DAPHNE~\cite{Aloisio2002b,AloIsio2002} on $\phi\to\eta\pi^0\gamma$ and
        $\phi\to\pi^0\pi^0\gamma$ favor these states to be four-quark $(q^2\bar{q}^2)$ 
        states~\cite{Close:2002zu,Achasov:2008ut}. In a more sophisticated picture
        \cite{Close:2002zu}, the two mesons are mostly $(qq)_{\bar{3}}(\bar{q}\bar{q})_3$ 
        in $S$-wave near the center, but further out they rearrange as $(q\bar{q})(q\bar{q})$ 
        and finally as meson-meson states. A Dalitz plot analysis of $D_S^+\to\pi^-\pi^+\pi^+$ 
        by the E791 collaboration finds a dominant contribution from $f_0(980)\pi^+$ pointing 
        to a large $s\bar{s}$ component in the wave function~\cite{Aitala:2000xt}. The charm 
        meson decay $D_S^+\to\pi^-\pi^+\pi^+$ is Cabibbo-favored, but has no strange meson 
        in the final state. 
  \item Is the $f_0(1370)$ a true $q\bar{q}$ resonance or of a different nature, e.g.
        generated by $\rho\rho$ molecular dynamics as suggested in~\cite{Klempt:2007cp}.
        The latter is supported by the huge $\Gamma_{\rho\rho}$, but very small 
        $\Gamma_{\eta\eta}$ and $\Gamma_{\sigma\sigma}$ partial widths observed in central 
        production (Table~\ref{tab:wa102_scalar-decays}). Recently, some authors have 
        shown that $\rho\rho$ interaction in the hidden gauge formalism can dynamically 
        generate the $f_0(1370)$ and $f_2(1270)$ meson states~\cite{Molina:2008jw,Geng:2008gx}.
        Unfortunately, central-production results completely disagree with Crystal-Barrel 
        results on the decay width of this state (Table~\ref{tab:cb_scalar-decays}). The 
        experimental situation is unclear. The Particle Data Group has however accepted the  
        $f_0(1370)$ as an established resonance.

        In the {\it red dragon} interpretation~\cite{Minkowski:1998mf}, the 
        $f_0(1370)$ is also not considered a genuine resonance, but part of the broad 
        background amplitude, which itself is associated with the expected glueball. In 
        contrast, a recent {\it study in depth of $f_0(1370)$}~\cite{Bugg:2007ja} 
        shows a highly significant 
        improvement of $\chi^2$ with the state included as resonance in fits to the five 
        primary sets of data requiring its existence. These sets include Crystal-Barrel 
        data on $p\bar{p}\to3\pi^0$ at rest, data on $p\bar{p}\to\pi^0\pi^0\eta$, BES-II 
        data on $J/\psi\to\phi\pi^+\pi^-$, and the CERN-Munich data for $\pi\pi$ 
        elastic scattering.
  \item Though the $f_0(1500)$ cannot be accommodated easily in $q\bar{q}$ nonets and
        exhibits reduced $\gamma\gamma$~couplings -- all signatures expected for 
        glueballs -- data on $J/\psi\to\gamma f_0(1500)$ is still statistically limited. 
        More data from the BES-III experiment is certainly required to show its enhanced 
        production in gluon-rich radiative $J/\psi$ decays.
  \item Are the two resonances listed in Table~\ref{tab:scalarmesons}, $f_0(1710)$ and 
        $f_0(1790)$, distinct states? Experimental evidence is claimed only by the 
        BES collaboration and has not been confirmed by others. The Particle Data 
        Group does not list the $f_0(1790)$ as resonance and also does not include
        it in averages, fits, limits, etc.
\end{itemize}

Other pictures have emerged for the assignment of scalar mesons to SU(3) flavor 
nonets based on different approaches to the questions above. Minkowski and Ochs
have classified the isoscalar resonances $f_0(980)$ and $f_0(1500)$ together with 
$a_0(980)$ and $K^\ast_0(1430)$ as members of the $0^{++}$ nonet~\cite{Minkowski:1998mf}. 
They claim a mixing of the isoscalar states, which is similar to that of the pseudoscalar 
$\eta$ and $\eta\,^\prime$. In this scenario, the ($\eta\,^\prime,~f_0(980)$) pair forms
a parity doublet, which is approximately degenerate in mass. Moreover, the $f_0(600)$
and $f_0(1370)$ are interpreted as different signals of the same broad resonance, 
which is associated with the lowest-lying $0^{++}$ glueball. Very similar nonet 
assignments are discussed in a quark model which is based on short-range instanton
effects~\cite{Klempt:1995ku}, also not considering the $f_0(1370)$ as $q\bar{q}$
resonance.

Moreover, the Gatchina group performed a $K$-matrix analysis of the isoscalar $0^{++}$ 
waves in the invariant mass range 280-1900~MeV/$c^2$ based on a large variety of different
data sets including data from GAMS, BNL on $\pi^- p$, Crystal Barrel, and CERN-Munich 
on $\pi^+\pi^-\to\pi^+\pi^-$. In their own notation, the ground-state $q\bar{q}$ scalar 
nonet is suggested to consist of $a_0(980)$, $f_0(1300)$ and $f_0(980)$, whereas
$a_0(1450)$, $f_0(1760)$ and $f_0(1500)$ form the nonet of the first radial 
excitation~\cite{Anisovich:2002ij}. There are also models~\cite{Beveren:2008} that 
predict many more scalar $q\bar{q}$ states than have been observed, thus negating 
the need to invoke a glueball explanation.

If we adopt the point of view of the Particle Data Group, then five isoscalar 
resonances are well established: $f_0(600)$, $f_0(980)$, $f_0(1370)$, $f_0(1500)$, 
and $f_0(1710)$. The following section discusses possible mixing scenarios of the
$J^{PC} = 0^{++}$ $q\bar{q}$ nonet with the lowest-mass $0^{++}$ glueball resulting 
in the three established $I=0$ resonances above 1~GeV/$c^2$.

\subsection{\label{sec:mixing}\emph{Mixing in the Scalar Sector}}
There have been many authors who have considered that the $f_{0}(1370)$, the 
$f_{0}(1500)$ and the $f_{0}(1710)$ are the physical manifestations of the underlying 
$0^{++}$ quarkonium nonet and the lowest mass glueball. It is generally assumed that 
the three bare states mix to yield the three physical states. Inputs to such calculations 
include the masses of the physical states as well as their decay rates into pairs of 
pseudoscalar mesons. The masses of the bare states and the mixing of the physical states 
then result. In addition to the choice of physical states, there appears to also be some
dependence on whether one assumes the bare glueball is more or less massive than the 
mostly $s\bar{s}$ state.

These discussions are driven by the properties of the $f_{0}(1500)$ -- both its relatively
narrow width and its decay pattern. The latter strongly disfavors an $s\bar{s}$ interpretion
of the state, but a pure $u\bar{u}/d\bar{d}$ interpretation appears inconsistent with 
its width. This was first addressed by Amsler and Close~\cite{Amsler:1995tu,Amsler:1995td}
who allowed for an admixture of glueball into the normal meson spectrum, but did not allow
for the direct decay of the glueball. This work was extended by Close and 
colleagues~\cite{Close:1996yc} to allow for the direct decay. The parameters from their
work seem to show that the direct glueball decay dominates over the normal quarkonia
decays. Finally, Giacosa and colleagues~\cite{Giacosa:2005qr} considered both neglecting 
and allowing the direct glueball decay. The results of these are then interpreted in terms 
of mixing between the $q\bar{q}$ and glueball states.

In the literature, the mixing is typically written in terms of equation~\ref{eq:mixing_matrix},
where the physical states, $f_{1,2,3}$ are identified with the $f_{0}(1370)$,
$f_{0}(1500)$ and $f_{0}(1710)$ respectively, and the bare states are parameterized in terms
of the ideally mixed $q\bar{q}$ states:
\begin{eqnarray}
\label{eq:nnbar}
n\bar{n} & = & \frac{1}{\sqrt{2}}\left(u\bar{u}+d\bar{d}\right) \\
\label{eq:ssbar}
s\bar{s}\,. & & 
\end{eqnarray}
However, it is also useful to look at these in terms of the SU(3) symmetric states:
\begin{eqnarray}
\label{eq:singlet}
\mid 1 ~\rangle & = & \frac{1}{\sqrt{3}}\left( u\bar{u} + d\bar{d} + s\bar{s} \right ) \\[1ex]
\label{eq:octet}
\mid 8 ~\rangle & = & \frac{1}{\sqrt{6}}\left( u\bar{u} + d\bar{d} - 2s\bar{s} \right )\,.
\end{eqnarray}
The physical states ($f_0$ mesons) can be written in terms of both the ideally mixed $q\bar{q}$ 
states (equations~\ref{eq:nnbar} and~\ref{eq:ssbar}) as in equation~\ref{eq:mixing_matrix}, or 
in terms of the SU(3) states (equations \ref{eq:singlet} and~\ref{eq:octet}) where a similar 
expression can be written:
\begin{eqnarray}
\label{eq:mixing_matrix}
\left ( \begin{array}{c}
\mid f_{1} ~\rangle \\
\mid f_{2} ~\rangle \\
\mid f_{3} ~\rangle \\
\end{array} \right )
& = &
\left( \begin{array}{ccc}
M_{1n} & M_{1s} & M_{1g} \\
M_{2n} & M_{2s} & M_{2g} \\
M_{3n} & M_{3s} & M_{3g} \\
\end{array} \right )
\cdot
\left ( \begin{array}{c}
\mid n\bar{n} ~\rangle \\
\mid s\bar{s} ~\rangle \\
\mid G ~\rangle \\
\end{array} \right )
\end{eqnarray}
In the following, we look at several different models for mixing in the scalar sector. In
each case, there are various model dependences included in the analysis. We also present
the results of the mixing analysis in terms of both representations of the underlying
$q\bar{q}$ states. In some schemes, there is a natural separation by flavor, and in
others by the SU(3) representation.  


Broadly speaking, the results divide into two categories. The first in which
the bare glueball comes out lighter than the $s\bar{s}$ state, and the 
second in which the bare glueball comes out heavier than the $s\bar{s}$ state.
We summarize these by first looking at the case where the bare glueball
is lighter. One of the earliest of these came from Close and 
Amsler~\cite{Amsler:1995tu,Amsler:1995td}. They allowed for the rate of $s\bar{s}$
quark production from the vacuum to be different from the $u$ and $d$ quarks. They
also allowed for the bare glueball to have a different coupling to $s\bar{s}$ than
to $u\bar{u}$ and $d\bar{d}$. In this work, they did not allow for the direct decay of
the glueball. However, in their work, the resulting couplings
were generally consistent with the flavor-blind assumption. Based mostly on 
results from the Crystal Barrel experiment, they found a mixing as given in
equation~\ref{eq:mixing_amslerclose}:
\begin{eqnarray}
\begin{array}{c}
\mid f_{0}(1370) ~\rangle \\
\mid f_{0}(1500) ~\rangle \\
\mid f_{0}(1710) ~\rangle \\
\end{array} 
& = &
\left( \begin{array}{ccc}
-0.91 & -0.07 &  0.40 \\
-0.41 &  0.35 & -0.84 \\
 0.09 &  0.93 &  0.36 \\
\end{array} \right )
\cdot
\begin{array}{c}
\mid n\bar{n} ~\rangle \\
\mid s\bar{s} ~\rangle \\
\mid G ~\rangle \\
\end{array} \nonumber \\[0.5ex]
\label{eq:mixing_amslerclose}
\begin{array}{c}
\mid f_{0}(1370) ~\rangle \\
\mid f_{0}(1500) ~\rangle \\
\mid f_{0}(1710) ~\rangle \\
\end{array} 
& = &
\left( \begin{array}{ccc}
-0.78 & -0.47 &  0.40 \\
-0.13 & -0.52 & -0.84 \\
 0.61 & -0.71 &  0.36 \\
\end{array} \right )
\cdot
\begin{array}{c}
\mid 1 ~\rangle \\
\mid 8 ~\rangle \\
\mid G ~\rangle \\
\end{array}
\end{eqnarray}
Close and colleagues~\cite{Close:1996yc} later extended this by including additional
data and including the direct decay of the glueball in their model. They found the 
mixing scheme as given in equation~\ref{eq:mixing_close_li}:
\begin{eqnarray}
\begin{array}{c}
\mid f_{0}(1370) ~\rangle \\
\mid f_{0}(1500) ~\rangle \\
\mid f_{0}(1710) ~\rangle \\
\end{array} 
& = &
\left( \begin{array}{ccc}
 0.86 &  0.13 & -0.50 \\
 0.43 & -0.61 & -0.61 \\
 0.22 &  0.76 &  0.60 \\
\end{array} \right )
\cdot
\begin{array}{c}
\mid n\bar{n} ~\rangle \\
\mid s\bar{s} ~\rangle \\
\mid G ~\rangle \\
\end{array} \nonumber \\[0.5ex]
\label{eq:mixing_close_li}
\begin{array}{c}
\mid f_{0}(1370) ~\rangle \\
\mid f_{0}(1500) ~\rangle \\
\mid f_{0}(1710) ~\rangle \\
\end{array} 
& = &
\left( \begin{array}{ccc}
0.78 & 0.39 & -0.50 \\
0.00 & 0.75 &  0.61 \\
0.62 &-0.49 &  0.60 \\
\end{array} \right )
\cdot
\begin{array}{c}
\mid 1 ~\rangle \\
\mid 8 ~\rangle \\
\mid G ~\rangle \\
\end{array}
\end{eqnarray}

Giacosa~\cite{Giacosa:2005zt} looked at mixing in an effective chiral approach. They 
took the couplings to be flavor blind, and carried out their analysis both with and
without a direct decay of the glueball. In the case without direct decay, the decays
proceed via the $q\bar{q}$ content of the states. If the mass of the bare glueball 
was assumed lighter than the $s\bar{s}$ state, they found the solution in 
equation~\ref{eq:mixing_giacosa_s1} for the case without a direct glueball decay and
that in equation~\ref{eq:mixing_giacosa_s3} for the case with a direct glueball
decay:
\begin{eqnarray}
\begin{array}{c}
\mid f_{0}(1370) ~\rangle \\
\mid f_{0}(1500) ~\rangle \\
\mid f_{0}(1710) ~\rangle \\
\end{array} 
& = &
\left( \begin{array}{ccc}
 0.86 & 0.24 &  0.45 \\
-0.45 &-0.06 &  0.89 \\
-0.24 & 0.97 & -0.06 \\
\end{array} \right )
\cdot
\begin{array}{c}
\mid n\bar{n} ~\rangle \\
\mid s\bar{s} ~\rangle \\
\mid G ~\rangle \\
\end{array} \nonumber \\[0.5ex]
\label{eq:mixing_giacosa_s1}
\begin{array}{c}
\mid f_{0}(1370) ~\rangle \\
\mid f_{0}(1500) ~\rangle \\
\mid f_{0}(1710) ~\rangle \\
\end{array} 
& = &
\left( \begin{array}{ccc}
 0.84 & 0.30 &  0.45 \\
-0.40 &-0.21 &  0.89 \\
 0.36 &-0.93 & -0.06 \\
\end{array} \right )
\cdot
\begin{array}{c}
\mid 1 ~\rangle \\
\mid 8 ~\rangle \\
\mid G ~\rangle \\
\end{array}
\end{eqnarray}

\begin{eqnarray}
\begin{array}{c}
\mid f_{0}(1370) ~\rangle \\
\mid f_{0}(1500) ~\rangle \\
\mid f_{0}(1710) ~\rangle \\
\end{array} 
& = &
\left( \begin{array}{ccc}
 0.79 & 0.26 &  0.56 \\
-0.58 & 0.02 &  0.81 \\
-0.20 & 0.97 & -0.16 \\
\end{array} \right )
\cdot
\begin{array}{c}
\mid n\bar{n} ~\rangle \\
\mid s\bar{s} ~\rangle \\
\mid G ~\rangle \\
\end{array} \nonumber \\[0.5ex]
\label{eq:mixing_giacosa_s3}
\begin{array}{c}
\mid f_{0}(1370) ~\rangle \\
\mid f_{0}(1500) ~\rangle \\
\mid f_{0}(1710) ~\rangle \\
\end{array} 
& = &
\left( \begin{array}{ccc}
 0.80 & 0.24 &  0.56 \\
 0.46 &-0.35 &  0.81 \\
 0.40 &-0.91 & -0.16 \\
\end{array} \right )
\cdot
\begin{array}{c}
\mid 1 ~\rangle \\
\mid 8 ~\rangle \\
\mid G ~\rangle \\
\end{array}
\end{eqnarray}
The common features of these mixing schemes are that the $f_{0}(1710)$ has a
very large $s\bar{s}$ component, while the largest part of the glueball tends
to be on the $f_{0}(1500)$; it is generally split over at least two of the three
states. One also sees that the $f_{0}(1370)$ has the largest SU(3) singlet
component.

The other situation is that in which the bare glueball comes out heavier than the 
bare $s\bar{s}$ state. The first of these was carried out by Weingarten and
Lee~\cite{Lee:1999kv}. They computed both the masses of the bare states as well 
as information on the decay of these states on the lattice. In their model,
they had a decay rate that favored heavier quarks, thus enhancing the coupling
to the $s\bar{s}$ states. They found the mixing scheme as in 
equation~\ref{eq:mixing_weingarten}:
\begin{eqnarray}
\begin{array}{c}
\mid f_{0}(1370) ~\rangle \\
\mid f_{0}(1500) ~\rangle \\
\mid f_{0}(1710) ~\rangle \\
\end{array} 
& = &
\left( \begin{array}{ccc}
 0.819 &  0.290 & -0.495 \\
-0.399 &  0.908 & -0.128 \\
 0.413 &  0.302 &  0.859 \\
\end{array} \right )
\cdot
\begin{array}{c}
\mid n\bar{n} ~\rangle \\
\mid s\bar{s} ~\rangle \\
\mid G ~\rangle \\
\end{array} \nonumber \\[0.5ex] 
\label{eq:mixing_weingarten}
\begin{array}{c}
\mid f_{0}(1370) ~\rangle \\
\mid f_{0}(1500) ~\rangle \\
\mid f_{0}(1710) ~\rangle \\
\end{array} 
& = &
\left( \begin{array}{ccc}
0.836 & 0.236 & -0.495 \\
0.198 & -0.972 &  0.128 \\
0.512 & -0.008 &  0.859 \\
\end{array} \right )
\cdot
\begin{array}{c}
\mid 1 ~\rangle \\
\mid 8 ~\rangle \\
\mid G ~\rangle \\
\end{array}
\end{eqnarray}
Giacosa~\cite{Giacosa:2005zt} et al. also have two solutions in which the bare glueball mass
is heavier than the $s\bar{s}$ mass. For the case of no direct glueball decay,
the mixing is given in equation~\ref{eq:mixing_giacosa_s2}, while for the case
of the direct glueball decay included, the nixing is given in equation~\ref{eq:mixing_giacosa_s4}.
\begin{eqnarray}
\begin{array}{c}
\mid f_{0}(1370) ~\rangle \\
\mid f_{0}(1500) ~\rangle \\
\mid f_{0}(1710) ~\rangle \\
\end{array} 
& = &
\left( \begin{array}{ccc}
 0.81 & 0.19 &  0.54 \\
-0.49 & 0.72 &  0.49 \\
-0.30 & 0.67 & -0.68 \\
\end{array} \right )
\cdot
\begin{array}{c}
\mid n\bar{n} ~\rangle \\
\mid s\bar{s} ~\rangle \\
\mid G ~\rangle \\
\end{array} \nonumber \\[0.5ex] 
\label{eq:mixing_giacosa_s2}
\begin{array}{c}
\mid f_{0}(1370) ~\rangle \\
\mid f_{0}(1500) ~\rangle \\
\mid f_{0}(1710) ~\rangle \\
\end{array} 
& = &
\left( \begin{array}{ccc}
 0.77 & 0.31 &  0.54 \\
 0.02 &-0.87 &  0.49 \\
 0.14 &-0.72 & -0.68 \\
\end{array} \right )
\cdot
\begin{array}{c}
\mid 1 ~\rangle \\
\mid 8 ~\rangle \\
\mid G ~\rangle \\
\end{array}
\end{eqnarray}
\begin{eqnarray}
\begin{array}{c}
\mid f_{0}(1370) ~\rangle \\
\mid f_{0}(1500) ~\rangle \\
\mid f_{0}(1710) ~\rangle \\
\end{array} 
& = &
\left( \begin{array}{ccc}
 0.82 & 0.57 & -0.07 \\
-0.57 & 0.82 &  0.00 \\
-0.06 & 0.04 & -0.99 \\
\end{array} \right )
\cdot
\begin{array}{c}
\mid n\bar{n} ~\rangle \\
\mid s\bar{s} ~\rangle \\
\mid G ~\rangle \\
\end{array} \nonumber \\[0.5ex] 
\label{eq:mixing_giacosa_s4}
\begin{array}{c}
\mid f_{0}(1370) ~\rangle \\
\mid f_{0}(1500) ~\rangle \\
\mid f_{0}(1710) ~\rangle \\
\end{array} 
& = &
\left( \begin{array}{ccc}
 1.00 & 0.01 & -0.07 \\
 0.01 &-1.00 &  0.00 \\
-0.03 &-0.07 & -0.99 \\
\end{array} \right )
\cdot
\begin{array}{c}
\mid 1 ~\rangle \\
\mid 8 ~\rangle \\
\mid G ~\rangle \\
\end{array}
\end{eqnarray}
Finally, Cheng~\cite{Cheng:2006hu} used lattice calculations for the mass of the 
scalar glueball to set the starting values for their fit to the exiting data. They 
also limited the input data on decay rates. In particular, they did not use the strong 
coupling of the $f_{0}(1500)$ to $\eta\eta\,^\prime$ due to the complications of the 
threshold opening. They found the mixing scheme as given in 
equation~\ref{eq:mixing_cheng}:
\begin{eqnarray}
\begin{array}{c}
\mid f_{0}(1370) ~\rangle \\
\mid f_{0}(1500) ~\rangle \\
\mid f_{0}(1710) ~\rangle \\
\end{array} 
& = &
\left( \begin{array}{ccc}
 0.78 &  0.51 & -0.36 \\
-0.54 &  0.84 &  0.03 \\
 0.32 &  0.18 &  0.93 \\
\end{array} \right )
\cdot
\begin{array}{c}
\mid n\bar{n} ~\rangle \\
\mid s\bar{s} ~\rangle \\
\mid G ~\rangle \\
\end{array} \nonumber \\[0.5ex]
\label{eq:mixing_cheng}
\begin{array}{c}
\mid f_{0}(1370) ~\rangle \\
\mid f_{0}(1500) ~\rangle \\
\mid f_{0}(1710) ~\rangle \\
\end{array} 
& = &
\left( \begin{array}{ccc}
0.93 & 0.03 & -0.36 \\
0.04 & -0.99 &  0.03 \\
0.37 &  0.04 &  0.93 \\
\end{array} \right )
\cdot
\begin{array}{c}
\mid 1 ~\rangle \\
\mid 8 ~\rangle \\
\mid G ~\rangle \\
\end{array}
\end{eqnarray}

While the amount of information from mixing, and the various models that went into
creating these are a bit overwhelming, there are some interesting trends in the results.
In particular, in the case where one assumes that the bare glueball is heavier than the 
$s\bar{s}$ state, all of the models support a picture in which the $f_{0}(1370)$ is 
mostly and SU(3) singlet state, the $f_{0}(1500)$ is mostly and SU(3) octet state, 
and the glueball is dominantly in the $f_{0}(1710)$ state. In the case where the 
bare glueball is lighter than the $s\bar{s}$ state, the octet and glueball assignments
switch, but there is stronger mixing between the components.

Under the assumptions that the $f_{0}(1370)$ exists, and the scalar states
with masses below $1$~GeV/c$^{2}$ are of a different origin, the mixing scenario
can provide a good description of the data, and makes it very likely that the 
scalar glueball exists and is manifested in these states. Unfortunately, the 
exact mixing scheme depends on the models used to describe glueball decays. 
There is some hope that better information on two-photon couplings to the 
scalar states as well as more information from heavier systems decaying into 
scalars may provide some additional insights on the problem, our feeling is that
things are likely to remain somewhat unclear. 
  

%
\section{Planned Experiments}
\subsection{\emph{BES-III}}
The BES-III experiment at BEPCII in Beijing started operation in summer of 2008.
It will ultimately accumulate data samples on the order of $10\times 10^9$ $J/\psi$,
$3\times 10^9$~$\psi(2S)$, and 30 million $D\bar{D}$ per running year. The amount of
expected $J/\psi$ data is nearly 200 times as large as the BES-II $58\times 10^6$ 
$J/\psi$ data sample. Such large data sets will make it possible to study hadron 
spectroscopy in the decays of the charmonium states and the charmed mesons. Of 
particular interest for glueball searches is the fact that $J/\psi$ decays have always 
been viewed as one of the best places to look for these states. The physics at BES-III 
is fully described in~\cite{Asner:2008nq}. Discussed are also detailed simulation
studies of $J/\psi\to\gamma\eta\eta$ and $\gamma\eta\eta\,^\prime$ to investigate the
BES-III sensitivity for the $2^{++}$~glueball candidate $f_J(2220)$. The authors point 
out that a partial wave analysis will be needed to resolve ambiguities.

\begin{figure}[tb!]
\begin{center}
\includegraphics*[angle=90,width=0.75\textwidth]{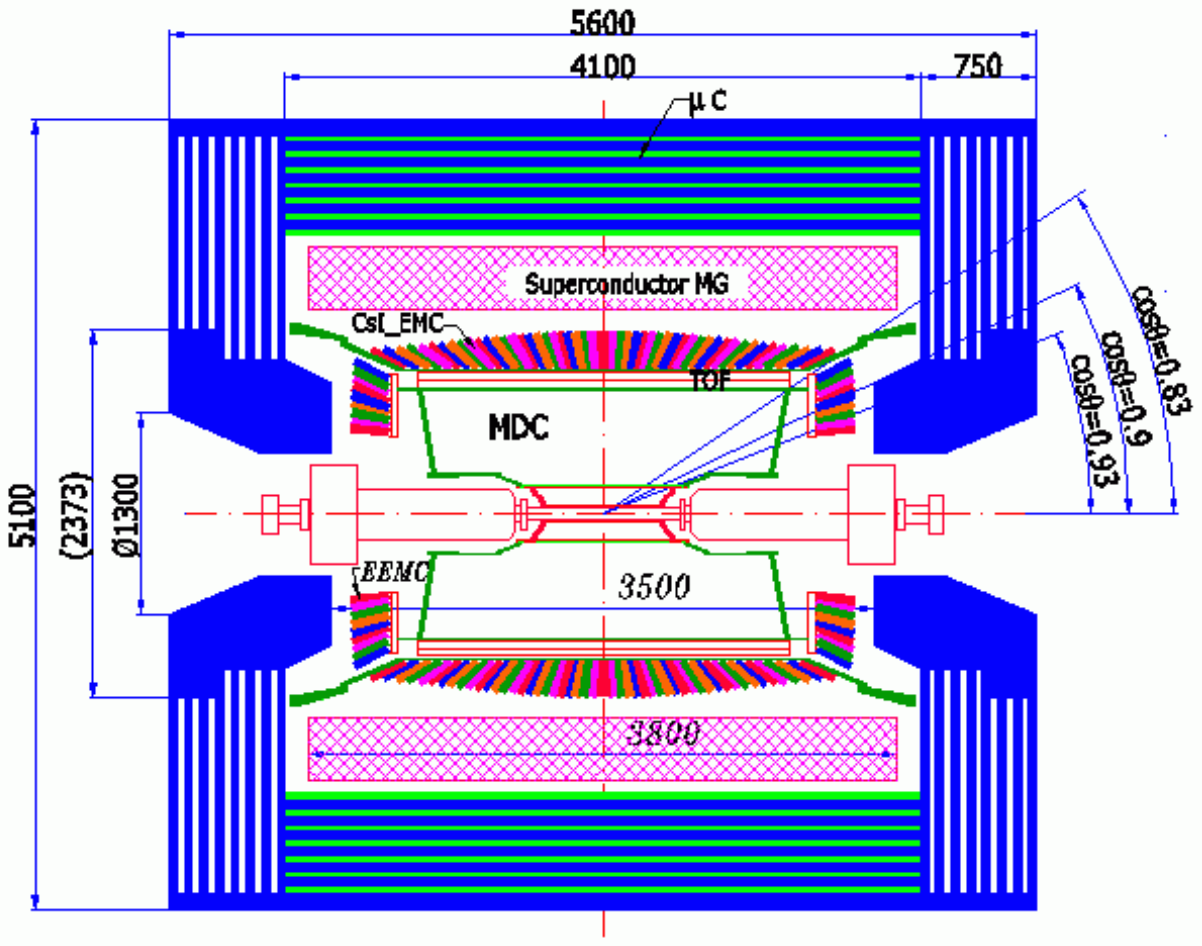}
\end{center}
\vspace{-1cm}
\caption[]{\label{fig:bes3} Layout of the BES-III Experiment at BEPCII~\cite{Asner:2008nq}.}
\end{figure}

The BES-III detector is shown schematically in Figure~\ref{fig:bes3}. In capability,
it will be very similar to the CLEO-c detector -- being able to accurately reconstruct
both charged particles and photons. The main detector components are a Helium-based drift 
chamber with a $dE/dx$~resolution that is better than 6\,\%, a CsI(Tl) crystal calorimeter,
and a Time-of-Flight system. A super-conducting solenoid magnet provides a central field
of 1.0~T.

\subsection{\emph{The COMPASS Experiment}}
The goal of the COMPASS experiment is the investigation of hadron structure and hadron 
spectroscopy, both manifestations of non-perturbative QCD. Key scientific issues addressed 
at COMPASS are nuclear spin structure and hadron spectroscopy. In 2004, a pilot hadron run 
using a 190~GeV $\pi^-$ beam on nuclear targets has been performed. Production of mesons
proceeds via diffractive pion dissociation. The incident particles only graze the target,
which remains intact. The pion beam itself is excited to some resonance $X^-$, which
subsequently decays: $\pi^- Pb\to X^- Pb\to \pi^-\pi^-\pi^+ Pb$, for instance. A dedicated
spectroscopy run using liquid hydrogen has been planned for 2008. The experiment may also
focus on the glue-rich environment produced in central production.

COMPASS (COmmon Muon and Proton Apparatus for Structure and Spectroscopy) is a fixed-target 
experiment approved by CERN in February 1997, which started data taking in 2002. The
detector is a two-stage magnetic spectrometer with a flexible setup to allow for physics
programs with different beams. The spectrometer is equipped with tracking systems based
on silicon detectors for high-precision tracking in the target region, micromega and
GEM detectors for small area and wire/drift chambers as well as straw tubes for large
area tracking. Particle identification is provided by a large-acceptance ring imaging
\v{C}erenkov detector and by hadronic (HCAL) and electromagnetic (ECAL) calorimeters.

\subsection{\emph{The GlueX Experiment at Jefferson Laboratory}}
The GlueX experiment is part of the Jefferson Lab 12-GeV upgrade -- and energy doubling
upgrade of the CEBAF accelerator. GlueX will be housed in a new photon-only experimental 
building (Hall D). Electrons of energy $12$~GeV will impinge in a thin diamond 
target and via coherent bremsstrahlung, produce a linearly-polarized, $8.4-9.0$~GeV
photon beam that interacts in the experimental target. 

The main physics program of GlueX will be to search for light-quark hybrid mesons
and to map out the spectrum of the exotic-quantum-number states. In order to do
this, it will be necessary to reconstruct final states with several charged particles
and photons. The GlueX experiment has been designed to have nearly full solid angle
coverage for these particles with sufficient energy and momentum resolution to
exclusively identify the desired final states. While the primary physics of GlueX
will be the search for light-quark hybrids, this will imply a fairly significant
program in the spectroscopy of light-quark mesons as well. In terms of glueballs, 
direct production of glueballs in photoproduction is supposed to be supressed. However,
some models suggest that decays of hybrid mesons via the glueball component of lighter
mesons may be enhanced.

\begin{figure}[tb!]
\begin{center}
\includegraphics[width=0.7\textwidth]{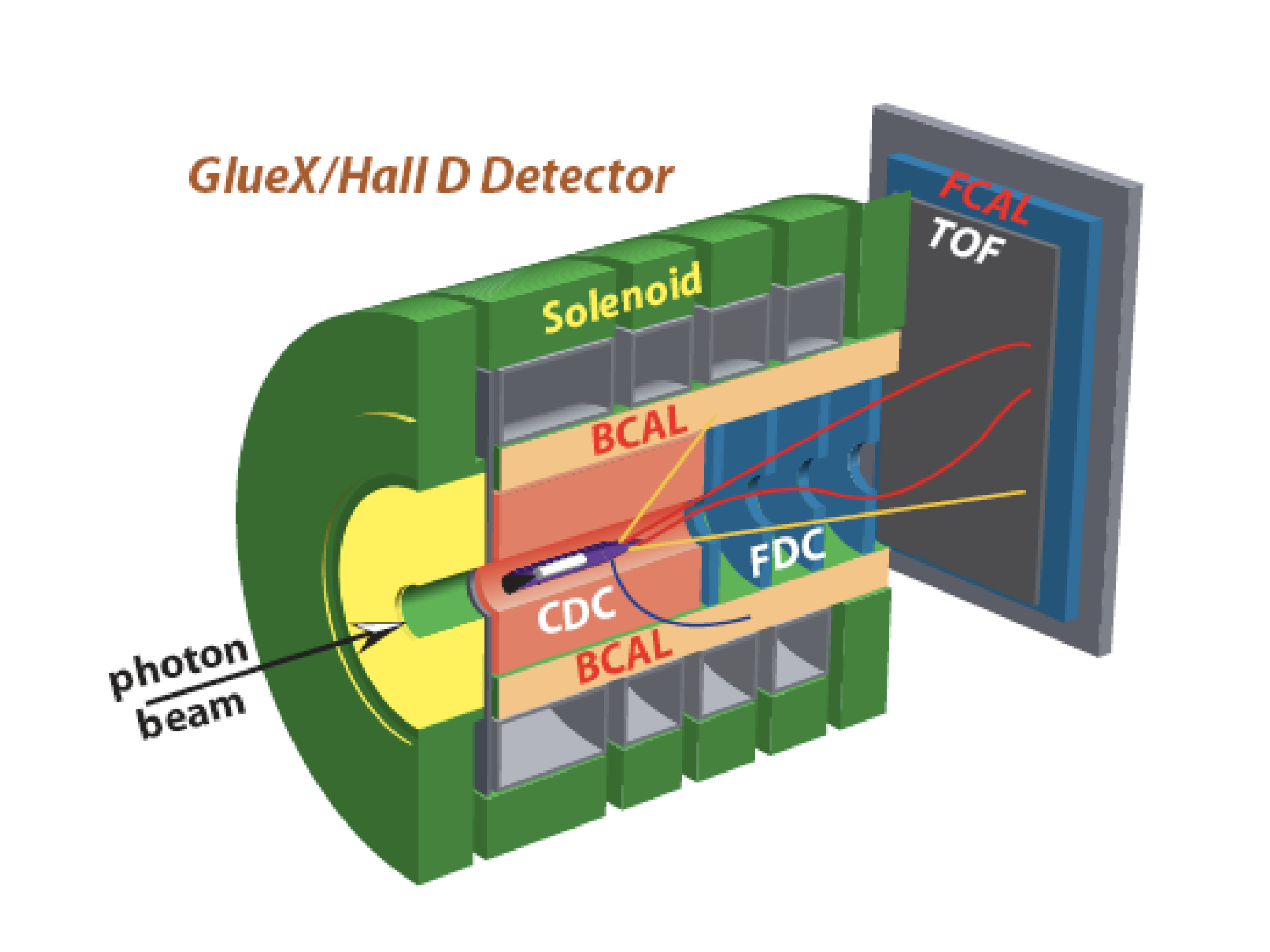}
\end{center}
\vspace{-0.5cm}
\caption[]{\label{fig:gluex} Layout of the GlueX Experiment at Jefferson Laboratory}
\end{figure}

The base-line detector is shown in~\ref{fig:gluex}. The detectors are in a $2.2$~T
solenoidal magnet which was originally used for the LASS experiment at SLAC. The all
solenoidal design is well matched to the $9$~GeV photon energy and the final states
with $4-6$ particles. 

The 12-GeV upgrade received U.S. Department of Energy approval and construction 
is expected to commence in 2009. First beam on target GlueX is expected in 2014.
 
\subsection{\emph{The PANDA Experiment at GSI}}
The PANDA (antiProton ANnihilation at DArmstadt) research program will be conducted 
at the Facility for Antiproton and Ion Research (FAIR), localized at GSI near Frankfurt 
in Germany. The heart of the new accelerator complex is a super-conducting High-Energy
Storage Ring (HESR) for antiprotons with a circumference of about 1,100 meters. A system of 
cooler-storage rings for effective beam cooling at high energies and various experimental 
halls will be connected to the facility. 

At the new FAIR facility, $1.5 - 15$~GeV/$c$ antiproton beams will allow high-precision 
hadron physics using the small energy spread available with antiproton beams (cooled to
$\Delta p/p\approx 10^{-5}$). The energy range has been chosen to allow detailed studies 
of hadronic systems up to charmonium states. The use of antiprotons allows to directly
form all states with non-exotic quantum numbers in formation experiments. Cross sections 
are considered higher compared to producing additional particles in the final state. 
Moreover, the observation of resonances in production but not in formation will indicate 
exotic physics. The main goals of the scientific program at PANDA are charmonium 
spectroscopy, the search for gluonic excitations, open and hidden charm in nuclei, and
also $\gamma$-ray spectroscopy of hypernuclei. There is some suggestion that if heavier 
glueballs exist, then $\bar{p}p$ annihilation may be a good place to look for them. 
Certainly, the information in the scalar sector was significantly enhanced by studies of 
$\bar{p}p$ annihilations at rest. In the search for glueballs, the collaboration will study
final states including $\phi\phi$ or $\phi\eta$ for states below 3.6~GeV/$c^2$ and $J/\psi
\eta$ and $J/\psi\phi$ for more massive states.

\begin{figure}[tb!]
\begin{center}
\includegraphics*[angle=-90,viewport=150 150 430 650,width=1.0\textwidth]{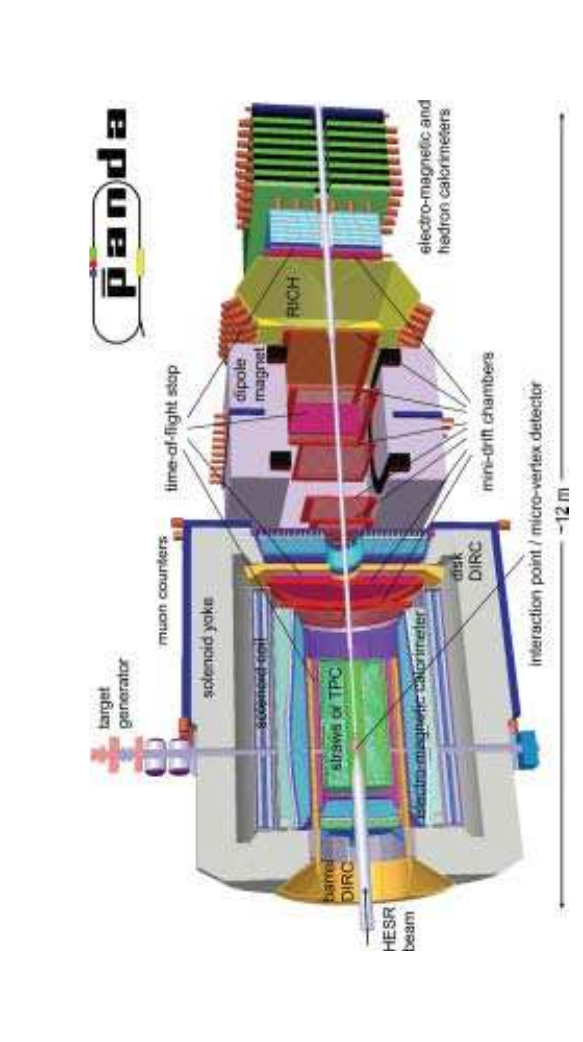}
\end{center}
\vspace{-0.2cm}
\caption[]{\label{fig:panda} Layout of the PANDA experiment at GSI: antiprotons from the
HESR enter from the left and hit clusters or pellets of hydrogen at the interaction point,
which is surrounded by detectors inside a solenoid magnet.}
\end{figure}

The physics will be done with the PANDA multipurpose detector located inside the HESR. 
Fig.~\ref{fig:panda} shows the general layout of the detector. The PANDA collaboration 
consists of $\sim 350$ physicists from 48 institutions worldwide. The general-purpose 
detector allows the detection and identification of neutral and charged particles over 
the relevant angular and energy range. The inner part of the detector can be modified 
for the needs of individual physics programs. Angular coverage up to $10^{\circ}$ is provided
by a magnetic forward spectrometer.

\section{Summary}
Over the last 20 years, a significant amount of data has been collected that is
relevant to the search for glueballs. It is also anticipated that more data will
be forthcoming from experiments that are just starting to take data now, and others
that will come online in the middle of the next decade. These new data have driven 
a significant effort in the theoretical understanding of existing data, with very 
sophisticated models being developed to try and explain the diverse phenomena that
have been observed. There has also been significant progress on lattice calculations,
and work is now moving toward unquenched calculations with the hope of reaching
close to the physical pions masses. While all of this has certainly improved our 
detailed understanding of QCD, it is not obvious that we are able to definitively 
answer the question of ``has a glueball been observed?''.

In the scalar sector, there is no question that more states than can be accommodated
by a single meson nonet have been found. However, the nature of all of these states
is still open for discussion. All pictures which claim a glueball require mixing
with the normal $q\bar{q}$ states, and various levels of interference to accommodate
the observations. Perhaps most significant is the question of the $f_{0}(1370)$.
Is this a true state, in which case the the $f_{0}(1370)$, $f_{0}(1500)$ and
$f_{0}(1710)$ are likely the scalar nonet mixed with the glueball? However, the 
$f_{0}(1370)$ may simply be part of the broad scalar signal associated with the 
$f_{0}(600)$ ($\sigma$). In this case, the glueball may be mixed into the broad
background and the mesons may be less coupled with the glueball component. However,
the nature of the broad scalar background may not necessarily be associated with
glue. Disentangling this situation is likely to be a challenge which may not have
a universally accepted answer.

The pseudoscalar sector may have an apparent extra state in the $1.4$~GeV/$c^{2}$ 
mass region. However, this appears to be relatively far removed from the expected 
$2$~GeV/$c^{2}$ mass of the $0^{-+}$ glueball. It is also crucial to examine
carefully the evidence for the $\eta(1295)$. The only real evidence appears to
come from a single experiment where the state is produced in conjunction with the
axial states ($f_{1}$ mesons). Nearly all other experiments that observe pseudoscalar
states see the two states around $1.4$~GeV/$c^{2}$, but not the lighter state.
For higher mass states, too little is currently known to make definitive statements.
However, given the small number of expected resonances in this sector, it is hoped 
that progress can be made here. In particular, results for BES-III may be able to
both produce and map the decays of the higher mass states in $J/\psi$ decays.

In the tensor sector, the lightest states are well established and clear. While there 
is a large number of other observed states, too little is known about individual
states to allow us to draw definitive conclusions about the true assignment of these
resonances. While it may eventually be possible to disentangle this sector, it is 
likely to require a rather large and dedicated effort to more clearly resolve the 
nature of many of the observed states. While some of the observed signals may have 
a glueball component, there is currently no definitive evidence for this.
\section{Acknowledgement}
The authors are indebted to many of our colleagues for useful discussions that
helped to define this paper. In particular, we would like to thank Ted Barnes, 
Simon Capstick, Frank Close, Paul Eugenio, Eberhard Klempt, Colin Morningstar 
and Eric Swanson for useful insight into many of the issues addressed in this paper.

This work was supported by the Department of Energy under contracts DE-FG02-87ER40315
and DE-FG02-92ER40735 and by the National Science Foundation under award numbers 
0653316 and 0754674. 
\clearpage
\bibliographystyle{h-elsevier3}
\bibliography{glueball08}

\begin{thebibliography}{100}

\bibitem{0034-4885-51-6-002}
F.E. Close, {\em Gluonic hadrons},
\newblock Reports on Progress in Physics 51 (1988) 833.

\bibitem{Amsler:1997up}
C. Amsler, {\em {Proton antiproton annihilation and meson spectroscopy with the
  Crystal Barrel}},
\newblock Rev. Mod. Phys. 70 (1998) 1293, hep-ex/9708025.

\bibitem{Godfrey:1998pd}
S. Godfrey and J. Napolitano, {\em {Light-meson spectroscopy}},
\newblock Rev. Mod. Phys. 71 (1999) 1411, hep-ph/9811410.

\bibitem{Amsler:2004ps}
C. Amsler and N.A. Tornqvist, {\em {Mesons beyond the naive quark model}},
\newblock Phys. Rept. 389 (2004) 61.

\bibitem{Masoni:2006rz}
A. Masoni, C. Cicalo and G.L. Usai, {\em {The case of the pseudoscalar
  glueball}},
\newblock J. Phys. G32 (2006) R293.

\bibitem{Klempt:2007cp}
E. Klempt and A. Zaitsev, {\em {Glueballs, Hybrids, Multiquarks. Experimental
  facts versus QCD inspired concepts}},
\newblock Phys. Rept. 454 (2007) 1, arXiv:0708.4016 [hep-ph].

\bibitem{Mathieu:2008me}
V. Mathieu, N. Kochelev and V. Vento, {\em {The Physics of Glueballs}},
\newblock (2008), arXiv:0810.4453.

\bibitem{Amsler:2008zz}
C. Amsler et~al., [\,Particle Data Group Collaboration\,], {\em {Review of
  Particle Physics}},
\newblock Phys. Lett. B667 (2008) 1.

\bibitem{Amsler:quarkmodel:2006}
C. Amsler, T. De-Grand and B. Krusche, {\em {Review of Particle Physics}},
\newblock J. Phys. G33 (2006) 165.

\bibitem{Amsler:1995tu}
C. Amsler and F.E. Close, {\em {Evidence for a scalar glueball}},
\newblock Phys. Lett. B353 (1995) 385, hep-ph/9505219.

\bibitem{Amsler:1995td}
C. Amsler and F.E. Close, {\em {Is $f_0(1500)$ a scalar glueball?}},
\newblock Phys. Rev. D53 (1996) 295, hep-ph/9507326.

\bibitem{PhysRevD.9.3471}
A. Chodos et~al., {\em New extended model of hadrons},
\newblock Phys. Rev. D 9 (1974) 3471.

\bibitem{Jaffe:1975fd}
R.L. Jaffe and K. Johnson, {\em {Unconventional States of Confined Quarks and
  Gluons}},
\newblock Phys. Lett. B60 (1976) 201.

\bibitem{Close:2000yk}
F.E. Close and A. Kirk, {\em {The mixing of the $f_0(1370)$, $f_0(1500)$ and
  $f_0(1710)$ and the search for the scalar glueball}},
\newblock Phys. Lett. B483 (2000) 345, hep-ph/0004241.

\bibitem{Close:2005vf}
F.E. Close and Q. Zhao, {\em {Production of $f_0(1710)$, $f_0(1500)$, and
  $f_0(1370)$ in $J/\psi$ hadronic decays}},
\newblock Phys. Rev. D71 (2005) 094022, hep-ph/0504043.

\bibitem{Chanowitz:1984cb}
M.S. Chanowitz, {\em {Resonances in Photon-Photon Scattering}},
\newblock Presented at 6th Int. Workshop on Photon-Photon Collisions, Lake
  Tahoe, CA, Sep 10-13, 1984.

\bibitem{Boglione:1997aw}
M. Boglione and M.R. Pennington, {\em {Unquenching the scalar glueball}},
\newblock Phys. Rev. Lett. 79 (1997) 1998, hep-ph/9703257.

\bibitem{Pennington:2008xd}
M.R. Pennington et~al., {\em {Amplitude Analysis of High Statistics Results on
  $\gamma\gamma\to\pi^+\pi^-$ and the Two Photon Width of Isoscalar States}},
\newblock Eur. Phys. J. C56 (2008) 1, 0803.3389.

\bibitem{Cakir:1994jf}
M.B. Cakir and G.R. Farrar, {\em {Radiative decay of vector quarkonium:
  Constraints on glueballs and light gluinos}},
\newblock Phys. Rev. D50 (1994) 3268, hep-ph/9402203.

\bibitem{Close:1996yc}
F.E. Close, G.R. Farrar and Z.p. Li, {\em {Determining the gluonic content of
  isoscalar mesons}},
\newblock Phys. Rev. D55 (1997) 5749, hep-ph/9610280.

\bibitem{Lee:1999kv}
W.J. Lee and D. Weingarten, {\em {Scalar quarkonium masses and mixing with the
  lightest scalar glueball}},
\newblock Phys. Rev. D61 (2000) 014015, hep-lat/9910008.

\bibitem{Carlson:1980kh}
C.E. Carlson et~al., {\em {Glueballs and Oddballs: Their Experimental
  Signature}},
\newblock Phys. Lett. B99 (1981) 353.

\bibitem{Chanowitz:2005du}
M. Chanowitz, {\em {Chiral suppression of scalar glueball decay}},
\newblock Phys. Rev. Lett. 95 (2005) 172001, hep-ph/0506125.

\bibitem{Isgur:1983wj}
N. Isgur and J.E. Paton, {\em {A Flux Tube Model for Hadrons}},
\newblock Phys. Lett. B124 (1983) 247.

\bibitem{Isgur:1984bm}
N. Isgur and J.E. Paton, {\em {A Flux Tube Model for Hadrons in QCD}},
\newblock Phys. Rev. D31 (1985) 2910.

\bibitem{PhysRevD.68.074007}
M. Iwasaki et~al., {\em Flux tube model for glueballs},
\newblock Phys. Rev. D 68 (2003) 074007.

\bibitem{Morningstar:1997ff}
C.J. Morningstar and M.J. Peardon, {\em {Efficient glueball simulations on
  anisotropic lattices}},
\newblock Phys. Rev. D56 (1997) 4043, hep-lat/9704011.

\bibitem{Faddeev:2003aw}
L. Faddeev, A.J. Niemi and U. Wiedner, {\em {Glueballs, closed flux tubes and
  $\eta(1440)$}},
\newblock Phys. Rev. D70 (2004) 114033, hep-ph/0308240.

\bibitem{PhysRevD.60.074024}
Y. Koma, H. Suganuma and H. Toki, {\em Flux-tube ring and glueball properties
  in the dual Ginzburg-Landau theory},
\newblock Phys. Rev. D 60 (1999) 074024.

\bibitem{Novikov:1979va}
V.A. Novikov et~al., {\em {In a Search for Scalar Gluonium}},
\newblock Nucl. Phys. B165 (1980) 67.

\bibitem{Kisslinger:2001pk}
L.S. Kisslinger and M.B. Johnson, {\em {Scalar mesons, glueballs, instantons
  and the glueball/sigma}},
\newblock Phys. Lett. B523 (2001) 127, hep-ph/0106158.

\bibitem{Narison:2005wc}
S. Narison, {\em {QCD tests of the puzzling scalar mesons}},
\newblock Phys. Rev. D73 (2006) 114024, hep-ph/0512256.

\bibitem{Mennessier:2008kk}
G. Mennessier, S. Narison and W. Ochs, {\em {Glueball nature of the
  $\sigma/f_{0}(600)$ from $\pi -\pi$ and $\gamma -\gamma$ scatterings}},
\newblock Phys. Lett. B665 (2008) 205, 0804.4452.

\bibitem{Szczepaniak:2003mr}
A.P. Szczepaniak and E.S. Swanson, {\em {The low lying glueball spectrum}},
\newblock Phys. Lett. B577 (2003) 61, hep-ph/0308268.

\bibitem{Chen:2005mg}
Y. Chen et~al., {\em {Glueball spectrum and matrix elements on anisotropic
  lattices}},
\newblock Phys. Rev. D73 (2006) 014516, hep-lat/0510074.

\bibitem{Heller:1995bz}
U.M. Heller, {\em {SU(3) lattice gauge theory in the fundamental adjoint plane
  and scaling along the Wilson axis}},
\newblock Phys. Lett. B362 (1995) 123, hep-lat/9508009.

\bibitem{Michael:1987wf}
C. Michael and M. Teper, {\em {The Glueball Spectrum and Scaling in SU(3)
  Lattice Gauge Theory}},
\newblock Phys. Lett. B206 (1988) 299.

\bibitem{Michael:1988jr}
C. Michael and M. Teper, {\em {The Glueball Spectrum in SU(3)}},
\newblock Nucl. Phys. B314 (1989) 347.

\bibitem{Bali:1993fb}
G.S. Bali et~al., [\,UKQCD Collaboration\,], {\em {A Comprehensive lattice
  study of SU(3) glueballs}},
\newblock Phys. Lett. B309 (1993) 378, hep-lat/9304012.

\bibitem{Sexton:1995kd}
J. Sexton, A. Vaccarino and D. Weingarten, {\em {Numerical Evidence for the
  Observation of a Scalar Glueball}},
\newblock Phys. Rev. Lett. 75 (1995) 4563, hep-lat/9510022.

\bibitem{Weingarten:1996pp}
D. Weingarten, {\em {Scalar quarkonium and the scalar glueball}},
\newblock Nucl. Phys. Proc. Suppl. 53 (1997) 232, hep-lat/9608070.

\bibitem{Close:2001ga}
F.E. Close and A. Kirk, {\em {Scalar glueball $q\bar{q}$ mixing above 1 GeV and
  implications for lattice QCD}},
\newblock Eur. Phys. J. C21 (2001) 531, hep-ph/0103173.

\bibitem{Giacosa:2004ug}
F. Giacosa, T. Gutsche and A. Faessler, {\em {A covariant constituent
  quark-gluon model for the glueball-quarkonia content of scalar-isoscalar
  mesons}},
\newblock Phys. Rev. C71 (2005) 025202, hep-ph/0408085.

\bibitem{Morningstar:1999rf}
C.J. Morningstar and M.J. Peardon, {\em {The Glueball spectrum from an
  anisotropic lattice study}},
\newblock Phys. Rev. D60 (1999) 034509, hep-lat/9901004.

\bibitem{Hart:2001fp}
A. Hart and M. Teper, [\,UKQCD Collaboration\,], {\em {On the glueball spectrum
  in O(a)-improved lattice QCD}},
\newblock Phys. Rev. D65 (2002) 034502, hep-lat/0108022.

\bibitem{McNeile:2007fu}
C. McNeile, {\em {Hard hadron spectroscopy}},
\newblock PoS LATTICE2007 (2007) 019, 0710.0985.

\bibitem{Hart:2006ps}
A. Hart et~al., [\,UKQCD Collaboration\,], {\em {A lattice study of the masses
  of singlet $0^{++}$ mesons}},
\newblock Phys. Rev. D74 (2006) 114504, hep-lat/0608026.

\bibitem{Gregory:2005yr}
E.B. Gregory et~al., {\em {Scalar glueball and meson spectroscopy in unquenched
  lattice QCD with improved staggered quarks}},
\newblock PoS LAT2005 (2006) 027, hep-lat/0510066.

\bibitem{Close:1987rj}
F.E. Close, {\em {Quarks in Hadrons and Nuclei}},
\newblock Prog. Part. Nucl. Phys. 20 (1988) 1.

\bibitem{Abele:2000xt}
A. Abele et~al., [\,Crystal Barrel Collaboration\,], {\em {Test of $N\bar{N}$
  potential models: Isospin relations in $\bar{p}d$ annihilations at rest and
  the search for quasinuclear bound states}},
\newblock Eur. Phys. J. C17 (2000) 583.

\bibitem{Adamo:1992bb}
A. Adamo et~al., [\,OBELIX Collaboration\,], {\em {First physics results from
  OBELIX}},
\newblock Sov. J. Nucl. Phys. 55 (1992) 1732.

\bibitem{Bertin:1995fx}
A. Bertin et~al., [\,OBELIX Collaboration\,], {\em {$E/\iota$ decays to
  $K\bar{K}\pi$ in $\bar{p}p$ annihilation at rest}},
\newblock Phys. Lett. B361 (1995) 187.

\bibitem{Bertin:1996nq}
A. Bertin et~al., [\,OBELIX Collaboration\,], {\em {Measurement of the
  $\eta(1440)\rightarrow K^{\pm}K^{0}_{L}\pi^{\mp}$ production rates from
  $\bar{p}p$ annihilation at rest at three different hydrogen target
  densities}},
\newblock Phys. Lett. B385 (1996) 493.

\bibitem{Bertin:1998sb}
A. Bertin et~al., [\,OBELIX Collaboration\,], {\em {Study of the isovector
  scalar mesons in the channel $\bar{p}p\rightarrow K^{\pm}K^{0}_{S}\pi^{\mp}$
  at rest with initial angular momentum state selection}},
\newblock Phys. Lett. B434 (1998) 180.

\bibitem{Cicalo:1999sn}
C. Cicalo et~al., [\,OBELIX Collaboration\,], {\em {Evidence for two
  pseudoscalar states in the 1.4 to 1.5 GeV mass region}},
\newblock Phys. Lett. B462 (1999) 453.

\bibitem{Nichitiu:2002cj}
F. Nichitiu et~al., [\,OBELIX Collaboration\,], {\em {Study of the
  $K^{+}K^{-}\pi^{+}\pi^{-}\pi^{0}$ final state in anti-proton annihilation at
  rest in gaseous hydrogen at NTP with the OBELIX spectrometer}},
\newblock Phys. Lett. B545 (2002) 261.

\bibitem{Bargiotti:2003ev}
M. Bargiotti et~al., [\,OBELIX Collaboration\,], {\em {Coupled channel analysis
  of $\pi^{+}\pi^{-}\pi^{0}$, $K^{+}K^{-}\pi^{0}$ and
  $K^{\pm}K^{0}_{S}\pi^{\mp}$ from $\bar{p}p$ annihilation at rest in hydrogen
  targets at three densities}},
\newblock Eur. Phys. J. C26 (2003) 371.

\bibitem{Salvini:2004gz}
P. Salvini et~al., [\,OBELIX Collaboration\,], {\em {$\bar{p}p$ annihilation
  into four charged pions at rest and in flight}},
\newblock Eur. Phys. J. C35 (2004) 21.

\bibitem{Aker:1992ny}
E. Aker et~al., [\,Crystal Barrel Collaboration\,], {\em {The Crystal Barrel
  spectrometer at LEAR}},
\newblock Nucl. Instrum. Meth. A321 (1992) 69.

\bibitem{Amsler:1992rx}
C. Amsler et~al., [\,Crystal Barrel Collaboration\,], {\em {Proton-antiproton
  annihilation into $\eta\eta\pi$: Observation of a scalar resonance decaying
  into $\eta\eta$}},
\newblock Phys. Lett. B291 (1992) 347.

\bibitem{Amsler:1993pr}
C. Amsler et~al., [\,Crystal Barrel Collaboration\,], {\em {Antiproton-proton
  annihilation at rest into $\omega\pi^{0}\pi^{0}$}},
\newblock Phys. Lett. B311 (1993) 362.

\bibitem{Amsler:1993xd}
C. Amsler et~al., [\,Crystal Barrel Collaboration\,], {\em {Protonium
  annihilation into $K^{0}_{L}K^{0}_{S}\pi^{0}$ and $K^{0}_{L}K^{0}_{S}\eta$}},
\newblock Phys. Lett. B319 (1993) 373.

\bibitem{Amsler:1994ah}
C. Amsler et~al., [\,Crystal Barrel Collaboration\,], {\em {$\eta\eta\,^\prime$
  threshold enhancement in $\bar{p}p$ annihilations into
  $\pi^{0}\eta\eta\,^\prime$ at rest}},
\newblock Phys. Lett. B340 (1994) 259.

\bibitem{Amsler:1994rv}
C. Amsler et~al., [\,Crystal Barrel. Collaboration\,], {\em {Observation of a
  scalar resonance decaying to $\pi^{+}\pi^{-}\pi^{0}\pi^{0}$ in $\bar{p}p$
  annihilation at rest}},
\newblock Phys. Lett. B322 (1994) 431.

\bibitem{Amsler:1995bf}
C. Amsler et~al., [\,Crystal Barrel Collaboration\,], {\em {Coupled channel
  analysis of $\bar{p}p$ annihilation into $\pi^{0}\pi^{0}\pi^{0}$ ,
  $\pi^{0}\eta\eta$ and $\pi^{0}\pi^{0}\eta$}},
\newblock Phys. Lett. B355 (1995) 425.

\bibitem{Amsler:1995bz}
C. Amsler et~al., [\,Crystal Barrel Collaboration\,], {\em {High statistics
  study of $f_{0}(1500)$ decay into $\eta\eta$}},
\newblock Phys. Lett. B353 (1995) 571.

\bibitem{Amsler:1995gf}
C. Amsler et~al., [\,Crystal Barrel Collaboration\,], {\em {High statistics
  study of $f_{0}(1500)$ decay into $\pi^{0}\pi^{0}$}},
\newblock Phys. Lett. B342 (1995) 433.

\bibitem{Amsler:1995wz}
C. Amsler et~al., [\,Crystal Barrel Collaboration\,], {\em {$E$ decay to
  $\eta\pi\pi$ in $\bar{p}p$ annihilation at rest}},
\newblock Phys. Lett. B358 (1995) 389.

\bibitem{Abele:1996fr}
A. Abele et~al., [\,Crystal Barrel Collaboration\,], {\em {A Study of
  $f_{0}(1500)$ decays into $4\pi^{0}$ in $\bar{p}p\to 5\pi^{0}$ at rest}},
\newblock Phys. Lett. B380 (1996) 453.

\bibitem{Abele:1996nn}
A. Abele et~al., [\,Crystal Barrel Collaboration\,], {\em {Observation of
  $f_{0}(1500)$ decay into $K_{L}K_{L}$}},
\newblock Phys. Lett. B385 (1996) 425.

\bibitem{Abele:1997dz}
A. Abele et~al., [\,Crystal Barrel Collaboration\,], {\em {Study of the
  $\pi^{0}\pi^{0}\eta\,^\prime$ final state in $\bar{p}p$ annihilation at
  rest}},
\newblock Phys. Lett. B404 (1997) 179.

\bibitem{Abele:1997qy}
A. Abele et~al., [\,Crystal Barrel Collaboration\,], {\em {High-mass
  $\rho$-meson states from $\bar{p}d$ annihilation at rest into
  $\pi^{-}\pi^{0}\pi^{0}p_{spectator}$}},
\newblock Phys. Lett. B391 (1997) 191.

\bibitem{Abele:1998kv}
A. Abele et~al., [\,Crystal Barrel Collaboration\,], {\em {Study of
  $\bar{p}p\rightarrow\eta\pi^{0}\pi^{0}\pi^{0}$ at rest}},
\newblock Nucl. Phys. B514 (1998) 45.

\bibitem{Abele:1999en}
A. Abele et~al., [\,Crystal Barrel Collaboration\,], {\em {Antiproton-proton
  annihilation at rest into $K^{+}K^{-}\pi^{0}$}},
\newblock Phys. Lett. B468 (1999) 178.

\bibitem{Abele:1999fw}
A. Abele et~al., [\,Crystal Barrel Collaboration\,], {\em {$\bar{p}d$
  annihilation at rest into $\pi^{+}\pi^{-}\pi^{-}p_{spectator}$}},
\newblock Phys. Lett. B450 (1999) 275.

\bibitem{Abele:2001js}
A. Abele et~al., [\,Crystal Barrel Collaboration\,], {\em {Study of $f_{0}$
  decays into four neutral pions}},
\newblock Eur. Phys. J. C19 (2001) 667.

\bibitem{Abele:2001pv}
A. Abele et~al., [\,Crystal Barrel Collaboration\,], {\em {$4\pi$ decays of
  scalar and vector mesons}},
\newblock Eur. Phys. J. C21 (2001) 261.

\bibitem{Amsler:2004kn}
C. Amsler et~al., [\,Crystal Barrel Collaboration\,], {\em {Study of antiproton
  annihilation on neutrons into $\omega\pi^{-}\pi^{0}$}},
\newblock Nucl. Phys. A740 (2004) 130.

\bibitem{Adomeit:1996nr}
J. Adomeit et~al., [\,Crystal Barrel Collaboration\,], {\em {Evidence for two
  isospin zero $J^{PC}=2^{-+}$ mesons at 1645 MeV and 1875 MeV}},
\newblock Z. Phys. C71 (1996) 227.

\bibitem{Abele:1999pf}
A. Abele et~al., [\,Crystal Barrel Collaboration\,], {\em {Observation of
  resonances in the reaction $\bar{p}p\rightarrow\pi^{0}\eta\eta$ at 1.94
  GeV/$c$}},
\newblock Eur. Phys. J. C8 (1999) 67.

\bibitem{Abele:2000qq}
A. Abele et~al., [\,Crystal Barrel Collaboration\,], {\em {$\bar{p}p$
  annihilation into $\omega\pi^{0}$, $\omega\eta$ and $\omega\eta\,^\prime$ at
  600 MeV, 1200 MeV and 1940 MeV/$c$}},
\newblock Eur. Phys. J. C12 (2000) 429.

\bibitem{Seth:2000tm}
K.K. Seth, [\,Crystal Barrel Collaboration\,], {\em {A high resolution search
  for the tensor glueball}},
\newblock Nucl. Phys. A663 (2000) 600.

\bibitem{Amsler:2002qq}
C. Amsler et~al., [\,Crystal Barrel Collaboration\,], {\em {Proton-antiproton
  annihilation at 900 MeV/$c$ into $\pi^{0}\pi^{0}\pi^{0}$,
  $\pi^{0}\pi^{0}\eta$ and $\pi^{0}\eta\eta$}},
\newblock Eur. Phys. J. C23 (2002) 29.

\bibitem{Amsler:2006du}
C. Amsler et~al., [\,Crystal Barrel Collaboration\,], {\em {Study of $K\bar{K}$
  resonances in $\bar{p}p\rightarrow K^{+}K^{-}\pi^{0}$ at 900 MeV/$c$ and 1640
  MeV/$c$}},
\newblock Phys. Lett. B639 (2006) 165.

\bibitem{Ahmad:1989gx}
S. Ahmad et~al., [\,ASTERIX Collaboration\,], {\em {The ASTERIX Spectrometer at
  LEAR}},
\newblock Nucl. Instrum. Meth. A286 (1990) 76.

\bibitem{Athar:2005nu}
S.B. Athar et~al., [\,CLEO Collaboration\,], {\em {Radiative decays of the
  $\Upsilon$(1S) to a pair of charged hadrons}},
\newblock Phys. Rev. D73 (2006) 032001, hep-ex/0510015.

\bibitem{Besson:2005ud}
D. Besson et~al., [\,CLEO Collaboration\,], {\em {Radiative Decays of the
  $\Upsilon$(1S) to $\gamma\pi^0\pi^0$, and $\gamma\eta\eta$ and
  $\gamma\pi^0\eta$}},
\newblock Phys. Rev. D75 (2007) 072001, hep-ex/0512003.

\bibitem{Asner:2008nq}
D.M. Asner et~al., {\em {Physics at BES-III}},
\newblock (2008), 0809.1869.

\bibitem{Geiger:1992va}
P. Geiger and N. Isgur, {\em {When can hadronic loops scuttle the OZI rule?}},
\newblock Phys. Rev. D47 (1993) 5050.

\bibitem{Lipkin:1996ny}
H.J. Lipkin and B.s. Zou, {\em {Comment on 'When can hadronic loops scuttle the
  Okubo- Zweig-Iizuka rule?'}},
\newblock Phys. Rev. D53 (1996) 6693.

\bibitem{Li:1996yn}
X.Q. Li, D.V. Bugg and B.S. Zou, {\em {A possible explanation of the *$\rho\pi$
  puzzle* in $J/\psi$, $\psi\,^\prime$ decays}},
\newblock Phys. Rev. D55 (1997) 1421.

\bibitem{Close:2002sf}
F.E. Close, A. Donnachie and Y.S. Kalashnikova, {\em {Radiative decays: A new
  flavor filter}},
\newblock Phys. Rev. D67 (2003) 074031, hep-ph/0210293.

\bibitem{Bai:2001dw}
J.Z. Bai et~al., [\,BES Collaboration\,], {\em {The BES upgrade}},
\newblock Nucl. Instrum. Meth. A458 (2001) 627.

\bibitem{Bai:1996wm}
J.Z. Bai et~al., [\,BES Collaboration\,], {\em {Studies of $\xi(2230)$ in $J /
  \psi$ radiative decays}},
\newblock Phys. Rev. Lett. 76 (1996) 3502.

\bibitem{Bai:1998tx}
J.Z. Bai et~al., [\,BES Collaboration\,], {\em {Experimental study of $J/\psi$
  radiative decay to $\pi^0 \pi^0$}},
\newblock Phys. Rev. Lett. 81 (1998) 1179.

\bibitem{Bai:2000}
J.Z. Bai et~al., [\,BES Collaboration\,], {\em {Partial Wave Analysis of
  $J/\psi\rightarrow\gamma \pi^{+}\pi^{-}\pi^{+}\pi^{-}$}},
\newblock Phys. Lett. B 472 (2000) 207.

\bibitem{Bai:2003}
J.Z. Bai et~al., [\,BES Collaboration\,], {\em {Partial Wave Analysis of
  $J/\psi\rightarrow\gamma K^{+}K^{-}$ and $\gamma K^{0}_{S}K^{0}_{S}$}},
\newblock Phys. Rev. D 68 (2003) 052003, hep-ex/0307058.

\bibitem{Ablikim:2004aa}
M. Ablikim et~al., [\,BES Collaboration\,], {\em {The $\sigma$ pole in
  $J/\psi\rightarrow\omega\pi^{+}\pi^{-}$}},
\newblock Phys. Lett. B 598 (2004) 149, hep-ex/0406038.

\bibitem{Ablikim:2004bb}
M. Ablikim et~al., [\,BES Collaboration\,], {\em {Study of
  $J/\psi\rightarrow\omega K^{+}K^{-}$}},
\newblock Phys. Lett. B 603 (2004) 138, hep-ex/0409007.

\bibitem{Bai:2004qj}
J.Z. Bai et~al., [\,BES Collaboration\,], {\em {A Study of
  $J/\psi\to\gamma\gamma V(\rho,\phi)$ Decays with the BES-II Detector}},
\newblock Phys. Lett. B594 (2004) 47, hep-ex/0403008.

\bibitem{Ablikim:2005aa}
M. Ablikim et~al., [\,BES Collaboration\,], {\em {Resonances in
  $J/\psi\rightarrow\phi\pi^{+}\pi^{-}$ and $\phi K^{+}K^{-}$}},
\newblock Phys. Lett. B 607 (2005) 243, hep-ex/0411001.

\bibitem{Ablikim:2006aw}
M. Ablikim et~al., [\,BES Collaboration\,], {\em {Partial Wave Analysis of
  $J/\psi\rightarrow\gamma \pi^{+}\pi^{-}$ and $\gamma \pi^{0}\pi^{0}$}},
\newblock Phys. Lett. B 642 (2006) 441.

\bibitem{Ablikim:2006dw}
M. Ablikim et~al., [\,BES Collaboration\,], {\em {Observation of a
  near-threshold enhancement in the $\omega\phi$ mass spectrum from the doubly
  OZI suppressed decay $J/\psi\rightarrow\gamma\omega\phi$}},
\newblock Phys. Rev. Lett. 96 (2006) 162002, hep-ex/0602031.

\bibitem{Kopp:1996kg}
S.E. Kopp, [\,CLEO Collaboration\,], {\em {The CLEO III Detector}},
\newblock Nucl. Instrum. Meth. A384 (1996) 61.

\bibitem{Kopp:2000gv}
S. Kopp et~al., [\,CLEO Collaboration\,], {\em {Dalitz analysis of the decay
  $D^0\to K^-\pi^+\pi^0$}},
\newblock Phys. Rev. D63 (2001) 092001, hep-ex/0011065.

\bibitem{Muramatsu:2002jp}
H. Muramatsu et~al., [\,CLEO Collaboration\,], {\em {Dalitz analysis of $D^0\to
  K^0_S \pi^+\pi^-$}},
\newblock Phys. Rev. Lett. 89 (2002) 251802, hep-ex/0207067.

\bibitem{Benslama:2002pa}
K. Benslama et~al., [\,CLEO Collaboration\,], {\em {Anti-search for the
  glueball candidate $f_J(2220)$ in two-photon interactions}},
\newblock Phys. Rev. D66 (2002) 077101, hep-ex/0204019.

\bibitem{Rubin:2004cq}
P. Rubin et~al., [\,CLEO Collaboration\,], {\em {First observation and Dalitz
  analysis of the $D^0\to K^0_S\eta\pi^0$ decay}},
\newblock Phys. Rev. Lett. 93 (2004) 111801, hep-ex/0405011.

\bibitem{CroninHennessy:2005sy}
D. Cronin-Hennessy et~al., [\,CLEO Collaboration\,], {\em {Searches for CP
  violation and $\pi\pi$ $S$-wave in the Dalitz- plot of
  $D^0\to\pi^+\pi^-\pi^0$}},
\newblock Phys. Rev. D72 (2005) 031102, hep-ex/0503052.

\bibitem{Cawlfield:2006hm}
C. Cawlfield et~al., [\,CLEO Collaboration\,], {\em {Measurement of interfering
  $K^{*+} K^-$ and $K^{*-} K^+$ amplitudes in the decay $D^0\to K^+ K^-
  \pi^0$}},
\newblock Phys. Rev. D74 (2006) 031108, hep-ex/0606045.

\bibitem{Bonvicini:2007tc}
G. Bonvicini et~al., [\,CLEO Collaboration\,], {\em {Dalitz plot analysis of
  the $D^+\to\pi^-\pi^+\pi^+$ decay}},
\newblock Phys. Rev. D76 (2007) 012001, 0704.3954.

\bibitem{:2008zi}
P. Rubin et~al., [\,CLEO Collaboration\,], {\em {Search for CP Violation in the
  Dalitz-Plot Analysis of $D^\pm\to K^+K^-\pi^\pm$}},
\newblock Phys. Rev. D78 (2008) 072003, 0807.4545.

\bibitem{Close:2001ay}
F.E. Close and A. Kirk, {\em {Large isospin mixing in $\phi$ radiative decay
  and the spatial size of the $f_0(980) - a_0(980)$ meson}},
\newblock Phys. Lett. B515 (2001) 13, hep-ph/0106108.

\bibitem{AloIsio2002}
A. Aloisio et~al., {\em Study of the decay $\phi \to \pi^0\pi^0\gamma$ with the
  {KLOE} detector},
\newblock Phys. Lett. B537 (2002) 21.

\bibitem{Aloisio2002b}
A. Aloisio et~al., {\em Study of the decay $\phi \to \eta\pi^0\gamma$ with the
  {KLOE} detector},
\newblock Phys. Lett. B536 (2002) 209.

\bibitem{Ambrosino:2005wk}
F. Ambrosino et~al., [\,KLOE Collaboration\,], {\em {Study of the decay
  $\phi\to f_0(980)\gamma\to\pi^+ \pi^-\gamma$ with the KLOE detector}},
\newblock Phys. Lett. B634 (2006) 148, hep-ex/0511031.

\bibitem{Robson:1977pm}
D. Robson, {\em {A Basic Guide for the Glueball Spotter}},
\newblock Nucl. Phys. B130 (1977) 328.

\bibitem{Kirk:1999av}
A. Kirk, [\,WA102 Collaboration\,], {\em {A glueball $q\bar{q}$ filter in
  central production}},
\newblock (1999), hep-ph/9908253.

\bibitem{Klempt:1998wr}
E. Klempt, {\em {The status of scalar mesons: Evidence for glueballs}},
\newblock Acta Phys. Polon. B29 (1998) 3367.

\bibitem{Armstrong:1991ch}
T.A. Armstrong et~al., [\,WA76 Collaboration\,], {\em {Study of the centrally
  produced $\pi\pi$ and $K\bar{K}$ systems at 85~GeV/$c$ and 300~GeV/$c$}},
\newblock Z. Phys. C51 (1991) 351.

\bibitem{Binon:1985ag}
F. Binon et~al., [\,GAMS-2000 Collaboration\,], {\em {Hodoscope Multi-Photon
  Spectrometer Gams-2000}},
\newblock Nucl. Instrum. Meth. A248 (1986) 86.

\bibitem{Alde:1985rr}
D. Alde et~al., [\,GAMS-4000 Collaboration\,], {\em {Acquisition System for the
  Hodoscope Spectrometer Gams-4000}},
\newblock Nucl. Instrum. Meth. A240 (1985) 343.

\bibitem{Abatzis:1995xx}
S. Abatzis et~al., [\,WA102 Collaboration\,], {\em {A further study of the
  centrally produced $\pi^{+}\pi^{-}$ and $\pi^{+}\pi^{-}\pi^{+}\pi^{-}$
  channels in $pp$ interactions at 300 and 450 GeV/c}},
\newblock Phys. Lett. B353 (1995) 589.

\bibitem{Barberis:1997ve}
D. Barberis et~al., [\,WA102 Collaboration\,], {\em {A study of the centrally
  produced $\pi^{+}\pi^{-}\pi^{+}\pi^{-}$ channel in $p p$ interactions at 450
  GeV/$c$}},
\newblock Phys. Lett. B413 (1997) 217, hep-ex/9707021.

\bibitem{Barberis:1997vf}
D. Barberis et~al., [\,WA102 Collaboration\,], {\em {A study of the
  $K\bar{K}\pi$ channel produced centrally in $p p$ interactions at
  450~GeV/$c$}},
\newblock Phys. Lett. B413 (1997) 225, hep-ex/9707022.

\bibitem{Barberis:1998bq}
D. Barberis et~al., [\,WA102 Collaboration\,], {\em {A study of the centrally
  produced $\phi\phi$ system in $p p$ interactions at 450~GeV/$c$}},
\newblock Phys. Lett. B432 (1998) 436, hep-ex/9805018.

\bibitem{Barberis:1998in}
D. Barberis et~al., [\,WA102 Collaboration\,], {\em {A study of the centrally
  produced $\pi^{+}\pi^{-}\pi^{0}$ channel in $p p$ interactions at
  450~GeV/$c$}},
\newblock Phys. Lett. B422 (1998) 399, hep-ex/9801003.

\bibitem{Barberis:1998tv}
D. Barberis et~al., [\,WA102 Collaboration\,], {\em {A study of the centrally
  produced $K^{*}(892)\bar{K}^{*}(892)$ and $\phi\omega$ systems in $p p$
  interactions at 450~GeV/$c$}},
\newblock Phys. Lett. B436 (1998) 204, hep-ex/9807021.

\bibitem{Barberis:1999wn}
D. Barberis et~al., [\,WA102 Collaboration\,], {\em {A spin analysis of the
  $4\pi$ channels produced in central $p p$ interactions at 450 GeV/$c$}},
\newblock Phys. Lett. B471 (2000) 440, hep-ex/9912005.

\bibitem{Barberis:1999ap}
D. Barberis et~al., [\,WA102 Collaboration\,], {\em {A partial wave analysis of
  the centrally produced $\pi^{0}\pi^{0}$ system in $p p$ interactions at 450
  GeV/$c$}},
\newblock Phys. Lett. B453 (1999) 325, hep-ex/9903044.

\bibitem{Barberis:1999an}
D. Barberis et~al., [\,WA102 Collaboration\,], {\em {A partial wave analysis of
  the centrally produced $\pi^{+}\pi^{-}$ system in $p p$ interactions at 450
  GeV/$c$}},
\newblock Phys. Lett. B453 (1999) 316, hep-ex/9903043.

\bibitem{Barberis:1999am}
D. Barberis et~al., [\,WA102 Collaboration\,], {\em {A partial wave analysis of
  the centrally produced $K^{+}K^{-}$ and $K^{0}_{S}K^{0}_{S}$ systems in $p p$
  interactions at 450 GeV/$c$ and new information on the spin of the
  $f_{J}(1710)$}},
\newblock Phys. Lett. B453 (1999) 305, hep-ex/9903042.

\bibitem{Barberis:1999cq}
D. Barberis et~al., [\,WA102 Collaboration\,], {\em {A coupled channel analysis
  of the centrally produced $K^{+}K^{-}$ and $\pi^{+}\pi^{-}$ final states in
  $p p$ interactions at 450 GeV/$c$}},
\newblock Phys. Lett. B462 (1999) 462, hep-ex/9907055.

\bibitem{Barberis:1999id}
D. Barberis et~al., [\,WA102 Collaboration\,], {\em {A study of the
  $\eta\eta\,^\prime$ and $\eta\,^\prime\eta\,^\prime$ channels produced in
  central $p p$ interactions at 450 GeV/$c$}},
\newblock Phys. Lett. B471 (2000) 429, hep-ex/9911041.

\bibitem{Barberis:1999be}
D. Barberis et~al., [\,WA102 Collaboration\,], {\em {A study of the
  $\eta\pi^{+}\pi^{-}$ channel produced in central $p p$ interactions at
  450~GeV/$c$}},
\newblock Phys. Lett. B471 (2000) 435, hep-ex/9911038.

\bibitem{Barberis:2000em}
D. Barberis et~al., [\,WA102 Collaboration\,], {\em {A study of the
  $f_{0}(1370)$, $f_{0}(1500)$, $f_{0}(2000)$ and $f_{2}(1950)$ observed in the
  centrally produced $4\pi$ final states}},
\newblock Phys. Lett. B474 (2000) 423, hep-ex/0001017.

\bibitem{Barberis:2000cd}
D. Barberis et~al., [\,WA102 Collaboration\,], {\em {A study of the $\eta\eta$
  channel produced in central $p p$ interactions at 450~GeV/$c$}},
\newblock Phys. Lett. B479 (2000) 59, hep-ex/0003033.

\bibitem{Barberis:2000kc}
D. Barberis et~al., [\,WA102 Collaboration\,], {\em {A study of the
  $\omega\omega$ channel produced in central $p p$ interactions at
  450~GeV/$c$}},
\newblock Phys. Lett. B484 (2000) 198, hep-ex/0005027.

\bibitem{Barberis:2000cx}
D. Barberis et~al., [\,WA102 Collaboration\,], {\em {A study of the centrally
  produced $\eta\pi^{0}$ and $\eta\pi^{-}$ systems in $p p$ interactions at
  450~GeV/$c$}},
\newblock Phys. Lett. B488 (2000) 225, hep-ex/0007019.

\bibitem{Barberis:2001bs}
D. Barberis et~al., [\,WA102 Collaboration\,], {\em {A study of the centrally
  produced $\pi^{0}\pi^{0}\pi^{0}$ channel in $p p$ interactions at
  450~GeV/$c$}},
\newblock Phys. Lett. B507 (2001) 14, hep-ex/0104017.

\bibitem{Cooper:1988km}
S. Cooper, {\em {Meson Production in Two-Photon Collisions}},
\newblock Ann. Rev. Nucl. Part. Sci. 38 (1988) 705.

\bibitem{Yang:1950rg}
C.N. Yang, {\em {Selection Rules for the Dematerialization of a Particle into
  Two Photons}},
\newblock Phys. Rev. 77 (1950) 242.

\bibitem{Ahohe:2005ug}
R. Ahohe et~al., [\,CLEO Collaboration\,], {\em {The search for $\eta(1440)\to
  K^0_S K^\pm \pi^\mp$ in two-photon fusion at CLEO}},
\newblock Phys. Rev. D71 (2005) 072001, hep-ex/0501026.

\bibitem{Decamp:1990jra}
D. Decamp et~al., [\,ALEPH Collaboration\,], {\em {ALEPH: A Detector for
  Electron-Positron Annihilations at LEP}},
\newblock Nucl. Instrum. Meth. A294 (1990) 121.

\bibitem{:1989kxa}
[\,L3 Collaboration\,], {\em {The Construction of the L3 Experiment}},
\newblock Nucl. Instrum. Meth. A289 (1990) 35.

\bibitem{Barate:1999ze}
R. Barate et~al., [\,ALEPH Collaboration\,], {\em {Search for the glueball
  candidates $f_0(1500)$ and $f_J(1710)$ in $\gamma\gamma$ collisions}},
\newblock Phys. Lett. B472 (2000) 189, hep-ex/9911022.

\bibitem{Acciarri:2000ex}
M. Acciarri et~al., [\,L3 Collaboration\,], {\em {$K^0_{S} K^0_{S}$ final state
  in two-photon collisions and implications for glueballs}},
\newblock Phys. Lett. B501 (2001) 173, hep-ex/0011037.

\bibitem{Acciarri:2000ev}
M. Acciarri et~al., [\,L3 Collaboration\,], {\em {Light resonances in $K^0_{S}
  K^\pm \pi^\mp$ and $\eta \pi^{+} \pi^{-}$ final states in $\gamma \gamma$
  collisions at LEP}},
\newblock Phys. Lett. B501 (2001) 1, hep-ex/0011035.

\bibitem{Aston:1986jb}
D. Aston et~al., {\em {The Strange Meson Resonances Observed in the Reaction
  $K^- p \to\bar{K}^0\pi^+\pi^- n$ at 11~GeV/$c$}},
\newblock Nucl. Phys. B292 (1987) 693.

\bibitem{Bityukov:1991ri}
S.I. Bityukov et~al., {\em {Observation of resonance with mass M = 1814 MeV,
  decaying into $\pi^-\eta\eta$}},
\newblock Phys. Lett. B268 (1991) 137.

\bibitem{Teige:1996fi}
S. Teige et~al., [\,E852 Collaboration\,], {\em {Properties of the $a_0(980)$
  meson}},
\newblock Phys. Rev. D59 (1999) 012001, hep-ex/9608017.

\bibitem{Aubert:2001tu}
B. Aubert et~al., [\,BABAR Collaboration\,], {\em {The BaBar Detector}},
\newblock Nucl. Instrum. Meth. A479 (2002) 1, hep-ex/0105044.

\bibitem{Aubert:2005wb}
B. Aubert et~al., [\,BABAR Collaboration\,], {\em {Measurements of neutral $B$
  decay branching fractions to $K^0_S \pi^+ \pi^-$ final states and the charge
  asymmetry of $B^0 \to K^{*+} \pi^-$}},
\newblock Phys. Rev. D73 (2006) 031101, hep-ex/0508013.

\bibitem{Iijima:2000cq}
T. Iijima and E. Prebys, [\,BELLE Collaboration\,], {\em {Commissioning and
  first results from BELLE}},
\newblock Nucl. Instrum. Meth. A446 (2000) 75.

\bibitem{Garmash:2004wa}
A. Garmash et~al., [\,BELLE Collaboration\,], {\em {Dalitz analysis of the
  three-body charmless decays $B^+\to K^+ \pi^+ \pi^-$ and $B^+\to K^+ K^+
  K^-$}},
\newblock Phys. Rev. D71 (2005) 092003, hep-ex/0412066.

\bibitem{Garmash:2006fh}
A. Garmash et~al., [\,Belle Collaboration\,], {\em {Dalitz analysis of
  three-body charmless $B^0\to K^0\pi^+\pi^-$ decay}},
\newblock Phys. Rev. D75 (2007) 012006, hep-ex/0610081.

\bibitem{pdg_scalar1}
S. Spanier and N.A. Tornqvist, {\em {Scalar mesons}},
\newblock Eur. Phys. J. C15 (2000) 437.

\bibitem{pdg_scalar2}
S. Spanier and N.A. Tornqvist, {\em {Note on scalar mesons}},
\newblock J. of Phys. G33 (2006) 546.

\bibitem{pdg_scalar3}
S. Spanier, N.A. Tornqvist and C. Amsler, {\em {Note on scalar mesons}},
\newblock Phys. Lett. B667 (2008) 1.

\bibitem{pdg_pseudoscalar}
C. Amsler and A. Masoni, {\em {The $\eta(1405)$, $\eta(1475)$, $f_{1}(1420)$,
  and $f_{1}(1510)$}},
\newblock Phys. Lett. B667 (2008) 1.

\bibitem{pdg_tensors2}
C. Amsler, {\em {Note on non-$q\bar{q}$ Candidates}},
\newblock Phys. Lett. B592 (2004) 1.

\bibitem{Devons:1973rs}
S. Devons et~al., {\em {Observations of $\bar{p}p\to 3\pi^0$, $2\pi^0\eta$ at
  Rest}},
\newblock Phys. Lett. B47 (1973) 271.

\bibitem{Gray:1983cw}
L. Gray et~al., {\em {Evidence for a $\pi\pi$ Isoscalar Resonance Degenerate
  with the $f\,^\prime$ Produced in $\bar{p}n$ Annihilations at Rest}},
\newblock Phys. Rev. D27 (1983) 307.

\bibitem{Binon:1983ny}
F.G. Binon et~al., [\,Serpukhov-Brussels-Annecy(LAPP) Collaboration\,], {\em
  {$G(1590)$: A Scalar Meson Decaying Into Two $\eta$ Mesons}},
\newblock Nuovo Cim. A78 (1983) 313.

\bibitem{Binon:1984ip}
F.G. Binon et~al., [\,Serpukhov-Brussels-Annecy(LAPP) Collaboration\,], {\em
  {Study of $\pi^- p\to\eta\,^\prime \eta n$ in a Search for Glueballs}},
\newblock Nuovo Cim. A80 (1984) 363.

\bibitem{Gaspero:1992gu}
M. Gaspero, {\em {Evidence for the dominance of an $I^G~J^{PC} = 0^+~0^{++}$
  resonance in $\bar{p} n\to 2\pi^+ 3\pi^-$ annihilation at rest}},
\newblock Nucl. Phys. A562 (1993) 407.

\bibitem{StrohmeierPresicek:1998fi}
M. Strohmeier-Presicek et~al., {\em {$4\pi$ decay modes of the $f_0(1500)$
  resonance}},
\newblock Phys. Lett. B438 (1998) 21, hep-ph/9808228.

\bibitem{Ellis:1994ww}
J.R. Ellis et~al., {\em {Abundant $\phi$ meson production in $\bar{p}p$
  annihilation at rest and strangeness in the nucleon}},
\newblock Phys. Lett. B353 (1995) 319, hep-ph/9412334.

\bibitem{Locher:2001zc}
M.P. Locher, V.E. Markushin and S. von Rotz, {\em {Antiproton-nucleon
  annihilation, multistep processes and the OZI rule}},
\newblock Nucl. Phys. A684 (2001) 414.

\bibitem{Edwards:1981ex}
C. Edwards et~al., {\em {Observation of an $\eta\eta$ Resonance in $J/\psi$
  Radiative Decays}},
\newblock Phys. Rev. Lett. 48 (1982) 458.

\bibitem{Groom:2000in}
D.E. Groom et~al., [\,Particle Data Group Collaboration\,], {\em {Review of
  Particle Physics}},
\newblock Eur. Phys. J. C15 (2000) 1.

\bibitem{Aston:1987am}
D. Aston et~al., {\em {A Study of the $K^0_S K^0_S$ System in the Reaction $K^-
  p \to K^0_S K^0_S \Lambda$ at 11 GeV/$c$}},
\newblock Nucl. Phys. B301 (1988) 525.

\bibitem{Reinnarth:2003}
J. Reinnarth, {\em {Exotische Mesonen im Endzustand $2\pi^+ 2\pi^-\eta$ in der
  Antiproton-Proton-Vernichtung in Ruhe}},
\newblock Ph.D. Thesis, University of Bonn  (2003).

\bibitem{Bai:1990hs}
Z. Bai et~al., [\,MARK-III Collaboration\,], {\em {Partial wave analysis of
  $J/\psi \to \gamma K^{0}_{S} K^{\pm} \pi^{\mp}$}},
\newblock Phys. Rev. Lett. 65 (1990) 2507.

\bibitem{Amsler:2004rd}
C. Amsler et~al., {\em {Production and decay of $\eta\,^\prime(958)$ and
  $\eta(1440)$ in $\bar{p}p$ annihilation at rest}},
\newblock Eur. Phys. J. C33 (2004) 23.

\bibitem{Kirk:2000ws}
A. Kirk, {\em {Resonance production in central $p p$ collisions at the CERN
  Omega spectrometer}},
\newblock Phys. Lett. B489 (2000) 29, hep-ph/0008053.

\bibitem{Yuan:2005es}
C.Z. Yuan, {\em {Hadron Spectroscopy from BES and CLEO-c}},
\newblock AIP Conf. Proc. 814 (2006) 65, hep-ex/0510062.

\bibitem{Antinori:1995wz}
F. Antinori et~al., [\,WA91 Collaboration\,], {\em {A Further study of the
  centrally produced $\pi^+\pi^-$ and $\pi^+ \pi^- \pi^+ \pi^-$ channels in $p
  p$ interactions at 300 and 450~GeV/$c$}},
\newblock Phys. Lett. B353 (1995) 589.

\bibitem{Buccella:1996uy}
F. Buccella, M. Lusignoli and A. Pugliese, {\em {Charm non-leptonic decays and
  final state interactions}},
\newblock Phys. Lett. B379 (1996) 249, hep-ph/9601343.

\bibitem{pdg_charm}
D. Asner, {\em {Note on Charm Dalitz Plot Analyses}},
\newblock J. of Phys. G33 (2006) 716.

\bibitem{Aitala:2000xu}
E.M. Aitala et~al., [\,E791 Collaboration\,], {\em {Experimental evidence for a
  light and broad scalar resonance in $D^+\to\pi^-\pi^+\pi^+$ decay}},
\newblock Phys. Rev. Lett. 86 (2001) 770, hep-ex/0007028.

\bibitem{Malvezzi:2003jp}
S. Malvezzi, {\em {Light quark and charm interplay in the Dalitz-plot analysis
  of hadronic decays in FOCUS}},
\newblock AIP Conf. Proc. 688 (2004) 276, hep-ex/0307055.

\bibitem{Minkowski:2004xf}
P. Minkowski and W. Ochs, {\em {$B$ decays into light scalar particles and
  glueball}},
\newblock Eur. Phys. J. C39 (2005) 71, hep-ph/0404194.

\bibitem{Amsler:2002ey}
C. Amsler, {\em {Further evidence for a large glue component in the $f_0(1500)$
  meson}},
\newblock Phys. Lett. B541 (2002) 22, hep-ph/0206104.

\bibitem{Ambrosino:2006hb}
F. Ambrosino et~al., [\,KLOE Collaboration\,], {\em {Dalitz plot analysis of
  $e^+ e^- \to\pi^0\pi^0\gamma$ events at $\sqrt{s}\cong M_\phi$ with the KLOE
  detector}},
\newblock Eur. Phys. J. C49 (2007) 473, hep-ex/0609009.

\bibitem{Achasov:2005hm}
N.N. Achasov and A.V. Kiselev, {\em {Properties of the light scalar mesons face
  the experimental data on the $\phi\to\pi^0\pi^0\gamma$ decay and the $\pi\pi$
  scattering}},
\newblock Phys. Rev. D73 (2006) 054029, hep-ph/0512047.

\bibitem{Boglione:2003xh}
M. Boglione and M.R. Pennington, {\em {Towards a model independent
  determination of the $\phi\to f_0\gamma$ coupling}},
\newblock Eur. Phys. J. C30 (2003) 503, hep-ph/0303200.

\bibitem{Oller:2002na}
J.A. Oller, {\em {Finite width effects in $\phi$ radiative decays}},
\newblock Nucl. Phys. A714 (2003) 161, hep-ph/0205121.

\bibitem{Marco:1999df}
E. Marco et~al., {\em {Radiative decay of $\rho^0$ and $\phi$ mesons in a
  chiral unitary approach}},
\newblock Phys. Lett. B470 (1999) 20, hep-ph/9903217.

\bibitem{Palomar:2003rb}
J.E. Palomar et~al., {\em {Sequential vector and axial-vector meson exchange
  and chiral loops in radiative phi decay}},
\newblock Nucl. Phys. A729 (2003) 743, hep-ph/0306249.

\bibitem{Caprini:2005zr}
I. Caprini, G. Colangelo and H. Leutwyler, {\em {Mass and width of the lowest
  resonance in QCD}},
\newblock Phys. Rev. Lett. 96 (2006) 132001, hep-ph/0512364.

\bibitem{Oller:1997ti}
J.A. Oller and E. Oset, {\em {Chiral Symmetry Amplitudes in the $S$-Wave
  Isoscalar and Isovector Channels and the $\sigma$, $f_0(980)$, $a_0(980)$
  Scalar Mesons}},
\newblock Nucl. Phys. A620 (1997) 438, hep-ph/9702314.

\bibitem{Coffman:1989nk}
D. Coffman et~al., [\,MARK-III Collaboration\,], {\em {Study of the
  Doubly-Radiative Decay $J/\psi\to\gamma \gamma\rho^0$}},
\newblock Phys. Rev. D41 (1990) 1410.

\bibitem{Augustin:1989zf}
J.E. Augustin et~al., [\,DM2 Collaboration\,], {\em {Radiative Decay of
  $J/\psi$ into $\eta(1430)$ and Nearby States}},
\newblock Phys. Rev. D42 (1990) 10.

\bibitem{Jones:2000nq}
R.W.L. Jones, [\,ALEPH Collaboration\,], {\em {Search for the glueball
  candidates $f_0(1500)$ and $f_J(1710)$ in $\gamma\gamma$ collisions in
  ALEPH}},
\newblock Prepared for PHOTON 2000: International Workshop on Structure and
  Interactions of the Photon (Including 13th International Workshop on
  Photon-Photon Collisions), Ambleside, Lake District, England, 26-31 Aug 2000.

\bibitem{Baltrusaitis:1985pu}
R.M. Baltrusaitis et~al., [\,MARK-III Collaboration\,], {\em {Observation of a
  Narrow $K\bar{K}$ State in $J/\psi$ Radiative Decays}},
\newblock Phys. Rev. Lett. 56 (1986) 107.

\bibitem{Augustin:1987da}
J.E. Augustin et~al., [\,DM2 Collaboration\,], {\em {Radiative Decay of $J /
  \psi$ into $\gamma \pi^+ \pi^-$}},
\newblock Z. Phys. C36 (1987) 369.

\bibitem{Alde:1986nx}
D. Alde et~al., [\,Serpukhov-Brussels-Los Alamos-Annecy(LAPP) Collaboration\,],
  {\em {2.22 GeV $\eta \eta\,^\prime$ Structure Observed in 38~GeV/$c$ and
  100~GeV/$c$ $\pi^- p$ Collisions}},
\newblock Phys. Lett. B177 (1986) 120.

\bibitem{Hasan:1996jy}
A. Hasan and D.V. Bugg, {\em {A search for the $\xi(2235)$ in $\bar{p} p\to
  \pi^- \pi^+$}},
\newblock Phys. Lett. B388 (1996) 376.

\bibitem{Bardin:1987at}
G. Bardin et~al., {\em {Search for a Narrow Resonance about the $\xi(2230)$ in
  the Formation Channel $\bar{p}p\to K^+ K^-$}},
\newblock Phys. Lett. B195 (1987) 292.

\bibitem{Sculli:1987bg}
J. Sculli et~al., {\em {Limits on $\xi(2.2)$ Formation in $\bar{p}p\to K^+
  K^-$}},
\newblock Phys. Rev. Lett. 58 (1987) 1715.

\bibitem{Evangelista:1997be}
C. Evangelista et~al., {\em {Measurement of the $\bar{p}p\to K_S K_S$ reaction
  from 0.6~GeV/$c$ to 1.9~GeV/$c$}},
\newblock Phys. Rev. D56 (1997) 3803, hep-ex/9707041.

\bibitem{Evangelista:1998zg}
C. Evangelista et~al., [\,JETSET Collaboration\,], {\em {Study of the reaction
  $\bar{p}p\to\phi\phi$ from 1.1~GeV/$c$ to 2.0~GeV/$c$}},
\newblock Phys. Rev. D57 (1998) 5370, hep-ex/9802016.

\bibitem{Buzzo:1997vh}
A. Buzzo et~al., [\,JETSET Collaboration\,], {\em {Search for narrow $\bar{p}p$
  resonances in the reaction $\bar{p}p\to \bar{p} p \pi^+ \pi^-$}},
\newblock Z. Phys. C76 (1997) 475, hep-ex/9801015.

\bibitem{Stanton:1979ya}
N.R. Stanton et~al., {\em {Evidence for Axial Vector and Pseudoscalar
  Resonances Near 1.275~GeV in $\eta\pi^+\pi^-$}},
\newblock Phys. Rev. Lett. 42 (1979) 346.

\bibitem{Ando:1986bn}
A. Ando et~al., {\em {Evidence for two Pseudoscalar Resonances of
  $\eta\pi^+\pi^-$ System in the $D(1285)$ and $E/\iota$ Regions}},
\newblock Phys. Rev. Lett. 57 (1986) 1296.

\bibitem{Fukui:1991ps}
S. Fukui et~al., {\em {Study on the $\eta\pi^+\pi^-$ system in the $\pi^- p$
  charge exchange reaction at 8.95~GeV/$c$}},
\newblock Phys. Lett. B267 (1991) 293.

\bibitem{Alde:1997vq}
D. Alde et~al., [\,GAMS Collaboration\,], {\em {Partial-wave analysis of the
  $\eta\pi^0\pi^0$ system produced in $\pi^- p$ charge exchange collisions at
  100~GeV/$c$}},
\newblock Phys. Atom. Nucl. 60 (1997) 386.

\bibitem{Manak:2000px}
J.J. Manak et~al., [\,E852 Collaboration\,], {\em {Partial-wave analysis of the
  $\eta\pi^+\pi^-$ system produced in the reaction $\pi^- p\to\eta\pi^+\pi^- n$
  at 18 GeV/$c$}},
\newblock Phys. Rev. D62 (2000) 012003, hep-ex/0001051.

\bibitem{Adams:2001sk}
G.S. Adams et~al., [\,E852 Collaboration\,], {\em {Observation of pseudoscalar
  and axial vector resonances in $\pi^- p\to K^+ K^- \pi^0 n$ at 18 GeV}},
\newblock Phys. Lett. B516 (2001) 264, hep-ex/0107042.

\bibitem{Anisovich:2001jb}
A.V. Anisovich et~al., {\em {Resonances in
  $\bar{p}p\to\eta\pi^+\pi^-\pi^+\pi^-$ at rest}},
\newblock Nucl. Phys. A690 (2001) 567.

\bibitem{Edwards:1982nc}
C. Edwards et~al., {\em {Observation of a Pseudoscalar State at $1440$~MeV in
  $J/\psi$ Radiative Decays}},
\newblock Phys. Rev. Lett. 49 (1982) 259.

\bibitem{Scharre:1980zh}
D.L. Scharre et~al., {\em {Observation of the Radiative Transition
  $\psi\rightarrow\gamma E(1420)$}},
\newblock Phys. Lett. B97 (1980) 329.

\bibitem{Amsler:iota:2006}
C. Amsler and A. Masoni, {\em {Review of Particle Physics}},
\newblock J. Phys. G33 (2006) 591.

\bibitem{Dover:1990kn}
C.B. Dover, T. Gutsche and A. Faessler, {\em {The case for quasinuclear
  $N\bar{N}$ bound states}},
\newblock Phys. Rev. C43 (1991) 379.

\bibitem{Weinstein:1990gu}
J.D. Weinstein and N. Isgur, {\em {$K\bar{K}$ Molecules}},
\newblock Phys. Rev. D41 (1990) 2236.

\bibitem{Locher:1997gr}
M.P. Locher, V.E. Markushin and H.Q. Zheng, {\em {Structure of $f_0(980)$ from
  a coupled channel analysis of $S$-wave $\pi \pi$ scattering}},
\newblock Eur. Phys. J. C4 (1998) 317, hep-ph/9705230.

\bibitem{Close:2002zu}
F.E. Close and N.A. Tornqvist, {\em {Scalar mesons above and below 1~GeV}},
\newblock J. Phys. G28 (2002) R249, hep-ph/0204205.

\bibitem{Achasov:2008ut}
N.N. Achasov, A.V. Kiselev and G.N. Shestakov, {\em {Theory of Scalars}},
\newblock Nucl. Phys. B, Proc. Suppl. 181-182 2008 (2008) 169, 0806.0521.

\bibitem{Aitala:2000xt}
E.M. Aitala et~al., [\,E791 Collaboration\,], {\em {Study of the $D_s^+\to\pi^-
  \pi^+ \pi^+$ decay and measurement of $f_0$ masses and widths}},
\newblock Phys. Rev. Lett. 86 (2001) 765, hep-ex/0007027.

\bibitem{Molina:2008jw}
R. Molina, D. Nicmorus and E. Oset, {\em {The $\rho\rho$ interaction in the
  hidden gauge formalism and the $f_0(1370)$ and $f_2(1270)$ resonances}},
\newblock Phys. Rev. D78 (2008) 114018, 0809.2233.

\bibitem{Geng:2008gx}
L.S. Geng and E. Oset, {\em {Vector meson-vector meson interaction in a hidden
  gauge unitary approach}},
\newblock (2008), 0812.1199.

\bibitem{Minkowski:1998mf}
P. Minkowski and W. Ochs, {\em {Identification of the glueballs and the scalar
  meson nonet of lowest mass}},
\newblock Eur. Phys. J. C9 (1999) 283, hep-ph/9811518.

\bibitem{Bugg:2007ja}
D.V. Bugg, {\em {A Study in Depth of $f_{0}(1370)$}},
\newblock Eur. Phys. J. C52 (2007) 55, 0706.1341.

\bibitem{Klempt:1995ku}
E. Klempt et~al., {\em {Scalar mesons in a relativistic quark model with
  instanton induced forces}},
\newblock Phys. Lett. B361 (1995) 160, hep-ph/9507449.

\bibitem{Anisovich:2002ij}
V.V. Anisovich and A.V. Sarantsev, {\em {K-matrix analysis of the
  ($I\,J^{PC}=0\,0^{++}$)-wave in the mass region below $1900$~MeV}},
\newblock Eur. Phys. J. A16 (2003) 229, hep-ph/0204328.

\bibitem{Beveren:2008}
E. van Beveren and G. Rupp, {\em {The spectrum of scalar-meson nonets in the
  Resonance-Spectrum Expansion}},
\newblock (2008), arXiv:0804.2573.

\bibitem{Giacosa:2005qr}
F. Giacosa et~al., {\em {Scalar meson and glueball decays within a effective
  chiral approach}},
\newblock Phys. Lett. B622 (2005) 277, hep-ph/0504033.

\bibitem{Giacosa:2005zt}
F. Giacosa et~al., {\em {Scalar nonet quarkonia and the scalar glueball: Mixing
  and decays in an effective chiral approach}},
\newblock Phys. Rev. D72 (2005) 094006, hep-ph/0509247.

\bibitem{Cheng:2006hu}
H.Y. Cheng, C.K. Chua and K.F. Liu, {\em {Scalar glueball, scalar quarkonia,
  and their mixing}},
\newblock Phys. Rev. D74 (2006) 094005, hep-ph/0607206.

\end{thebibliography}
\end{document}